# THEORY AND APPLICATIONS OF FLUCTUATING–CHARGE MODELS

BY

JIAHAO CHEN

B.S., University of Illinois at Urbana-Champaign, 2002
M.S., University of Illinois at Urbana-Champaign, 2008

DISSERTATION

Submitted in partial fulfillment of the requirements
for the degree of Doctor of Philosophy in Chemical Physics
in the Graduate College of the
University of Illinois at Urbana-Champaign, 2009

Urbana, Illinois

Doctoral Committee:

Professor Todd Martínez, Chair
Professor Dirk Hundertmark
Professor Duane Johnson
Professor Nancy Makri

# Abstract

Fluctuating–charge models are computationally efficient methods of treating polarization and charge–transfer phenomena in molecular mechanics and classical molecular dynamics simulations. They are also theoretically appealing as they are minimally parameterized, with parameters corresponding to the chemically important concepts of electronegativities and chemical hardness. However, they are known to overestimate charge transfer for widely separated atoms, leading to qualitative errors in the predicted charge distribution and exaggerated electrostatic properties. We present the charge transfer with polarization current equilibration (QTPIE) model, which solves this problem by introducing distance–dependent electronegativities. A graph–theoretic analysis of the topology of charge transfer allows us to relate the fundamental quantities of charge transfer back to the more familiar variables that represent atomic partial charges. This allows us to formulate a unified theoretical framework for fluctuating–charge models and topological charge descriptors. We also demonstrate the important role of charge screening effects in obtaining correct size extensivity in electrostatic properties. Analyzing the spatial symmetries of these properties allows us to shed light on the role of charge conservation in the electronegativity equalization process. Finally, we develop a water model for use in classical molecular dynamics simulations that is capable of treating both polarization and charge transfer phenomena.

*In piam memoriam*

Chan Kim Leong (陳劍亮)

1948—2002

永垂不朽

Forever remembered

never diminished

## Acknowledgments

Firstly, my heartfelt gratitude to my advisor, Professor Todd “Genius” Martínez, who has wowed me many times over with his incredibly deep scientific intuition and broad familiarity with the field of computational and theoretical chemistry. Thanks also to Professor Nancy Makri, my undergraduate thesis advisor, who has inspired me with her distinctive synthesis of mathematical elegance and physical insight. Nancy and Todd have been my invaluable *in loco parenteis* at the University of Illinois at Urbana–Champaign, and to whom I owe much of my professional development as a scientist.

Thanks to the past and present members of the Martínez group whom I have had the pleasure to know personally and have taught me so much over the years—Joshua Coe, Lee Cremar, Hanneli Hudock, Benjamin “Prime” Kaduk, Chaehyuk Ko, Krissie Lamothe, Jeff Leiding, Ben Levine, Beth Lindquist, Crystal Manohar, Seth Olsen, Jane Owens, Mitchell Ong, Taras Pogorelov, Chutintorn “Por” Punwong, Christopher Schlosberg, Hongli Tao, Alexis Thompson, Ivan Ufimtsev, Kelly Van Haren, Sergey Varganov, Aaron Virshup, Siyang “Sandy” Yang, and Guishan Zheng. Special thanks to Aaron for being my long-suffering listening post, Ben for getting me started in the group, and Josh for patiently annotating my work with his patently stoic, no–nonsense attitude. Thanks also to Hongli and Guishan for teaching me many Chinese technical terms.

Thanks to Prof. Susan Atlas, Dr. Steven Valone and Professor Troy van Voorhis, with whom I have had the pleasure to discuss my thesis research and many other topics in computational chemistry.

Thanks to Professor Andrew Gewirth for his honest critiques, and whose interest in my work has sustained my own motivation when it has flagged.

Thanks to Professor Gregory Girolami, for his exemplary stewardship of the Chemistry department during his tenure as Chair, and for upholding the ideals of academic freedom in the face of external pressure.

Thanks to Professors Dirk Hundertmark and Richard Laugesen, who have taught me by example how mathematicians think, and how to conquer my fear of contemporary mathematics and the recital of definition, lemma and theorem.

Thanks to Professor Duane Johnson, who first taught me how to do computational science.

Thanks to Jeb Kegerreis, who has taken good care of my copy of Feynman and Hibbs.

Thanks to the members of the Theoretical and Computational Biophysics Group, particularly Peter Freddolino, Chris Harrison, John Stone, Professor Emad Takhorshjid, and Professor Klaus Schulten for our mutual scientific interests.

Thanks to the numerous Singaporeans and honorary Singaporeans over the years who have established an invaluable support network and fostered a second home in the cornfields; in particular Ray Chay, See Wee Chee, Kejia Chen, Esther Chiew, Qinwei Chow, Gavin Chua, Tuan Hoang, Muhammad Helmi Khaswan, Cindy Khoo, Freddy Lee, Sher– May Liew, Shirleen Lim, Christine Ng, Nam Guan Ng, Chin Chuan Ong, Willie Chuin Hong Ong, Zhun-Yong Ong, Joy Pang, Yee Lin Pow, Daniel Quek, Wenqiang Shen, Chong Kian Soh, Jeremy Tan, Louis Tay, Jason Teo, Kuan Khoon Tjan, Gerald Yew Tung Wan, Sang Woo, Michelle Toh–Wong, and Wee Hong Yeo.  Special thanks also go to Chiao-Lun Cheng and Shawn Lim.

Thanks to the Krannert Center Students' Association and Krannert Center Community Volunteers' Association, for making it possible to broaden my artistic horizons beyond anything I could have thought possible. Special mention must go to the loyal and dedicated volunteers: Joan and Richard Bazzetta, Adam Bussan, Felix Chan, Brian Cudiamat, Vasilica Crecea, Astrid Dussinger, Courtney Egg, Cyril Jacquot, Andy Mitofsky, Alexandra and Susan Wright, Meghan Wyllie, and Julie Yen, all of whom have made volunteering such a special experience.

Thanks to the outstanding support staff of the School of Chemical Sciences, particularly Connie Knight, Kathy Lankster, Beth Myler, Sandy Pijanowski, Lori Sage–Karlson, Theresa Struss and Vicki Tempel. Thanks for bringing the treats, stuffing the mailboxes and filing the paperwork.

Thanks to my past colleagues at DSO National Laboratories—especially Seok Khim Ang, Loke Yew Chew, Parh Jinh Chia, Wee Kang Chua, Cyrus Goh, Peter Goh, Kin Seng Lai, Ernest Wei–Pin Lau, Kenneth Eng Kian Lee, May Lee, Geok Kieng Lim, Han Chuen Lim, Jonathan Lim, Puay Ying Lim, Regine Oh, Selwyn Sean Scharnhorst, Beng Sing Tan, Joseph Soon–Thiam Tay, Kien Boon Teo, and Wee Hsiung Wee—who have never stopped believing in me.

Most importantly, I must thank my mother, Chiung–Li Chen, and my brother, Yuhao Chen, for surviving these difficult years together as an inalienable and inseparable family.

Special thanks must go to the many people who have helped me write the introductory review of electronegativity in Chapter 1, which has benefited immensely from the contributions of Dr. Steve Valone (LANL), who provided helpful advice and references that I had overlooked; Prof. William B. Jensen (McMicken College), who entertained a very insightful personal discussion over the premodern aspects of electronegativity; Prof. Thom Dunning, Jr. and Dr. David Woon, for discussions of the explorations of electronegativity effects in valence bond theory and generalized valence bond theory; Mr. Nicholas Cher Shiung Lee , who generously helped with the French translations; and Prof. Dirk Hundertmark, who contributed his native insight to the German literature and translations.

**Table of Contents**

# Chapter 1. Electronegativity and chemical hardness, and their relationship with fluctuating–charge models

## 1.1. The treatment of electrostatic phenomena in force fields

Molecular mechanics is one of the most successful computational tools available for the study of chemical systems, especially when used for classical molecular dynamics simulations. Molecular mechanics employs energy functions, more colloquially known as force fields, that contain various terms that represent the energetic contributions of various chemical phenomena such as chemical bond stretches, angle bending, torsional interactions, van der Waals interactions, and electrostatic interactions.[1] A typical force field takes the form

$$\begin{aligned} E = &\sum_{b \in \text{bonds}} k_b \left(R_b - \bar{R}_b\right)^2 + \sum_{a \in \text{angles}} \kappa_a \left(\theta_a - \bar{\theta}_a\right)^2 + \sum_{d \in \text{dihedrals}} \kappa_d^D \cos n_d \omega_d \\ &+ \sum_{i<j}^{N} 4\varepsilon_{ij} \left[\left(\frac{\sigma_{ij}}{R_{ij}}\right)^{12} - \left(\frac{\sigma_{ij}}{R_{ij}}\right)^{6}\right] + \sum_{i<j}^{N} \frac{q_i q_j}{R_{ij}} \end{aligned} \tag{1.1}$$

where $k_b$ is the bond stretch constant of the bond $b$, $R_b$ is the length of the bond $b$, $\bar{R}_b$ is the equilibrium length of the bond $b$, $\kappa_a$ is the angle bending constant of the angle $a$, $\theta_a$ is the value of the angle $a$, $\bar{\theta}_a$ is the equilibrium value of the angle $a$, $\kappa_d^D$ is the dihedral constant of the dihedral $d$, $n_d$ is the order of the dihedral interaction, $\omega_d$ is the value of the dihedral angle $d$, $\varepsilon_{ij}$ is the Lennard-Jones binding energy,[2] $\sigma_{ij}$ is the effective pairwise van der Waals interaction radius between atoms $i$ and $j$, $R_{ij}$ is the distance between atoms $i$ and $j$, and $q_i$ is the charge on atom $i$. Of particular interest in this work is the electrostatic interaction term, which most often takes the form appropriate for the electrostatic energy of two interacting point charges. In most force fields currently in use, a point charge is assigned to each atom, which is then treated as a parameter for that particular force field. The charges, once determined, remain unchanged over the course of the dynamical simulation. This is often referred to as the fixed–charge or frozen–charge approximation. This becomes problematic in many situations, e.g. when highly polarizable species come into close contact during the dynamics, which would in reality cause a distortion of the charge distribution.

Indeed, recent studies have found that such conventional force fields are increasingly inadequate for today's systems of interest, as phenomena such as polarization and charge transfer are neglected. The conventional wisdom is that such effects are small and negligible; however, there are well–documented examples whereby ignoring these effects can lead to qualitative errors in simulations.[3-7] Perhaps the most well–known example of polarization playing an important role in molecular modeling is Warshel and Levitt's seminal 1976 study of the lysozyme reaction, where it was discovered that the reaction intermediate was energetically unstable and did not form a local minimum unless polarization effects were taken into account.[6] The reaction intermediate was stabilized by the formation of induced dipoles in the enzyme–substrate complex, which would not have been treated correctly in a conventional force field. Another famous case study is Rick, Stuart and Berne's 1994 study of the hydration of the chloride ion in a small water droplet.[8] Using the nonpolarizable OPLS/AA force field, the chloride ion preferred to remain buried in the center of the droplet, which would be the result expected from the simple Onsager model.[9] However, the polarizable TIP4P-FQ water model developed in that work showed a clear preference for the ion to remain on the surface—an entropic effect that was consistent with the experimental evidence. Less appreciated, perhaps, is the importance of charge transfer effects. While they have long been recognized to be important for the modeling of ceramic materials and semiconductors,[10-13] charge transfer effects have thought to be of negligible importance in biomolecular systems. However, semiempirical energy decomposition studies by van der Vaart, Metz and coworkers appears to challenge this conventional wisdom: not only have they found that charge-transfer can be 3-5 times as important as polarization in contributing to protein-protein interactions,[14] but they have also found that the neglect of charge transfer can give rise to the wrong sign of the hydration energy in the solvated cold shock protein A system.[15] Consequentially, the literature on methods to incorporate polarization is more extensive than that for modeling charge transfer.

Two of the many popular types of methods for incorporating polarization are inducible dipoles,[5, 6, 16, 17] where additional variables are introduced to describe dipole moments induced by mutual polarization interactions; and Drude oscillators,[17-19] where polarization is described by the change in distance between the atomic nucleus and a fixed countercharge attached by a harmonic potential. However, neither of these methods are cannot provide a description of charge transfer, a process that is critical for charge defect reactions, charge migration or transport phenomena. This is in some sense surprising and contrary to physical intuition, as charge transfer is merely an extreme form of polarization: while polarization results in a redistribution of charge density within molecules, charge transfer is a redistribution of charge density across molecules.

In contrast, there are several classes of methods that exist for modeling both charge transfer and polarization effects: for example, fluctuating–charge models,[4, 10, 20, 21] which model polarization by recomputing the charge distribution in response to changes in geometry or external perturbations; empirical valence bond (EVB) methods,[7, 13, 22] which parameterize the energetic contributions of individual valence bond configurations; and effective fragment potential (EFP)–type methods,[23, 24] which use energy decompositions of *ab initio* data to construct parameterized effective potentials. We choose to study only fluctuating–charge models, as the other methods that treat both polarization and charge transfer are computationally far more costly. In EFPs, polarization is modeled using distributed, inducible dipoles while charge transfer is represented separately as a sum over antibonding orbitals of the electron acceptor. The latter necessitates *a priori* specification of the charge acceptors and donors, as well as the provision of parameters for every orbital being summed over. Not only is this description computationally expensive, but it also fails to provide a unified picture of polarization and charge transfer. In contrast, EVB does provide this unified treatment, but suffers from the exponential growth in the number of relevant valence bond configurations with system size. In contrast, fluctuating–charge models introduce only a modest computational cost over conventional fixed–charge force fields, even for large systems. Several of these methods have been used effectively in dynamics simulations for many different systems: *fluc*-q (FQ) in the TIP4P-FQ water model,[8, 25] and the charge response kernel for liquid–water interactions;[5, 26-33] Siepmann and Sprik's model for interfacial water;[34] the ES+ model[10], and QEq[35] in the universal force field (UFF)[36] and the reactive force field (ReaxFF)[11, 37] for oxides; EEM[38, 39] in the Delft molecular mechanics model (DFF)[40, 41] and the reactive force field (ReaxFF) method,[37] and the CHARMM C22 force field for biomolecular simulations;[42, 43] and the chemical potential equalization (CPE) method of York and Yang,[44] The wide variety of applications thus demonstrates their utility in describing polarization effects in classical molecular dynamics. In addition, fluctuating–charge models are theoretically appealing as they provide a unified treatment of polarization and charge transfer with only two parameters per atom. As will be discussed later, these parameters can be identified with the chemically important concepts of electronegativity[45-50] and (chemical) hardness,[51-54] which we will later see are the molecular analogues of the Fermi level and band gap respectively. These drive the redistribution of atomic charges in response to electrostatic interactions according to the principle of electronegativity equalization.[47-49, 55]

The development of fluctuating-charge models is closely intertwined with the history of the concepts of electronegativity and (chemical) hardness. The concept of electronegativity itself is old and arguably dates back almost to the dawn of modern chemistry.[56, 57] In fact, the early literature shows clear evidence for the rudiments of modern concepts such as electronegativity equalization and the early uses of quantitative electronegativity scales. To date, there has been no review of these concepts that shows the close relationship to fluctuating-charge models. For this reason, the rest of this chapter is dedicated to surveying the development of the concept of electronegativity, starting from its earliest recorded notions and eventually culminating in the modern formulation of electronegativity as understood in the context of quantum mechanical theories. We will also see how closely related concepts such as electronegativity equalization and chemical hardness play an integral role in the maturation of this concept, and understand the fundamental connection between these concepts and the development of fluctuating-charge models. Finally, we provide a generic formulation of fluctuating-charge models and survey the major extant models, both historical and modern.

## 1.2. Early notions of electronegativity and its equalization

On November 20, 1806, Sir Humphry Davy, FRS, MRIA, described the first major contribution of the English-speaking world to the nascent field of chemistry, namely that the electrical properties of matter are crucial to understanding chemical reactivity. The first of three Bakerian Lectures to the Royal Society of London, *On Some Chemical Agencies of Electricity*, describes Davy's many seminal experiments on electrochemistry, showing the power of electricity to break down water, minerals and other compounds into their elemental constituents.[58] Among other things, Davy's experiments decisively overturned the long-established notion of the elemental nature of water. It is in the later half of the Lecture, however, where Davy first speculated on the role of electricity in determining which chemicals react, and which do not. In Section VIII, *On the Relations between the Electrical Energies of Bodies, and their Chemical Affinities,* he compared the attraction of oppositely charged objects with the reaction of chemical substances:

> Amongst the substances that combine chemically, all those, the electrical energies of which are well known exhibit opposite states; thus, copper and zinc, gold and quicksilver, sulphur and the metals, the acid and alkaline substances, afford apposite instances; and supposing perfect freedom of motion in their particles or elementary matter, they ought, according to the principles laid down, to attract each other in consequence of their electrical powers… [The fact] different bodies, after being brought into contact, should be found differently electrified; [and] its relation to chemical affinity is, however, sufficiently evident. May it not be identical with it, and an essential property of matter?

Davy claimed that the similarities between electrical and chemical reactions are not only analogous, but are fundamentally related to each other. Davy's thesis was that electrical imbalance within matter is what causes substances to react with other, and furthermore was responsible for determining their relative tendencies toward chemical reaction:

> "Supposing two bodies, the particles of which are in different electrical states, and those states sufficiently exalted to give them an attractive force superior to the power of aggregation, a combination would take place which would be more or less intense according as the energies were more or less perfectly balanced; and the change of properties would be correspondently proportional.
>
> "This would be the simplest case of chemical union. But different substances have different degrees of the same electrical energy in relation to the same body[…]
>
> "When two bodies repellent of each other act upon the same body with different degrees of the same electrical attracting energy, the combination would be determined by the degree; and the substance possessing the weakest energy would be repelled; and this principle would afford an expression of the causes of elective affinity, and the decompositions produced in consequence."

Davy then went on to propose that if electrical imbalance were indeed the cause of chemical reactivity, it would be possible to classify and rank chemical substances by their electrical content. This would bring order and sensibility into the vast corpus of empirical chemical data for the first time in history. Again quoting from Davy's Lecture:

> "Allowing combination to depend upon the balance of the natural electrical energies of bodies, it is easy to conceive that a *measure* may be found of the artificial energies, as to intensity and quantity produced in the common electrical machine, or the Voltaic apparatus, capable of destroying this equilibrium; and such a measure would enable us to make a scale of electrical powers corresponding to degrees of affinity."

It is fascinating to note from the preceding paragraphs how Davy laid the foundations for our modern concept of electronegativity. In addition, Davy's thesis that chemical changes occur to restore electrical balance in matter is clearly a precursor to what we now call the principle of electronegativity equalization, namely that substances of different electronegativities react so as to produce compounds in which the elements have been in some sense electrically neutralized.

Davy's Bakerian Lecture influenced many of his contemporaries to focus on the fundamentals of chemical reactivity, and in particular to construct the classification of matter by its electrical content. The first major advance was made by Amadeo Avogadro, who created an oxygenicity scale that ranked chemical substances by their affinity with oxygen:[59]

> « Quoi-qu'il en soit de l'hypothèse sur l'identité de l'affinité avec l'action électrique, que l'auteur en déduit, elle nous montrent qu'il y a une étroite liaison entre l'antagonisme réciproque acide et alcalin, et al puissance motrice de l'électricité dans le contact de deux corps à la manière de Volta, l'acide prenant en ces cas l'électricité négative ou résineuse, et l'alcali, la positive ou vitreuse, et l'électricité, artificiellement communiquée à ces corps, favorisant ou empêchant leur combinaison, selon qu'elle concourt avec les électricités produits par le contact, ou qu'elle les contrarie ; et comme cette même faculté motrice de l'électricité a lieu entre tous les corps susceptibles de combinaison, que l'oxigène se comporte à la manière des acides, et l'hydrogène, à la manière des alcalis, et qu'en général les propriétés des composés à cet égard dépendent de celles de leurs composans, on ne peut guère douter qu'un antagonisme de même genre n'ait lieu entre tus ces corps, et de la manière que nous l'avons expliqué ci-dessus… Il est clair en effet que, d'après la correspondance indiquée, l'*hétérogénéité électrique*, par laquelle deux corps s'électrisent plus ou moins fortement dans le contact, devient la mesure de l'antagonisme ou affinité chimique entre ces corps...»

"Whatever the hypothesis on the identity of the link with electric action, the author deduces from the hypothesis that there is a close link between the mutual antagonism of acid and alkali, and the electromotive force of electricity in the contact between two bodies as described by Volta. In the present case, the acid takes the negative or "resinous" electricity and the alkali takes the positive or "vitreous"; the electricity, supplied artificially to these entities, either favors or prevents their reaction depending on whether it agrees with the electricity produced by the contact, or whether it opposes this electricity. And because the abovementioned electromotive property of electricity occurs in all entities that react, and because oxygen behaves like acids and hydrogen like alkalis and because the properties of compounds depend on those of their components with respect to this property, we can hardly doubt that an

> antagonism of a similar nature should take place between all these entities in the likes of what was explained above... it is obvious that, according to the link above, electrical heterogeneity, where two entities electrify each other (i.e. react with each other) to varying degrees while in contact, becomes the measure of the chemical affinity/antagonism between them..."

Avogadro's oxygenicity scale was based on his measurements of electrode potentials at which elemental deposits formed. In this respect, Avogadro's oxygenicity scale is not only a measure of acidity and alkalinity with respect to oxygen, but it is also a measure of electronegativities in the sense of Mulliken, as described below and as suggested by the term "*l'hétérogénéité électrique*" (electrical heterogeneity).

It is at this stage that two distinct notions of electronegativity developed, one electrical, the other thermochemical. Soon after Avogadro's oxygenicity scale was published, Berzelius coined the word 'electronegativity' in his influential essay of 1811, *Essai sur la nomenclature chimique* (*Essay on chemical nomenclature*).[60] In this and later essays, Berzelius constructed an electronegativity scale that could explain heats of reaction, or what we now call reaction enthalpies.[60, 61] However, the theory that he had developed to justify his scale turned out to be incompatible with the known laws of electrostatics, and was based on a summary of empirical data rather than any directly observable quantity.[56] Furthermore, Berzelius's theory could not be applied to the rapidly growing field of organic chemistry, causing chemists to lose interest.[57] Nevertheless, Berzelius's work exploring the relationship between electronegativity, chemical reactivity and enthalpy ultimately culminated in Pauling's thermochemical studies of electronegativity, which will be discussed shortly.[56, 57, 62, 63] Before we do so however, it should be noted that the discovery of the electron and atomic structure at the dawn of the twentieth century ushered in new perspectives on the nature of electricity. This allowed Avogadro's notion of oxygenicity, which was a property of the bulk element in an electrochemical apparatus, to be eventually related to an purely atomic characteristic, namely as a measure of the energetic ease of transferring charge. Johannes Stark was the first to point out this correlated well with atomic properties such as ionization potentials and electron affinities (then known as saturation tendencies):[57, 64]

> "We have been describing the tendency of the chemical elements to become saturated with respect to negative electrons. And this saturation tendency differs from element to element in keeping with the magnitude of its ionization energy. The greater the force with which a chemical atom holds on to its own electrons, the greater its ionization energy, and in general the greater its saturation tendency for additional negative electrons[...]
>
> Experience has shown that the ionization energy of the metals is smaller than that of hydrogen, and that this, in turn, is smaller than the metalloids. If one arranges the chemical elements in an increasing series according to their ionization energies, the so-called electropositive elements will be found at the beginning and the electronegative elements at the end."

In modern terms, Stark had proposed to quantify electronegativities using just ionization potentials, which measured how easy it was for chemical species to give up their electrons. Mulliken later proposed to use both pieces of information, the ionization potential and the electron affinity, to quantify electronegativities that would measure both electron losses and gains.[65, 66] We will see later how this insight developed into the modern view of electronegativity as the chemical potential of charge in molecular systems.[50, 67]

### 1.3. The Pauling and Mulliken scales of electronegativity

We now arrive at the modern era of negativity with Pauling's scale of 1923, which was designed to quantify "the power of an atom in a molecule to attract electrons to itself". Pauling's primary quantity of interest was the bond enthalpy, which he had shown to be approximately additive for covalent systems.[68] With this principle of additivity, Pauling reduced the study of thermochemistry to elementary thermodynamic cycles involving fundamental processes that broke or formed individual bonds.[45, 68] The thermodynamic energies of such elementary processes, termed bond enthalpies, were further broken down into covalent and ionic parts. Pauling reasoned that it was plausible to calculate the covalent term for a heteroatomic bond (A-B) as the average of the corresponding homoatomic bonds (A-A and B-B).[68] He then attributed the discrepancy between the pure covalent term and the actual bond enthalpy to an additional stabilization term $\Delta$ that quantified the contribution of ionic bonding, i.e.

$$\mathrm{BE}(\text{A-B}) = \frac{\mathrm{BE}(\text{A-A}) + \mathrm{BE}(\text{B-B})}{2} + \Delta(\mathrm{A,B}) \tag{1.2}$$

where BE is the bond enthalpy and $\Delta(\mathrm{A,B})$ is the stabilization term attributed to ionic interactions. Pauling further proposed that the ionic term could be analyzed in terms of electronegativity differences, and hence proposed an electronegativity scale based on a statistical analysis of the ionic terms using the empirical relation

$$\Delta(\mathrm{A,B}) = \left(\chi_A^{(\mathrm{P})} - \chi_B^{(\mathrm{P})}\right)^2 \tag{1.3}$$

where $\chi_A^{(\mathrm{P})}$ is a physical constant associated with atom A, and likewise for atom B. These constants are now known as the Pauling electronegativities. As Eq. (1.3) defines electronegativities up to a global constant, Pauling arbitrarily chose the electronegativity of hydrogen to be $\chi_H^{(\mathrm{P})} = 2.1$ so that his initial data set would have electronegativities that were all positive.

Pauling's work represents a significant development from Berzelius's original concept of electronegativity in two important respects. First, Berzelius's electronegativity refers to a property of the bulk element, whereas Pauling's electronegativity is a property of an atom in its molecular environment.[69] Second, Pauling's scale is quantitative and allowed for a semiquantitative explanation for periodic trends in bond enthalpy data.[62, 68] However, Pauling was unable to provide a complete theoretical foundation for his electronegativities. As discussed above, Pauling electronegativities are based on two principles, the additivity of bond enthalpies, and their partitioning into covalent and ionic terms. Pauling was able to justify the latter by considering the relative weights of the corresponding valence bond configurations,[68, 70] and Mulliken later provided another justification of Pauling's electronegativities in the context of molecular orbital theory.[66] Nevertheless, Pauling was unable to justify the former additivity principle of bond enthalpies from first principles. For this reason, it is sometimes said that Pauling's electronegativities are at best an empirical summary of the available thermochemical data of small compounds. Furthermore, Pauling later found that the ionic term $\Delta(\mathrm{A,B})$ in Eq. (1.2) was not always positive. This led him to propose the use of the geometric mean rather than the arithmetic mean to define a new ionic term[45]

$$\Delta'(\mathrm{A,B}) = \mathrm{BE}(\mathrm{A\text{-}B}) - \sqrt{\mathrm{BE}(\mathrm{A\text{-}A})\mathrm{BE}(\mathrm{B\text{-}B})} \qquad (1.4)$$

Pauling then defined a new scale of electronegativities using the formula

$$\left|\chi_A^{(\mathrm{P}')} - \chi_B^{(\mathrm{P}')}\right| = 0.208\sqrt{\Delta'(\mathrm{A,B})} \qquad (1.5)$$

along with the arbitrary choice of $\chi_H^{(\mathrm{P}')} = 2.1$, and where the arbitrary constant of 0.208 was chosen for maximal agreement with is original scale of electronegativities. Confusingly, the subsequent literature does not often distinguish between these P and $\mathrm{P}'$ scales.

Despite the progress in developing a quantitative scale, Pauling's electronegativity scale retains significant theoretical disadvantages. First, bond enthalpies themselves are not directly measurable quantities except only in very rare cases. Second, the underlying assumption of additivity[68] is at best approximate, as bond enthalpies exhibit significant variation over many chemical systems.[71] Third, the Pauling scale assigns a single value of electronegativity to each element regardless of the local environment and electronic structure. Such a scale, for example, conflates *sp*, $sp^2$ and $sp^3$ carbon systems, resulting in significant inaccuracies in the calculation of bond enthalpies. Later work by Hinze, Whitehead and Jaffé on orbital–specific electronegativities[48, 49, 72] showed that such calculations could be made more accurate by treating each hybridization state of carbon separately. However, this does not still account for variations in bond enthalpies on the order of tens of kilocalories per mole across a great many chemical systems, even when controlling for such differences in hybridization states.[71]

While Pauling's electronegativity scale is known to every chemist, it is Mulliken's scale that is theoretically better understood and is now taken to be the true measure of electronegativity by theoretical chemists. Unlike Pauling's scale, which is based on empirical relations in thermochemical data, Mulliken's scale is firmly rooted in intrinsic atomic electronic properties.[65] Following up from Stark's suggestion to use the ionization potential to quantify electronegativities, Mulliken proposed in 1934 to define electronegativities as a simple arithmetic mean of the ionization potential (IP) and electron affinity[73] (EA) of a species A, i.e.

$$\chi_{\mathrm{A}} = \frac{\mathrm{IP}(\mathrm{A}) + \mathrm{EA}(\mathrm{A})}{2} = \frac{E(A^{+}) - E(A^{-})}{2} \tag{1.6}$$

where $E(\mathrm{A})$ refers to the energy of the species A, and the second equality follows from the definitions of ionization potential and electron affinity.[73] In contrast to Stark's suggestion to use only the ionization potentials, which is a measure of the ease of losing an electron, Mulliken proposed to include also information about the ease of gaining an electron by averaging with the electron affinity as well. Mulliken termed them absolute electronegativities, or electroaffinities, so as to distinguish them from Pauling's electronegativities.

Mulliken's electronegativity has close relationships to other well-known observables in solid state physics. For example, the negative of the Mulliken electronegativity has been found to be an excellent approximation to the workfunction of a metal.[74] Also, if we assume that Koopmans's theorem is valid,[75, 76] it follows directly from the one–electron approximation and neglect of orbital relaxation effects that $\mathrm{IP} = -\varepsilon_{\mathrm{HOMO}}$ and $\mathrm{EA} = -\varepsilon_{\mathrm{LUMO}}$ where $\varepsilon_{\mathrm{HOMO}}$ is the energy of the highest occupied molecular orbital (HOMO) and $\varepsilon_{\mathrm{LUMO}}$ is the energy of the lowest unoccupied molecular orbital (LUMO). Then, Mulliken's electronegativity is equivalent to

$$\chi_A = -\frac{\varepsilon_{\mathrm{HOMO}} + \varepsilon_{\mathrm{LUMO}}}{2} \tag{1.7}$$

This is nothing more than the molecular analogue of the Fermi level. This insight bears special resonance with the modern treatment of electronegativity, which identifies it as the negative of the chemical potential of charge.[50] Importantly, Mulliken had defined his electronegativity scale using ionization potentials and electron affinities relative to specific electronic states as defined by van Vleck valence states,[65, 66, 77] a detail that has since been often overlooked. Later attempts to extend Mulliken's original calculations to other elements and chemical species have largely ignored the important question of the relevant electronic states of A, $A^+$ and $A^-$ being considered, instead relying solely on experimentally determined ionization potentials and electron affinities, which Mulliken had strenuously cautioned against in his original papers. Pritchard and Skinner had criticized the early literature for this oversight, pointing out that such studies might "be misleading in cases where the ionization potential of the ground–state of an atom is far removed (in energy) from the ionization potential of the atom in its appropriate valence–state (e.g. Zn, Cd, Hg)."[78] Identifying experimental quantities to the theoretical valence–state–specific counterparts is justifiable only in the case of some isolated atoms in their ground states, and it is not theoretically justifiable to relate the electronegativities of isolated atoms to those of atoms in molecules,[66, 79, 80] as their electronic states are different. However, the generalization of the van Vleck valence state[77] to arbitrary systems have proven difficult and complicated,[81-83] and the question of correctly defining atoms in molecules remains open.

Even with this gross simplification of using only experimental atomic ground state data, Mulliken's electronegativities were not widely used until recently, because electron affinities are notoriously difficult to measure accurately owing to the difficulty of producing stable anions in the gas phase.[84] The difficulty in determining Mulliken's electronegativity had significantly hampered its acceptance among experimentalists since its introduction. Even in the cases where experimental data were available, critics have commented on apparent discrepancies between the experimentally determined values and the expectations of chemical intuition.[82, 85-87] Arguably, it is precisely because of the difficulty of calculating and experimentally measuring Mulliken electronegativities, and the increasingly obvious flaws of Pauling's original electronegativity scale, that stimulated the proliferation of many electronegativity scales,[46, 55, 88-91] the discussion of which is beyond the scope of this discussion.

## 1.4. The variation of electronegativity with charge

Pauling was acutely aware that his electronegativity scales did not adequately capture the variation of atomic electronegativities with their charges. In *The Nature of the Chemical Bond*,[45] Pauling wrote on p. 65 about how the electronegativities provided needed corrections to estimate "the effect of formal charge on the [electronegativity] values". Pauling used the example of comparing an amine nitrogen (N in $NR_3$) and an ammonium nitrogen (N in $NR_4^+$, which he treated as a formal ion $N^+$). Pauling proposed that the electronegativity of $N^+$ should be very closely related that of the right neighbor in the periodic table, namely oxygen. However, the additional electron the neutral atom having the same had previously estimated from X-ray diffraction data that the screening effect of the valence electron lost when going from N to $N^+$ [92] to be $c$ = 0.4, being also an estimate of the ionic character of the bonds formed by the nitrogen atom. Based on this, he estimated $\chi_{N^+}^{(P)}$, the electronegativity of $N^+$, as a linear interpolation between neutral nitrogen and its neighboring element, oxygen:

$$\chi_{N^+}^{(P)} = (1-c)\,\chi_{N}^{(P)} + c\chi_{O}^{(P)} \tag{1.8}$$

Pauling's initial study was followed up in 1946, when Daudel and Daudel attempted to analyze the deficiencies of the Pauling electronegativity scale:[93]

> « Pauling admet cependant que l'électronégativité d'un atome dépend de la charge électrique qui entoure son noyau. Il admet, par exemple, que l'électronégativité de l'ion $N^+$ est 3,3, alors que celle de l'azote n'est que 3. Or, il est bien facile de voir que la charge entourant le noyau d'un élément donné dépend de la molécule à laquelle il appartient... On doit donc s'attendre à ce que l'électronégativité d'un élément varie avec la molécule dans laquelle il se trouve. D'un autre côté, Pauling remarque que sa méthode perd en précision dès qu'elle est appliquée à une molécule dans laquelle la différence d'électronégativité entre les éléments qui la constituent dépasse 1,5. Or, c'est précisément dans ce cas que l'effet de charge est important... »

> "Pauling admits that the electronegativity of an atom depends on the electrical charge that surrounds the nucleus. He admits, for example, that the electronegativity of the $N^+$ ion is 3.3, while that of nitrogen is only 3. However, it is easy to see that the charge surrounding the nucleus of a given element depends on the molecule to which it belongs... We must therefore expect that the electronegativity of an element varies with the molecule to which it is located. On the other hand, Pauling noted that his method loses accuracy when it is applied to a molecule in which the difference in electronegativity between the elements which constitute it exceeds 1.5. It is precisely in this case that the effect of charge is important."

In this same work, Daudel and Daudel wrote down what we now recognize today as the very first fluctuating-charge model in history. Their key observation was that ionic character of bonds varies simultaneously with the electronegativity differences of the participating atoms, and so the original calculations of Pauling need to be iterated to self–consistency.[93] For illustrative purposes, we consider a diatomic molecule AB with A more electronegative than B. In modern terms, Daudel and Daudel defined the charges in terms of the ionicity

$$\gamma_{AB} = -q_A = q_B \tag{1.9}$$

where the ionicity is related to the electronegativity difference using Pauling's empirical relation[68]

$$\gamma_{\mathrm{AB}} = 1 - \exp\left[-k\left(\chi^{(\mathrm{DD})}_{\mathrm{A}^{q_A}} - \chi^{(\mathrm{DD})}_{\mathrm{B}^{q_B}}\right)^2\right] \tag{1.10}$$

with the fitting parameter taking the value $k$ = 0.25. The electronegativity $\chi^{(\mathrm{DD})}_{\mathrm{A}^{q_A}}$ refers to the electronegativity of the partially charged species $A^{q_A}$, which are calculated using the formula

$$\chi^{(\mathrm{DD})}_{\mathrm{A}^{+q_A}} = \chi^{(\mathrm{DD})}_{\mathrm{A}^0} + q_A \delta^{(\mathrm{DD})}_{A^+} \tag{1.11}$$

where $\chi_{A^0}^{(DD)}$ is the reparameterized electronegativity of the neutral species $A^0$, and $\delta_{A^+}^{(DD)}$ is a new parameter quantifying the change in electronegativity per unit charge, and was calculated in a way similar to Pauling's original argument in Eq. (1.8). In other words, $\delta_{A^+}^{(DD)}$ is the (chemical) hardness of the species A, even though it was not recognized for a long time that the hardness is a fundamentally different parameter from the electronegativity. While Daudel and Daudel focused on the electronegativities that resulted from their calculation, the self–consistent nature of the calculation also means that their model is also a method for calculating the magnitudes of these atomic charges as well. It is straightforward to see that Daudel and Daudel's atomic electronegativities do not become equal, although the difference between them usually reduces in magnitude.

Independently, Sanderson proposed his famous electronegativity equalization principle in 1951, in which the atomic electronegativities are equalized when atoms interact to form stable molecules.[55] Using his own rather quirky electronegativity scale, Sanderson parlayed this notion of electronegativity into his later textbooks, and in this way was influential in bringing the concept of electronegativity equalization into the mainstream of inorganic chemistry.[94-102] As we shall see in the next section, this stimulated the further discussion and development of the principle of electronegativity equalization.[103] Del Re was the first to use this principle to construct a charge model.[104] Later, Parr and coworkers would prove that for any fermionic system in its ground electronic state, the principle of electronegativity equalization follows naturally from density functional theory. Furthermore, the electronegativity of every orbital is equal, not just the electronegativities of the valence orbitals.[50] These early charge models have the distinct disadvantage of equitably distributing charges among all atoms of the same elemental type, which is clearly not always applicable.[105] This prompted the work of Gasteiger and Marsili, who proposed a charge model based on a notion of partial equalization of orbital electronegativities (PEOE).[106] These models simulated the flow of charge in bond formation by imposing pairwise equalization of electronegativities.[106, 107] While still in use to day, this model is no longer considered to be theoretically sound due to Parr and coworker's proof of full equalization of electronegativities.[50] Furthermore, Nalewajski,[79, 108] and Mortier *et al.* [39] discovered that a fluctuating-charge model that also took into account off-diagonal terms corresponding to charge-charge electrostatic interactions could overcome the earlier unphysical effects of equitable charge distribution. This was seen to have resolved the debate about full vs. partial equalization.

### 1.5. Modern concepts of electronegativity and chemical hardness

We now regard electronegativities as the change in energy as the amount of charge on the system changes. This notion is given precise meaning in density functional theory, and it turns out to be very closely related to Mulliken's original notion of electronegativities. The earliest development toward our modern understanding of electronegativity came in 1961, when Iczkowski and Margrave proposed an expansion of the energy of an atom in a power series with respect to its charge:[47]

$$E(q) = E(0) + \left(\frac{\partial E}{\partial q}\right)_{q=0} q + \frac{1}{2}\left(\frac{\partial^2 E}{\partial q^2}\right)_{q=0} q^2 + \cdots \tag{1.12}$$

In particular, they truncated these series to second order and evaluated this energy expansion at $q = \pm 1$, leading to

$$E(1) = E(0) + \left(\frac{\partial E}{\partial q}\right)_{q=0} + \frac{1}{2}\left(\frac{\partial^2 E}{\partial q^2}\right)_{q=0} + \cdots \tag{1.13}$$

$$E(-1) = E(0) - \left(\frac{\partial E}{\partial q}\right)_{q=0} + \frac{1}{2}\left(\frac{\partial^2 E}{\partial q^2}\right)_{q=0} + \cdots \tag{1.14}$$

Solving these two equations for the first derivative then leads to

$$\left(\frac{\partial E}{\partial q}\right)_{q=0} = \frac{E(1) - E(-1)}{2} \tag{1.15}$$

which then shows that to $O\left(\left(\partial^3 E / \partial q^3\right)_{q=0}\right)$, the Mulliken electronegativity measures how the energy of an atom changes with its charge:

$$\chi_{\mathrm{A}} = \left(\frac{\partial E(A)}{\partial q}\right)_{q=0} \tag{1.16}$$

In other words, the Mulliken electronegativity is the atomic analogue of an electrical potential. In doing so, Iczkowski and Margrave rediscovered an earlier fact discovered by Pritchard and Sumner that the Mulliken electronegativity could be recovered by a three-point quadratic approximation to the derivative above.[109] Later, Perdew, Parr, Levy and Balduz would show that Eq. (1.15) is in fact exact for noninteracting systems in quantum mechanics, owing to derivative discontinuities in the energy function as a function of total particle number.[110-112]

Hinze, Whitehead and Jaffé made the next significant advance by extending the concept of electronegativities to individual orbitals.[48, 49, 72] They proposed to define orbital electronegativities as the change in the atomic energy as the occupation of that orbital was varied,[72] i.e.

$$\chi^{(\mathrm{HWJ})} = \frac{\partial E}{\partial n} \tag{1.17}$$

This notion bears very close relationship with the later developments of Janak's theorem in density functional theory,[113-116] and fractional occupation number methods.[117-129] In addition, Hinze and coworkers asserted that the electronegativities of valence orbitals participating in bonding interactions must be equal, and in so doing formulated the first modern version of the principle of electronegativity equalization.[72] However, the validity of the equalization of orbital electronegativities was not shown rigorously until 1995 by Liu and Parr.[115] In addition, neither Iczkowski and Margrave, nor Hinze and coworkers, provided an interpretation for the physical content of these electronegativity terms. This was provided by Klopman in 1964, who showed that in the framework of semiempirical theory, electronegativities are given by the one-electron nuclear-electron Coulomb integrals, plus the contribution of the Hartree-Fock mean field.[130] Klopman was also the first to point out that the energy of an atomic system could be nondifferentiable with respect to orbital occupations.[130] Nevertheless, he showed that despite this, the equalization of orbital electronegativities could be formulated successfully.[130-133] Next, Gyftopoulos and Hatsopoulos in 1968 showed that the orbital-based electronegativities of Hinze and Jaffé could be reconciled with a statistical mechanical treatment of the molecular environment with atoms being considered as part of a grand canonical ensemble.[67] Importantly, they showed from thermodynamic considerations that the chemical potential of charge at zero temperature is equal to the Mulliken electronegativity if the ground state is not degenerate.

The last major development in the concept of electronegativity leading up to current times occurred in 1978, when Parr and coworkers showed that the Mulliken definition, and its interpretation as the first derivative of the energy with respect to a change in charge, can be reconciled within the framework of Hohenberg-Kohn density functional theory as[50]

$$-\chi = \mu = \left(\frac{\partial E}{\partial N}\right)_v = v(r) + \frac{\delta F[\rho(r)]}{\delta \rho} \tag{1.18}$$

where $\mu$ is the chemical potential, $N$ is the number of electrons in the system, $v$ is the external potential of the nuclei felt by the electrons, $r$ is an arbitrary position in real space, $F$ is the Hohenberg-Kohn universal functional, and $\rho$ is the electronic charge density. The first two equalities are essentially those of Iczkowski and Margrave, whereas the third equality shows the separate contributions of nuclear-electronic attraction (in $v$) and the electron-electron interactions as given by the derivative of the universal functional. It is important to note that the third equality must hold everywhere in space, and therefore the chemical potential is everywhere constant. For this to be true the variation in the nuclear potential must be exactly canceled by a counteracting variation in the derivative $\delta F / \delta \rho$. The validity of this statement is closely related to the existence and treatment of the derivative discontinuity in density functional theory,[110, 111] a topic that remains controversial even today.[13, 134-140]

It was also at this time that the notion of (chemical) hardness matured into a quantitative concept. The notion of chemical hardness was first introduced in inorganic chemistry by Pearson, in the context of hard and soft acids and bases (HSAB).[51-53, 141, 142] Pearson used the term ‘hardness’ loosely to mean the relative ease of nucleophilic substitutions, which is very closely allied to the concept of Lewis acidity.[51, 141] It is interesting to note how the development of the concept of chemical hardness parallels that of the preceding discussion on electronegativity. When the term was first introduced, the concept of hardness and softness were applied to acids and bases, and in a qualitative fashion to classify substances into hard, soft and borderline, and all for the purpose of understanding the nature of chemical reactivity. The quantitative breakthrough came in 1983, when Parr and Pearson proposed a definition of the chemical hardness as[54]

$$\eta = \frac{1}{2}\frac{\partial^2 E}{\partial N^2} = \frac{1}{2}\left(\mathrm{IP} - \mathrm{EA}\right) \tag{1.19}$$

where the second equality holds in the same finite–difference sense as the Iczkowski and Margrave study of the Mulliken electronegativity. While the factor of ½ was initially chosen by Parr and Pearson to give a superficial symmetry to the Mulliken formula for electronegativity, it has turned out to be notationally far more convenient to drop this numerical prefactor, and the modern literature has overwhelmingly chosen to do so. Thus in line with modern usage, the prefactor of ½ is discarded in this work. Again, the Parr-Pearson definition is attractive due to its close relationship with experimental atomic observables. Again assuming Koopmans's theorem is valid,[75, 76] it is straightforward to see that the chemical hardness is equivalent to

$$\eta = \frac{\varepsilon_{\mathrm{LUMO}} - \varepsilon_{\mathrm{HOMO}}}{2} \tag{1.20}$$

This is the molecular analogue of half the band gap,[111] just as the electronegativity is the molecular analogue of the Fermi level. It is also not difficult to relate the hardness with the self-repulsion energy,[119, 143] $U$, an empirical parameter for electron correlation that plays a prominent role in density functional tight-binding (DFTB) theories,[144-146] the Hubbard model,[147] and DFT+$U$.[148] Huheey's early papers also show clear evidence of recognizing the significance of the chemical hardness by relating it to the notion of charge capacitance,[82, 94, 149-151] an observation that has also been made by others.[152-155] In Chapter 3, we will develop this notion further into a complete description of fluctuating–charge models as molecular versions of electrical circuits.[135, 150]

Parr and Pearson were not the first to have investigated the chemical hardness, as early evidence for chemical hardness was provided by Gyftopoulos and Hatsopoulos in 1968, [67] where they showed that in the context of Hartree-Fock theory, the energy of the valence orbital as a function of the mean occupation number is given by

$$\begin{aligned}\varepsilon_0\left(q\right) &= \frac{\mathrm{IP}+\mathrm{EA}}{2}q+\frac{\mathrm{IP}-\mathrm{EA}}{2}\left(1-\sqrt{1-q^2}\right)\\ &= \frac{\mathrm{IP}+\mathrm{EA}}{2}q+\frac{\mathrm{IP}-\mathrm{EA}}{2}\left(\frac{1}{2}q^2\right)+O\left(q^4\right)\end{aligned} \tag{1.21}$$

where on the second line we expanded the first line in a Maclaurin series. The coefficient in front of the quadratic term is none other than the Parr-Pearson formula for the chemical hardness.[54] However, Parr and Pearson were indeed the first to identify the notion of chemical hardness as separate and distinct from that of electronegativity. The concept of chemical hardness has also been given rigorous meaning in density functional theory as a second-order functional derivative.[79, 156, 157] It was also quickly recognized from these studies that the chemical hardness could be thought of as a diagonal analogue of the Coulomb interactions, as they are of the same order of expansion in the Iczkowski and Margrave series. It is therefore sometimes convenient to consider the chemical hardness as a response matrix that treats (and possibly generalizes) both the diagonal hardnesses and the charge–charge interactions.[79, 108, 157-159] This perspective of the hardness matrix as a linear response kernel that determines the first-order change in the charge density in response to a change in the external potential has been adopted by York and Yang in the parameterization their chemical potential equalization (CPE) method,[44]

Morita and Kato in their charge response kernel method,[29-32] and Banks *et al.* in the development of the OPLS-FQ polarizable force field.[160]

The modern era of fluctuating-charge models began with the discovery that considering charge-charge interactions in fluctuating-charge models eliminated the principal flaw of Sanderson's electronegativity equalization scheme, namely that all atoms of the same elemental type would receive the same charge regardless of molecular environment.[39, 108, 161, 162] It was also shown that the difference between partial equalization and full equalization models were very small once this refinement was made.[39, 79] Henceforth, the debate over full vs. partial equalization became largely irrelevant. It is at this point where the first modern fluctuating–charge model, EEM, was developed, taking into account the contributions of both electronegativity and hardness.[38, 39] We now proceed to the description of the general features of these models.

### 1.6. Formulation of modern fluctuating-charge models

In fluctuating-charge models, the energy is formally expanded in the charge distribution

$$E\left(q_1,\ldots,q_N\right) = E_0 + \sum_{i=1}^{N} q_i \chi_i + \tfrac{1}{2}\sum_{i=1}^{N}\sum_{j=1}^{N} q_i q_j \eta_{ij} + \cdots \tag{1.22}$$

where $E_0$ is the charge-independent component of the energy, $q_i$ is the partial charge on atom $i$, and $\chi_i$ and $\eta_{ij}$ are the first- and second-order coefficients of the expansion respectively. From the preceding discussion, we interpret the first-order expansion coefficient $\chi_i$ as the electronegativity of atom $i$, and the second-order expansion coefficients $\eta_{ij}$ as the hardness matrix, where the diagonal element $\eta_{ii}$ is the chemical hardness of atom $i$, and the off-diagonal element $\eta_{ij}$, $i \neq j$ represents pairwise interactions between atom $i$ and atom $j$.

In virtually all fluctuating-charge models, the expansion is truncated to second order in the charges.[21] In addition, the charge-independent term $E_0$ is often discarded for simplicity. The charge distribution is then obtained by minimizing the energy function with respect to the charge distribution, but subject to one constraint, namely that of charge conservation:

$$\sum_{i=1}^{N} q_i = Q \tag{1.23}$$

The advantage of truncating the energy expansion in Eq. (1.22) is that the constrained minimization problem can be formulated as a system of linear equations. In order to enforce the constraint Eq. (1.23), introduce a Lagrange multiplier $\mu$. As the Lagrange multiplier enforces a number conservation constraint, $\mu$ can be interpreted as the chemical potential, and its use to enforce charge conservation is equivalent to applying the principle of electronegativity equalization.[49, 50, 72] This transforms the constrained minimization problem into an unconstrained minimization problem in $N + 1$ variables, where the function to be minimized is

$$\begin{aligned} F\left(q_1, \ldots, q_N; \mu\right) &= E\left(q_1, \ldots, q_N\right) - \mu\left(\sum_{i=1}^{N} q_i - Q\right) \\ &= \sum_{i=1}^{N} q_i\left(\chi_i - \mu\right) + \tfrac{1}{2}\sum_{i=1}^{N}\sum_{j=1}^{N} q_i q_j \eta_{ij} + \mu Q \end{aligned} \tag{1.24}$$

The function $F$ is the Legendre transformation of the energy $E$, where the total charge $Q$ is replaced by the chemical potential $\mu$. Therefore we can interpret $F$ as a free energy.

Minimizing this free energy with respect to its variables leads to requiring that the first derivatives all vanish:

$$0 = \frac{\partial F}{\partial q_i} = \chi_i - \mu + \sum_{j=1}^{N} q_j \eta_{ij} \tag{1.25}$$

$$0 = \frac{\partial F}{\partial \mu} = -\sum_{i=1}^{N} q_i + Q \tag{1.26}$$

As expected, Eq. (1.26) is identical to the constraint equation Eq. (1.23). Solving this set of simultaneous equations thus yields the desired charge distribution $(q_1, \ldots, q_N)$. As a final detail, we note that for the charge distribution to be a minimal solution to this set of equations, rather than some other kind of extremum, the second derivative test requires that the matrix $\left(\partial^2 F / \partial q_i \partial q_j\right)_{i,j=1}^{N} = \boldsymbol{\eta}$ has eigenvalues that are all non-negative.

It is illustrative to rewrite the problem in matrix-vector notation. Introduce the column vectors $\mathbf{q} = (q_1, \ldots, q_N)^T$ and $\mathbf{1} = (1, \ldots, 1)^T$ as well as the hardness matrix $\boldsymbol{\eta} = (\eta_{ij})_{i,j=1}^{N}$. Then the energy functions are

$$E(\mathbf{q}) = \mathbf{q} \cdot \boldsymbol{\chi} + \tfrac{1}{2} \mathbf{q}^T \boldsymbol{\eta} \mathbf{q} \tag{1.27}$$

$$F(\mathbf{q}, \mu) = \mathbf{q} \cdot (\boldsymbol{\chi} - \mu \mathbf{1}) + \tfrac{1}{2} \mathbf{q}^T \boldsymbol{\eta} \mathbf{q} \tag{1.28}$$

and Eqs. (1.25) and (1.26) that determine the charge distribution **q** and the chemical potential $\mu$ can be written as the matrix equation

$$\begin{pmatrix} \boldsymbol{\eta} & \mathbf{1} \\ \mathbf{1}^T & 0 \end{pmatrix} \begin{pmatrix} \mathbf{q} \\ \mu \end{pmatrix} = \begin{pmatrix} -\boldsymbol{\chi} \\ Q \end{pmatrix} \tag{1.29}$$

The major extant fluctuating-charge models differ mostly in minor details in the specification of the expansion coefficients $\chi_i$ and $\eta_{ij}$. Of these, the electronegativities and diagonal hardnesses are almost always given as parameters and the variations are almost exclusively in the functional form of the charge–charge interactions. The original electronegativity equalization model, EEM, uses classical Coulomb interactions for the charge–charge interactions.[38, 39] However, it was quickly recognized that this led to numerical instabilities in the model, even when far away from the coincidence limit of zero interatomic distance. Therefore, the EEM model was quickly modified to incorporate screening effects in later applications, e.g. in ReaxFF.[37] The first such model was the charge equilibration (QEq) method of Rappé and Goddard,[35] where the off-diagonal hardnesses were given by two-electron Coulomb integrals over *s*-type Slater orbitals[163]

$$\eta_{ij} = J_{ij}\left(R_i, R_j\right) = \iint_{\mathbb{R}^{3\times 2}} \frac{\phi_i^2\left(r_1; R_i\right)\phi_j^2\left(r_2; R_j\right)}{\left|r_1 - r_2\right|} dr_1\, dr_2 \qquad (1.30)$$

$$\phi_i\left(r; R\right) = \phi_i^{\mathrm{S}}\left(r; R\right) = \frac{\left(2\zeta_i\right)^{n_i+\frac{1}{2}}}{\sqrt{4\pi\left(2n_i\right)!}}\left|r - R\right|^{n_i - 1} e^{-\zeta_i\left|r-R\right|} \qquad (1.31)$$

where $n_i$ is the principal quantum number of atom $i$ and $\zeta_i$ is the Slater orbital exponent. Similar integrals are used in the fluc-*q*[8, 25] and ES+ [10] models. The chemical potential equalization (CPE) model uses two-electron Coulomb integrals over Gaussian-type atomic orbitals with empirical parameters for Fukui function corrections, and can be extended to orbitals with higher angular momenta. [44] In the CHARMM C22 force field,[42, 43] the Coulomb interactions are screened with empirical functions. All of these schemes can be considered approximations to the exact integral

$$J_{ij} = \iint_{\mathbb{R}^{3\times 2}} \frac{f_i\left(r_1\right) f_j\left(r_2\right)}{\left|r_1 - r_2\right|} dr_1\, dr_2 \qquad (1.32)$$

where $f_i$ is the Fukui function[164] of atom (or orbital) $i$, that arises from a density functional treatment of electronegativity equalization.[165] This generalizes even the original formulation of the EEM model, as the classical Coulomb interaction is recovered with delta function basis functions.

The discussion of this introduction brings us reasonably close to the current state of the art in the field of fluctuating-charge models. However, there remain long-standing problems with these models that limit their total usefulness. One of the longest-standing problems, that of artificially high intermolecular charge transfer, will be discussed in Chapter 2. We analyze the origin of the problem and introduce the QTPIE charge model that we have developed that does not suffer from this problem. However, we have had to make a change of variables away from working with atomic charges as our fundamental quantity. Chapter 3 outlines the numerical difficulties faced in the computations and details computational algorithms and methods for reducing the computation cost of a factor of *ca.* 10. A deeper study of the relationship between the new charge transfer variables and the original atomic charge variables is also presented, uncovering a surprising isomorphism between the two representations that is made possible because of a symmetry of classical electrostatics. This has a practical consequence of drastically reducing the computational cost of QTPIE so that it is no more expensive than other fluctuating-charge models. Next, in Chapter 4, we investigate the calculation of electrostatic properties within fluctuating-charge models and present a partial solution toward another outstanding problem of fluctuating-charge models, namely that they exhibit superlinear polarizabilities. We also present initial results toward a water model capable of polarization and charge transfer. Finally in Chapter 5, we summarize our findings and comment on the issues raised that remain unresolved from this work.

# Chapter 2. The dissociation catastrophe in fluctuating–charge models

Portions of this chapter were adapted from

Chen, J.; Martinez, T. J. *Chem. Phys. Lett.* **2007**, *438*, 315—320.

## 2.1. The dissociation catastrophe and overestimation of charge transfer

In this chapter, we investigate one of the most well-known problems with fluctuating-charge models, which is their overestimation of charge transfer at large internuclear separations. It is instructive to consider the behavior of the typical fluctuating-charge model for a neutral diatomic molecule. Since we have $q_1 = -q_2$, it is possible to substitute the charge constraint directly into the energy function and write it as

$$E\left(q_1;\mathbf{R}\right) = \left(\chi_1 - \chi_2\right)q_1 + \tfrac{1}{2}\left(\eta_1 - 2J_{12}\left(\left|\mathbf{R}_1 - \mathbf{R}_2\right|\right) + \eta_2\right)q_1^2 \tag{2.1}$$

This is minimized by the analytic solution

$$q_1\left(\mathbf{R}\right) = \frac{\chi_2 - \chi_1}{\eta_1 - 2J_{12}\left(\left|\mathbf{R}_1 - \mathbf{R}_2\right|\right) + \eta_2} \tag{2.2}$$

We therefore see that this fluctuating-charge model always predicts a nonzero charge on each atom unless they have equal electronegativities or at least one atom has infinite hardness. While this is reasonable for chemically bonded systems, it fails to describe, even qualitatively, the charge transfer behavior at infinite separation. As the atoms are drawn ever further apart, $\left|\mathbf{R}_1 - \mathbf{R}_2\right| \to \infty$, the Coulomb interaction vanishes, so that the charge on atom 1 tends to the limit

$$\lim_{\left|\mathbf{R}_1 - \mathbf{R}_2\right| \to \infty} q_1\left(\mathbf{R}\right) = \frac{\chi_2 - \chi_1}{\eta_1 + \eta_2} \neq 0 \tag{2.3}$$

The model therefore predicts nonzero charge transfer even for dissociated systems, which is clearly unphysical for diatomic molecules in the gas phase. This leads to a dissociation catastrophe whereby intermolecular charge transfer is severely overestimated, causing electrostatic properties such as the dipole moment and the on-axis component of the polarizability to diverge. This renders such models useless for describing intermolecular charge transfer processes.

In practice, fluctuating-charge models require further constraints proscribing intermolecular charge transfer in practical simulations.[1-5] For example, the TIP4P-FQ water model of Rick and coworkers constrains the flow of charge to lie exclusively within each water molecule.[3] Similar constraints were found necessary for calculating size extensive polarizabilities in spatially extended systems.[1, 2] Without such constraints, the water model would predict unrealistically large dipole moments and polarizabilities, and produces large qualitative errors in dynamical simulations. Recent work has also shown that even with such constraints, and even for molecular geometries near equilibrium, fluctuating-charge models generally overestimate the propensity for charge flow in polyatomic molecules, giving rise to inflated values of molecular electrostatic properties such as dipole moments and polarizabilities.[1, 6-8]

The unphysical prediction of nonzero charge transfer at infinity can be understood by turning off the Coulomb interaction terms in the fluctuating-charge model. The energy function can then be written as the simple sum of noninteracting atomic energy functions

$$E(\mathbf{q};\mathbf{R}) = \sum_{i=1}^{N} E_i^{at}(q_i) \qquad (2.4)$$

and each of these atomic energies can be written in the form

$$E_i^{at}\left(q_i\right)=\tfrac{1}{2}\eta_i\left(q_i+\frac{\chi_i}{\eta_i}\right)^2-\frac{\chi_i^2}{2\eta_i}+E_i^0 \qquad (2.5)$$

Thus in the absence of any interatomic interactions, the charge predicted by fluctuating–charge models defaults to the solution $q_i=-\chi_i/\eta_i$, being the minimum point of the parabola of Eq. (2.5). As both the atomic electronegativity and atomic hardness are constants, it is unclear how this problem can be solved while remaining in atom space, i.e. the solution space spanned by the vector of atomic charges **q**.

The dissociation catastrophe can be interpreted as the consequence of an unrealistic assumption inherent in fluctuating-charge models, namely that pairs of atoms can exchange charge with equal facility regardless of their distance. This is true only in metallic phases, and therefore the extent to which this model fails to predict sensible charge distributions can be attributed to a fault in the underlying physics in assuming that molecular systems have metallic character.[1, 9] We therefore desire a fluctuating-charge model that can predict partial charges in such geometries without this implicit assumption of metallicity. Previous work by Morales and Martínez have analyzed charge equilibration methods from a wavefunction viewpoint to elucidate the important issues.[10, 11] First, the process of charge transfer is the fundamental process in fluctuating-charge models, and therefore measures of the charge transfer between pairs of atoms are in some sense more fundamental quantities than the atomic charges that are produced as a result of such charge flows. Second, Morales and Martínez found that electronegativities should depend on molecular geometries. These ideas guide our development of a new charge equilibration method.

## 2.2. The introduction of distance-dependent electronegativities

Previous work in the Martínez group has analyzed the behavior of charge equilibration methods and have addressed their shortcomings in the CC-QVB2 model,[10, 11] which was constructed and tested numerically for diatomic molecules. Here, we describe our generalization to polyatomic molecules and test the method's numerical accuracy. The fundamental variables of our new method are not atomic partial charges $\mathbf{q}$, but charge transfer variables $\mathbf{p}$ that describe a polarization current, i.e. a tendency for electronic density to migrate from one atom onto another. The method is thus named QTPIE, for charge transfer with polarization current equilibration.[12] The charge transfer variables are related to the atomic charges by continuity:

$$q_i = \sum_j p_{ji} \tag{2.6}$$

where $p_{ji}$ describes the amount of charge transferred from the *i*th atom to the *j*th atom. It is natural to assume that the charge transfer variables exhibit skew symmetry, i.e. $p_{ij} = -p_{ji}$. These charge transfer variables were first introduced in 1968 by Borkmann and Parr in the context of bond charges for diatomic potential energy curves.[13, 14] However, they were first used in their current form in 1983 by Allinger and coworkers in the Induced Dipole Moment and Energy (IDME) method,[15] an early polarizable force field where the charge transfer variables were integral in combining inducible dipoles[16, 17] and fluctuating charges via a reparameterized Del Re model.[18] This allowed the method to treat both through-bond and through-space polarization effects. Allinger and coworkers interpreted charge transfer variables as being responsible for the dipole moment of the bonds between pairs of atoms. Banks and coworkers have also found that these charge transfer variables, which they called bond-charge increments, were useful for the numerical fitting procedures for parameterizing charge equilibration models.[19] These variables were then used in the construction of the AACT model, which was found to improve the prediction of electrostatic properties in extended molecular systems.[1] In terms of these charge transfer variables, the energy expression for fluctuating charge models take the form:

$$E(\mathbf{p}) = \sum_{ij} \chi_i p_{ji} + \frac{1}{2} \sum_{ijkl} p_{ki} p_{lj} J_{ij} \tag{2.7}$$

The transformed variables allow us to modify the electronegativities to include distance dependence for every atom pair. It is only in this new representation that it is possible to introduce this distance dependence explicitly as an attenuation function $f_{ji} = f_{ji}(\mathbf{R})$ which penalizes long-range charge transfer between pairs of atoms by rescaling the potential difference between those pairs. This modified energy function is the central equation of QTPIE:

$$\begin{aligned} E(\mathbf{p}) &= \sum_{ij} \chi_i f_{ji} p_{ji} + \frac{1}{2} \sum_{ijkl} p_{ki} p_{lj} J_{ij} \\ &= \sum_{i<j} p_{ji} \left[ \left( \chi_j - \chi_i \right) f_{ji} + \frac{1}{2} \sum_{k<l} p_{lk} \left( J_{ik} - J_{jk} - J_{il} + J_{jl} \right) \right] \end{aligned} \tag{2.8}$$

On the second line of Eq. (2.8), we exploited the antisymmetry of the charge transfer variables and the symmetry of $f_{ij}$ to write the equation in skew-symmetric form.

As shown previously,[10] the attenuation function $f_{ji}$ should decay with distance on a length scale related to the orbitals involved on atoms *i* and *j*. Note that if the attenuation function $f_{ji}$ were chosen to be a constant independent of distance, the QTPIE model would reduce to the QEq model. This confirms our earlier claim that fluctuating-charge models like QEq belie an inherent assumption of metallicity, as with no typical length scale of potential rescaling, there is long-range order that facilitates charge transfer over macroscopic distances. Also, detailed balance requires $f_{ji}$ to be invariant under index exchange, i.e. $f_{ij} = f_{ji}$. The simplest choice of $f_{ij}$ is therefore an overlap integral between orbitals on the *i*th and *j*th atoms, as demonstrated by the previous maximum entropy studies.[10, 11] In this chapter, take this function to be a scaled overlap integral of the n*s*-type orbitals which are used to represent the screened Coulomb interaction, adopting the same choice of orbitals as was used in the QEq model,[20] i.e.

$$f_{ji} = k_{ji} S_{ji} = k_{ji} \left\langle \phi_j \middle| \phi_i \right\rangle \tag{2.9}$$

The scaling factors $k_{ji}$ could be optimized, even for different bond types; however, here we simply choose $k_{ji}$ to be unity for all atom pairs unless otherwise stated. The sum in Eq. (2.8) is *not* limited to bonded atom pairs — all information about molecular connectivity is embedded in the screened Coulomb interaction and the attenuation factor $f_{ij}$ — so bonding need not be specified *a priori*. We use the QEq parameters for electronegativities, hardnesses, and orbital radii without modification. Explicit reparameterization can thus be expected to improve all of the results reported in this chapter.

Minimizing the energy of Eq. (2.8) with respect to all charge transfer variables leads to the system of linear simultaneous equations

$$\forall i, j : 0 = \frac{\partial E}{\partial p_{ji}} = \left(\chi_j - \chi_i\right) k_{ji} S_{ji} + \sum_{k<l} p_{lk} \left(J_{ik} - J_{jk} - J_{il} + J_{jl}\right) \tag{2.10}$$

The QTPIE solution for a diatomic molecule is thus:

$$q_2 = p_{21} = \frac{\chi_2 - \chi_1}{J_{11} - 2J_{12} + J_{22}} k_{12} S_{12} \Rightarrow \lim_{R_{12} \to \infty} q_2 = 0 \tag{2.11}$$

In contrast to the QEq solution for the diatomic system, QTPIE correctly predicts vanishing charge transfer in the dissociation limit and should therefore provide a more accurate description of fluctuating charges for non-equilibrium geometries.

## 2.3. Results and discussion

The QEq and QTPIE methods were implemented in Scilab and solved in a linear algebraic representation in the space of unique atomic pairs. We did not implement the charge-dependent atomic radius for hydrogen atom described in the original QEq method.[20] Thus, the results presented here are denoted QEq(-H), indicating that the hydrogen correction is not employed.

The QTPIE method as formulated contains $O(n^2)$ charge transfer variables but charge transfer around closed loops does not influence the energy expression of Eq. (2.8). Both methods therefore only have $n$ - 1 independent variables. The linear system of Eq. (2.10) is therefore rank-deficient and hence singular. The system was solved by constructing the pseudoinverse from singular value decomposition.[21]

We performed calculations on three representative small molecules: sodium chloride, water and phenol. For each molecule, we compared the predictions of QEq(-H) and QTPIE with the results of *ab initio* calculations. Since atomic charges are not well-defined quantum-mechanical observables, we chose two distinct definitions for comparison, namely Mulliken population analysis[22, 23] and distributed multipole analysis (DMA).[24] The DMA calculation was restricted to monopoles on the atomic centers. The electronic structure calculations for these charge analyses were in general performed using multi-reference *ab initio* methods with small basis sets. We chose well-localized basis sets to facilitate comparisons between the *ab initio* and QTPIE/QEq methods, which implicitly use minimal basis sets.

For illustrative purposes, we present results of the QEq(-H) and QTPIE models applied to an isolated sodium chloride molecule at different internuclear distances. *Ab initio* results are obtained from a complete active space (CAS) calculation[25] using eight electrons in five orbitals, i.e. CAS(8/5), with a 3-21G basis set.[26] This full valence active space wavefunction describes both ionic and covalent characters. Because of the weakly avoided crossing between the covalent and ionic diabatic states, the transition from ionic to covalent character on the ground electronic state is quite rapid, as seen in both the Mulliken and DMA charges shown in Figure 2.1. The Mulliken and DMA definitions of the atomic charges give similar values throughout, indicating the robustness of the *ab initio* partial charges we are using for comparison with the QTPIE and QEq(-H) results.

**Figure 2.1.** Partial charge on the sodium atom in dissociating NaCl as computed using QEq(-H) (black solid line) and QTPIE (red solid line). CAS(8,5)/3-21G calculations were analyzed using Mulliken population analysis (blue dotted line) and distributed monopole analysis (DMA) (orange dotted line). At infinity, QEq(-H) predicts significant charge transfer while QTPIE predicts uncharged fragments in this limit, in agreement with the *ab initio* results. The experimentally-determined equilibrium bond length of NaCl is indicated on the graph ($R_{eq}$=2.361Å).

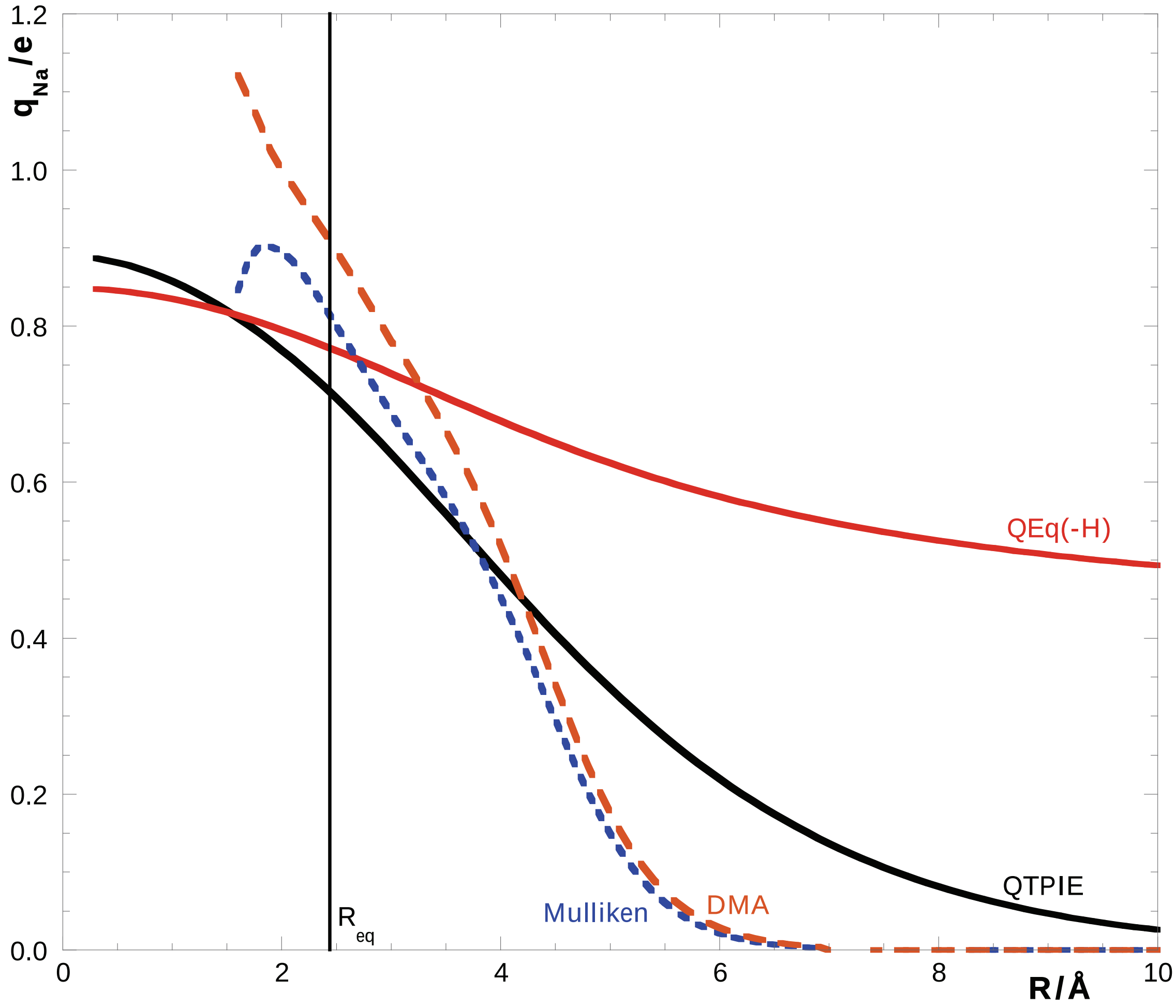

Figure 2.1 shows that, as expected from our earlier analysis, the QEq(-H) method exhibits a dissociation catastrophe, i.e. it predicts finite charge transfer at infinite separation: asymptotically $q_{\mathrm{Na}} = -q_{\mathrm{Cl}} = 0.394$. However, QTPIE correctly predicts no charge transfer at this dissociation limit. The QTPIE charges are not in quantitative agreement with the *ab initio* charges. This is expected, since only a fully quantum mechanical method is expected to capture the weakly avoided crossing (at large internuclear distance) between the covalent and ionic states.

We also calculated partial charges for asymmetrically dissociated water molecules. In this hypothetical reaction, the H-O-H internal bond angle was set to $\theta = 104.5°$ and one of the O-H bonds was kept fixed at 0.97 Å while the other O-H bond length was varied. The *ab initio* data were computed at the CAS(10,7)/STO-3G level of theory. In Figure 2.2, we show the atomic charges on the dissociating hydrogen and oxygen atom computed from *ab initio,* QEq(-H), and QTPIE methods. The atomic charge on the remaining hydrogen atom can be deduced by considering overall charge neutrality. Similar to the NaCl example, the QTPIE charges are asymptotically correct, unlike the QEq(-H) values. The QTPIE partial charge on the oxygen atom in the OH fragment is closer to the *ab initio* result than that predicted by QEq(-H). However, it is still too large, indicating an overestimation of the dipole moment of OH. Thus, we attempt the simplest reparameterization possible, namely varying $k_{OH}$ of Eq. (2.9), while demanding that $k_{OH}$ = $k_{HH}$. We chose the value for $k_{OH}$ which led to agreement of the partial charge on oxygen atom at the equilibrium geometry of the water molecule ($k_{OH}$ = $k_{HH}$ = 0.4072). With this modification, the QTPIE charges are in good agreement with the *ab initio* values across the whole range of O-H distances, as shown in Figure 2.3.

**Figure 2.2.** Partial charges (in atomic units) for a dissociating water molecule as predicted by QEq(-H) (red solid line) and QTPIE (black solid line). Positive values are charges on the dissociating hydrogen, and negative values are charges on the oxygen. Also shown are charges from distributed multipole analysis (DMA) (orange dotted line) as performed on a CAS(10/7) wavefunction in a STO-3G basis set. QTPIE without reparameterization reproduces the vanishing charge on the dissociating hydrogen atom at infinite separation predicted by the *ab initio* method. The equilibrium bond length of the O-H bond on water is indicated on the graph ($R_{eq}$ = 0.957Å).

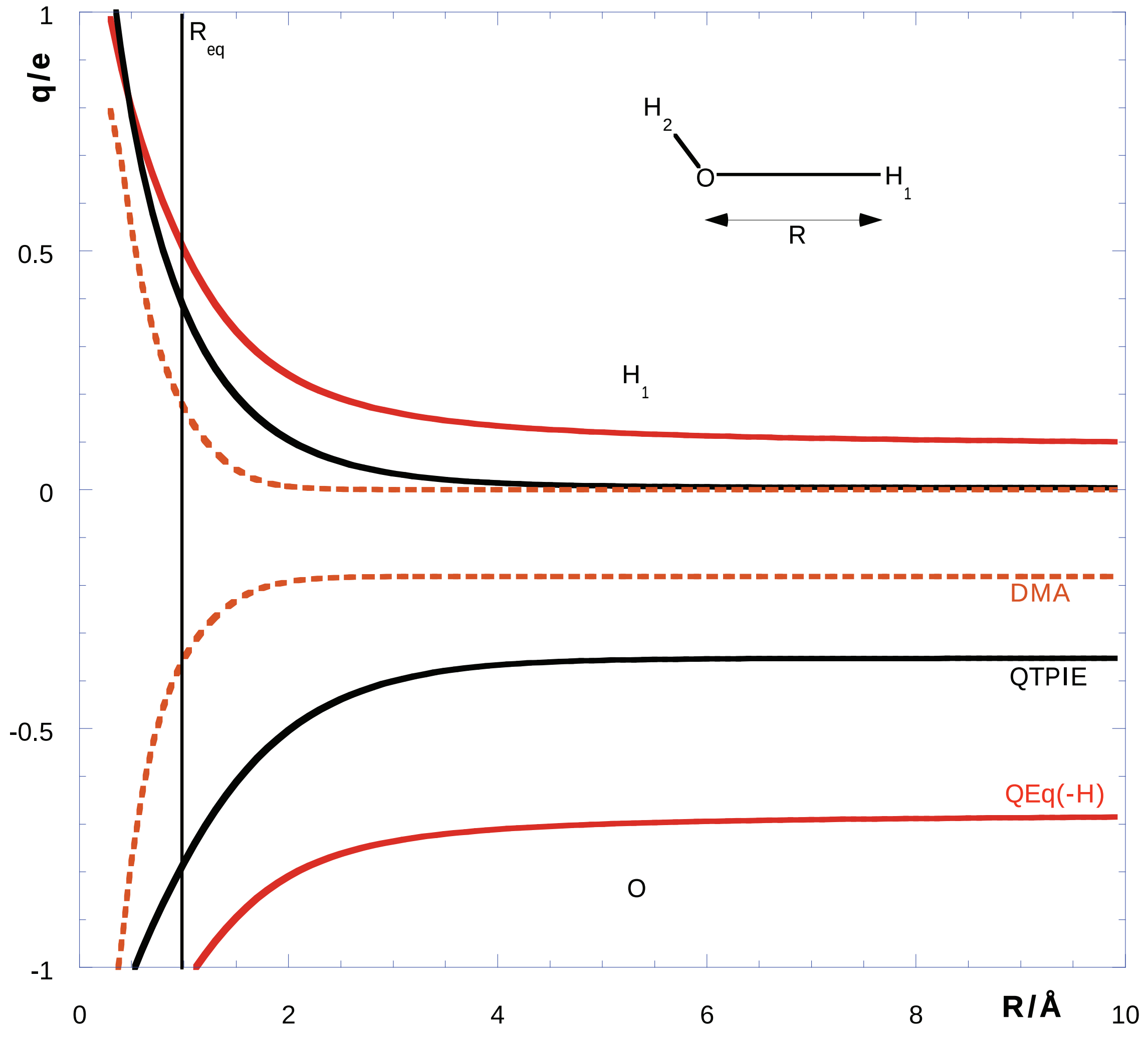

**Figure 2.3.** As in Figure 2.2, but using $k_{OH} = k_{HH} = k$ of Eq. 2.8 which is optimized ($k$ = 0.4072) to give agreement of QTPIE and DMA charges at the equilibrium geometry of the water molecule. With minimal reparameterization, the QTPIE method agrees well with *ab initio* charges throughout (except for very short bond distances, where the concept of partial charge breaks down).

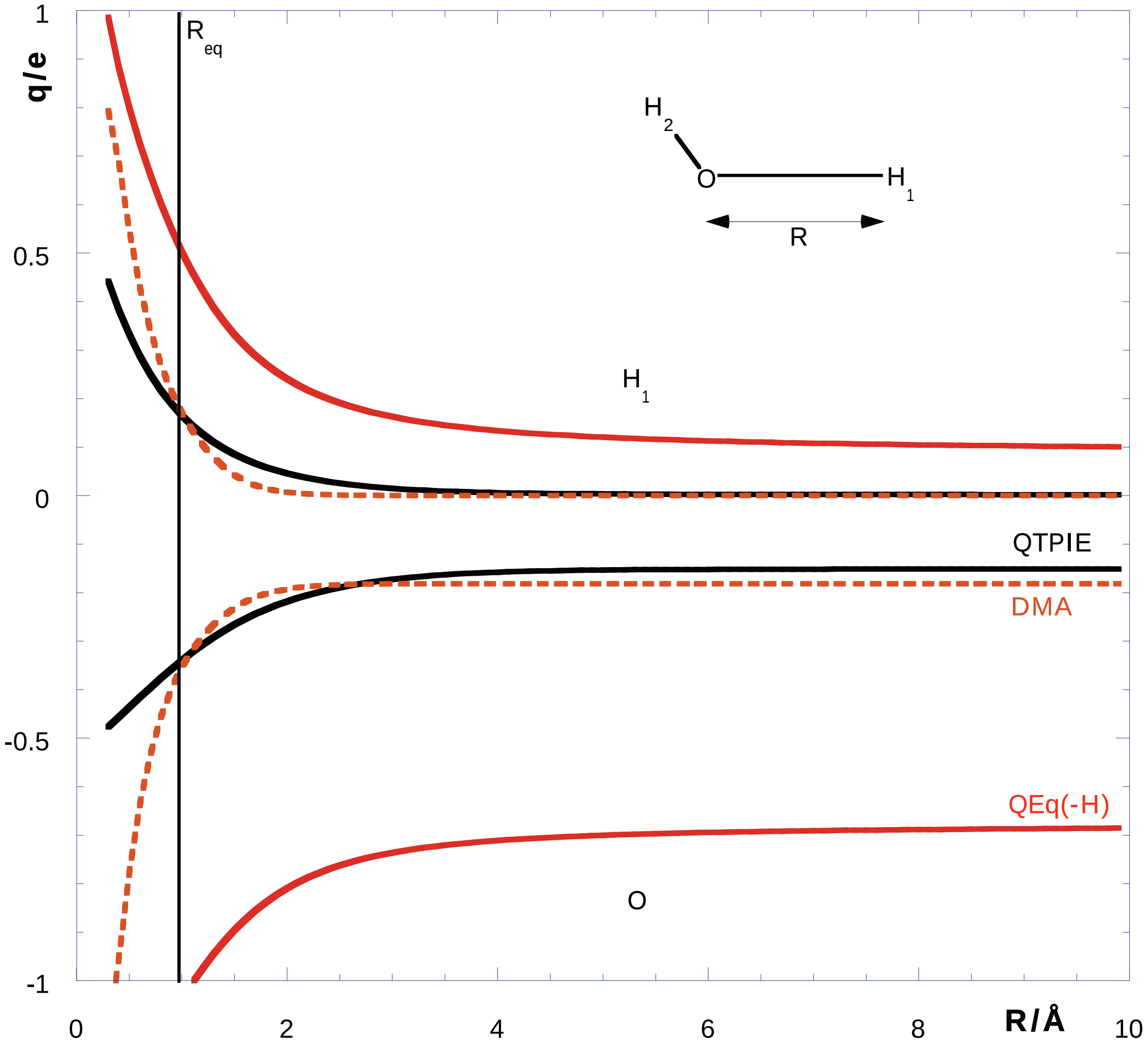

**Figure 2.4.** As in Figure 2.3, but for varying internal angles $\theta$.

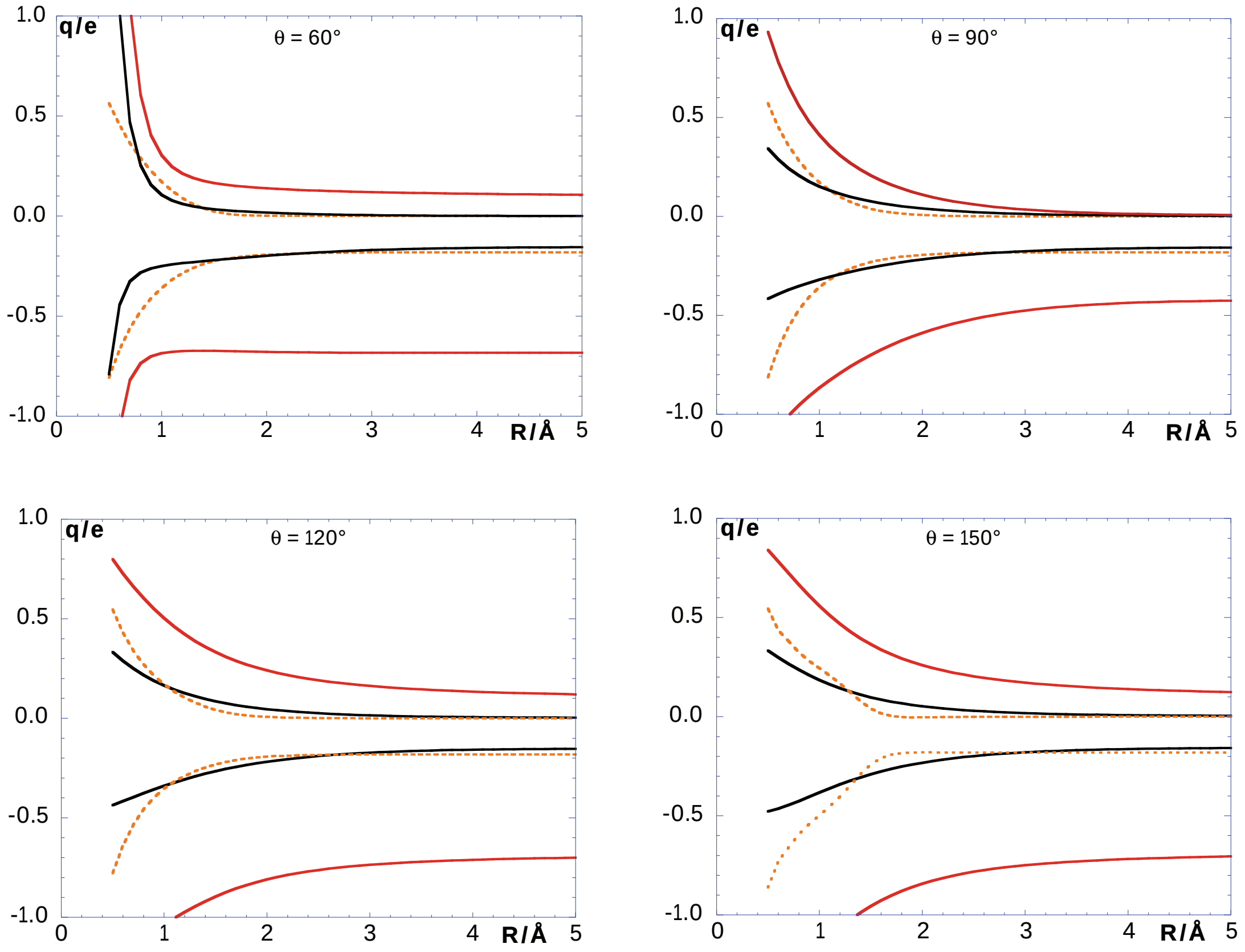

**Figure 2.5.** Atomic partial charges for phenol in the equilibrium geometry as computed with QTPIE, QEq(-H) (bold), and Mulliken population analysis on the MP2/cc-pVDZ wavefunction (italics).

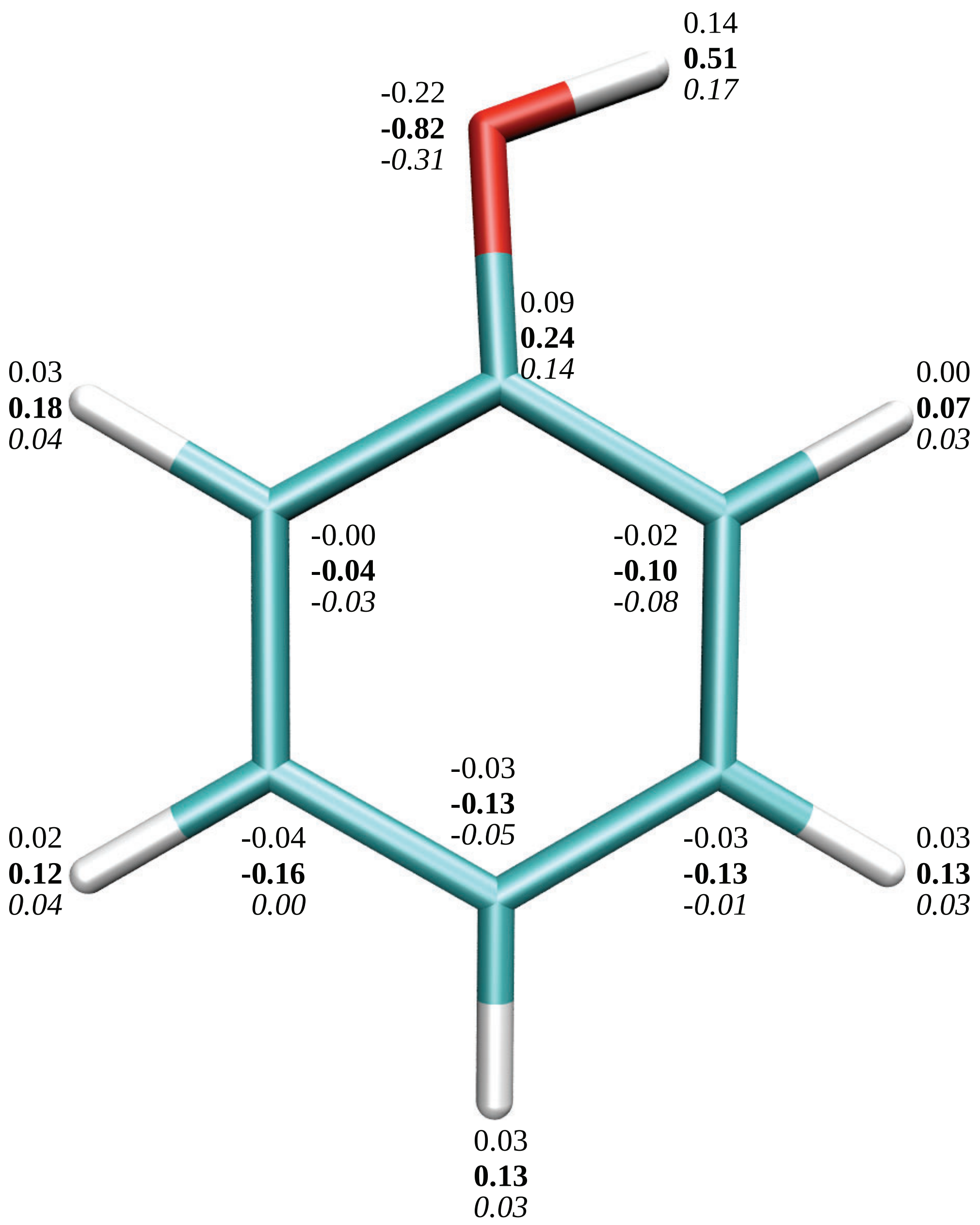

In order to explore the adequacy of a single set of QTPIE parameters for other molecular geometries, we computed similar dissociation curves with varying the internal angle ∠HOH in the range 60°-150°. The results (using $k_{OH}$=$k_{HH}$=0.4072, as discussed above) are compared with *ab initio* charges from Mulliken analysis on CASPT2(10,7)/STO-3G data in Figure 2.4. The results from QTPIE remain in similarly good agreement with the *ab initio* calculations for all of these geometries, particularly in the dissociation limit. In Figure 2.5, we show that this good agreement between *ab initio* and QTPIE charges persists for larger molecules such as phenol.

It is important that a fluctuating charge model be able to accurately model the change in atomic charges with response to an external electric field. Thus, we have also computed the molecular polarizability tensor using QEq and QTPIE. These results are again compared with *ab initio* calculations. The QEq model has two shortcomings when computing molecular polarizabilities. The first is a tendency to overestimate the in-plane components, which is related to the overestimation of charges for weakly interacting (i.e. widely separated) atoms. The second is its inability to calculate the out-of-plane component of the molecular polarizability tensor for planar molecules.[27] This latter deficiency arises because the model considers only atomic charges and not atomic dipoles or charge centers apart from the locations of the atoms. This makes it impossible to have charge fluctuations along any direction other than in directions directly leading to another point charge. In terms of molecular graphs, charge flow is restricted only to edges and therefore cannot flow out of the plane of the molecule. Similar restrictions apply in the QTPIE method as described here, and thus one might expect that QTPIE will also fail to describe the out-of-plane polarizabilities for planar molecules. Dummy atoms specified in the molecular geometry could conceivably improve matters, but only at the expense of additional parameters. Charge equilibration methods with expanded basis sets, which describes the charge fluctuations from a single *s* function per atomic site to include also *p*-type functions, are a more promising route to solve this problem.[28]

The QTPIE energy expression in an external electrostatic field $\vec{\varepsilon}$ is given by:

$$E\left(\mathbf{p};\vec{\varepsilon}\right)=\sum_{ij}\chi_i^0 f_{ji}p_{ji}+\frac{1}{2}\sum_{ijkl}p_{ki}p_{lj}J_{ij}+\sum_{ij}p_{ij}\vec{R}_i\cdot\vec{\varepsilon} \tag{2.12}$$

We compute the QTPIE polarizability numerically by the method of finite fields, being the fluctuation in the dipole moment with respect to changes in $\vec{\varepsilon}$. The dipole moment was recalculated with re-optimized charge transfer variables in Eq. (2.12) at each value of $\vec{\varepsilon}$. The scaling factor for the overlap, $k_{ij}$, was taken to be unity in all of these QTPIE calculations. Table 2.1 summarizes the results for sodium chloride, water and phenol. The *ab initio* polarizabilities were calculated as second derivatives of the second-order Møller-Plesset perturbation theory (MP2) energy, also using the method of finite fields. The *ab initio* calculations use an aug-cc-pVDZ basis set[29, 30] which includes the diffuse functions necessary for accurate calculations of polarizabilities. Ground state equilibrium geometries were optimized using MP2/aug-cc-pVDZ; the same geometries were used for all polarizability calculations.

Molecular polarizabilities calculated using the three methods above were found to be stable with respect to small perturbations in the nuclear geometries, so discrepancies in the eigenvalues due to geometric effects can be ruled out. As expected, both QEq(-H) and QTPIE incorrectly predict a vanishing out-of-plane component of the polarizability for these planar molecules. Interestingly, the eigenvalues of the polarizability tensor in QTPIE turn out to be identical to those from QEq(-H). We will discuss this later, and in great detail, in Chapter 4.

**Table 2.1.** Eigenvalues (sorted by descending magnitude) of the dipole polarizability tensor (in units of Å$^3$) for three molecules.

| | QEq(-H) | QTPIE | MP2/aug-cc-pVDZ |
|---|---|---|---|
| | 13.9474 | 13.9474 | 4.5042 |
| NaCl | 0.0000 | 0.0000 | 3.6932 |
| | 0.0000 | 0.0000 | 3.6931 |
| | 3.4653 | 3.4653 | 1.4502 |
| H2O | 1.2317 | 1.2317 | 1.3678 |
| | 0.0000 | 0.0000 | 1.2883 |
| | 24.6244 | 24.6244 | 13.6758 |
| Phenol | 20.3270 | 20.3270 | 12.3621 |
| | 0.0000 | 0.0000 | 6.9981 |

## 2.4. Reparameterization in terms of primitive s-type Gaussians

Our previous studies in this Chapter, and the QTPIE model as published,[12] uses two–electron Coulomb integrals over s–type primitive Slater type orbitals (STOs) of the form Eq. (1.30) in the calculation of the screened Coulomb interactions in Eq. (1.29). This was chosen in line with the QEq model,[20] which the QTPIE model can be considered a derivative of. It was originally claimed that the use of STOs introduced greater accuracy in the screening calculation.[20, 31] However, it is possible to substitute the use of two–electron Coulomb integrals over s–type primitive Gaussian orbitals, with orbital exponents fitted to reproduce the results from the much more expensive *s*–type Slater type orbitals used in QEq.

We construct these Gaussian orbitals by minimizing the norm of the $L^2$–difference between the homonuclear Coulomb integral over Slater orbitals and over Gaussian orbitals, i.e. given a Slater exponent $\zeta$, we want the Gaussian exponents $\alpha$ that minimizes

$$\left\| J^G(\alpha) - J^S(\zeta) \right\|_2^2 = \left\langle J^G(\alpha), J^G(\alpha) - 2J^S(\zeta) \right\rangle_2 + \left\| J^S(\zeta) \right\|_2^2 \tag{2.13}$$

where $\langle f,g \rangle_2 = \int_0^\infty f(x) g(x)\, dx$ is the inner product in the function space $L^2\left[0,\infty\right)$, $\|f\|_2 = \sqrt{\langle f, f \rangle_2}$ is the $L^2$–norm, $J^G$ is the two–electron Coulomb integral over *s*–type primitive Gaussian orbitals

$$J^G(R;\alpha) = \frac{2\alpha}{\pi} \iint_{\mathbb{R}^6} \frac{e^{-\alpha|r_1 - R|^2} e^{-\alpha|r_2|^2}}{|r_1 - r_2|} dr_1\, dr_2 = \frac{\operatorname{erf}\sqrt{\alpha} R}{R} \tag{2.14}$$

and $J^S$ is the two–electron Coulomb integral over *s*–type Slater orbitals

$$J^S(R;\zeta,n) = \frac{(2\zeta)^{4n+2}}{((2n)!)^2} \iint_{\mathbb{R}^6} \frac{|r_1 - R|^n |r_2|^n e^{-\zeta|r_1 - R|} e^{-\zeta|r_2|}}{|r_1 - r_2|} dr_1\, dr_2 \tag{2.15}$$

which is given in closed–form in the literature.[32] As the Slater exponent $\zeta$ is given for each minimization, the last term in Eq. (2.13) can be dropped without affecting the results of the minimization, and therefore the minimization problem is solved by the Gaussian exponent $\alpha$ that solves the equation

$$0 = \frac{\partial}{\partial \alpha} \left\langle J^G(\alpha), J^G(\alpha) - 2J^S(\zeta) \right\rangle_2 = \left\langle 2\frac{dJ^G(\alpha)}{d\alpha}, J^G(\alpha) - J^S(\zeta) \right\rangle_2 \tag{2.16}$$

We find the solution to Eq. (2.16) numerically using the secant method[33] with a trust radius of $\alpha/4$ at each iteration. The algorithm was terminated once the integral on the right hand side of Eq. (2.16) was less than $10^{-16}$ in absolute magnitude. The results are presented in Table 2.2, along with the maximum absolute error as defined by

$$\text{MAE} = \max_{0 \le R < \infty} \left| J^G(R;\alpha) - J^S(R;\zeta) \right| \tag{2.17}$$

**Table 2.2.** Exponents of atomic orbital exponents that best reproduce the two–electron Slater integrals over the QEq orbitals.

| Element | Slater exponent[a] | Gaussian exponent | Error[b] |
|---|---|---|---|
| H | 1.0698 | 0.5434 | 0.01696 |
| Li | 0.4174 | 0.1668 | 0.00148 |
| C | 0.8563 | 0.2069 | 0.00162 |
| N | 0.9089 | 0.2214 | 0.00166 |
| O | 0.9745 | 0.2240 | 0.00167 |
| F | 0.9206 | 0.2313 | 0.00169 |
| Na | 0.4364 | 0.0959 | 0.00085 |
| Si | 0.7737 | 0.1052 | 0.00088 |
| P | 0.8257 | 0.1085 | 0.00089 |
| S | 0.8690 | 0.1156 | 0.00092 |
| Cl | 0.9154 | 0.1137 | 0.00091 |
| K | 0.4524 | 0.0602 | 0.00125 |
| Br | 1.0253 | 0.0701 | 0.00133 |
| Rb | 0.5162 | 0.0420 | 0.00121 |
| I | 1.0726 | 0.0686 | 0.00127 |
| Cs | 0.5663 | 0.0307 | 0.00114 |

[a]From Ref. [20].

[b]Maximum absolute error as defined in Eq. (2.17).

As can be seen clearly from Table 2.2, the approximation of replacing *s*-type STOs with suitably parameterized *s*-type GTO primitives can be made to an accuracy of $10^{-3}$ atomic units (Hartree per electron charge squared per bohr). This results in significant computational savings, as it is well-known that the Coulomb integral over GTOs is much more easily calculated than over STOs.[23, 32]

## 2.5. Conclusions

We have defined a new fluctuating charge model, QTPIE, which defines atomic charges as sums over charge-transfer variables. This construction allowed us to create a simple fluctuating-charge model that exhibits correct asymptotic behaviors for weakly-interacting atoms, i.e. near dissociation. We did not make any significant attempt to optimize the parameters for QTPIE, but instead used parameters (electronegativities, hardnesses, and orbital radii for the shielded Coulomb interaction) optimized for the QEq method. Unfortunately, these improvements come at the expense of introducing a much less compact representation of the charge distribution when compared to atomic charges. We shall see in the next chapter how it is possible to reformulate the QTPIE model purely in terms of atomic charges.

# Chapter 3. The reformulation of QTPIE in terms of atomic charges

Portions of this chapter are adapted from

Chen, J.; Hundertmark, D.; Martínez, T. J. *J. Chem. Phys.* **129**, *2008*, 214113

## 3.1. The disadvantages of the charge-transfer variable representation

In the previous Chapter, we have demonstrated that introducing explicit geometry dependence into the electronegativities allowed us to solve the problem of nonvanishing charge transfer at infinite separation in fluctuating–charge models.[1, 2] In order to do so, we were required to make a change of variables from atomic charges to charge transfer variables in the formulation of the QTPIE model.[2] While this allowed us to attenuate long-distance charge transfer, this change in representation came at the price of introducing many more variables to solve for. For a system of $N$ atoms with a specified total charge, there are $N - 1$ linearly independent atomic charge variables, but $\frac{1}{2}N\left(N-1\right)=O\left(N^2\right)$ charge-transfer variables. This has important consequences when considering the computational cost of the QTPIE model and weighing its merits against other fluctuating-charge models. A numerical implementation of the QTPIE model based on naïve direct solvers that find the charge transfer variables would have a computational complexity of $O\left(N^6\right)$.[3] This cost can be reduced using iterative methods to $O\left(N^4\right)$,[4, 5] and exploiting sparsity could in principle reduce the cost further to $O\left(N^2\right)$.[6] However, we would expect the prefactor to be very large due to the long-range nature of the Coulomb interaction, which would severely limit the amount of sparsity that could be expected in the numerical system of equations. Furthermore, the charge-transfer – charge-transfer interaction matrix **A**, which is the analog of the hardness matrix in the space of charge-transfer variables, turns out to be rank-deficient, which necessitates using more costly numerical algorithms such as singular value decomposition (SVD)[3] to solve the QTPIE equations.

In this Chapter, we study how to reduce the computational cost of the QTPIE model, by investigating the origins of the rank deficiency of **A**, and presenting more practical methods to solve the QTPIE model. We show that in addition to SVD, a complete orthogonal decomposition (COD) technique exists for solving the rank-deficient QTPIE model as formulated in charge transfer variables.[3] We also study in great detail the charge continuity relation defined in Eq. (2.6), which is the transformation of variables that brings us to charges from charge-transfer variables. It turns out that there exists an information-preserving inverse transformation that allows us to map charge-transfer variables exactly onto specific linear combinations of atomic charge variables. This allows an exact reformulation of the QTPIE working equations in terms of atomic charges, which then produces a reformulated QTPIE model that is of the same computational complexity as other fluctuating-charge models that are expressed in terms of atomic charge variables.

### 3.2. Null modes of the capacitance matrix

Recall that the linear system of equations of Eq. (2.10) defines the QTPIE model. It is possible to interpret this equation either as a linear system involving the partial contraction of a four-tensor, or as a regular matrix-vector problem in a higher-dimensional space, which we term bond space which is spanned by linear combinations of charge-transfer variables. We will adopt the latter perspective in the rest of this work.

**Figure 3.1.** Visual representations of eigenvectors of the atom space hardness matrix **J** of Eq. (2.8) (A-1—A-3) and the bond space hardness matrix **A** of Eq. (3.2) (B-1—B-3) for a single water molecule. The δ+, δ- symbols represent increases and decreases in charge on the respective atoms, while arrows show the direction of charge transfer with relative magnitudes indicated by their thicknesses. The respective eigenvalues are 0.181 (A-1), 0.101 (A-2), 1.231 (A-3), 1.273 (B-1), 0.350 (B-2), and 0.000 (B-3). Although the magnitude of the nonzero eigenvalues depends on the choice of parameters used to construct **J** and **A**, the presence of the zero eigenvalue and the character of the corresponding eigenvector shown in B-3 do not depend on such details of parameterization.

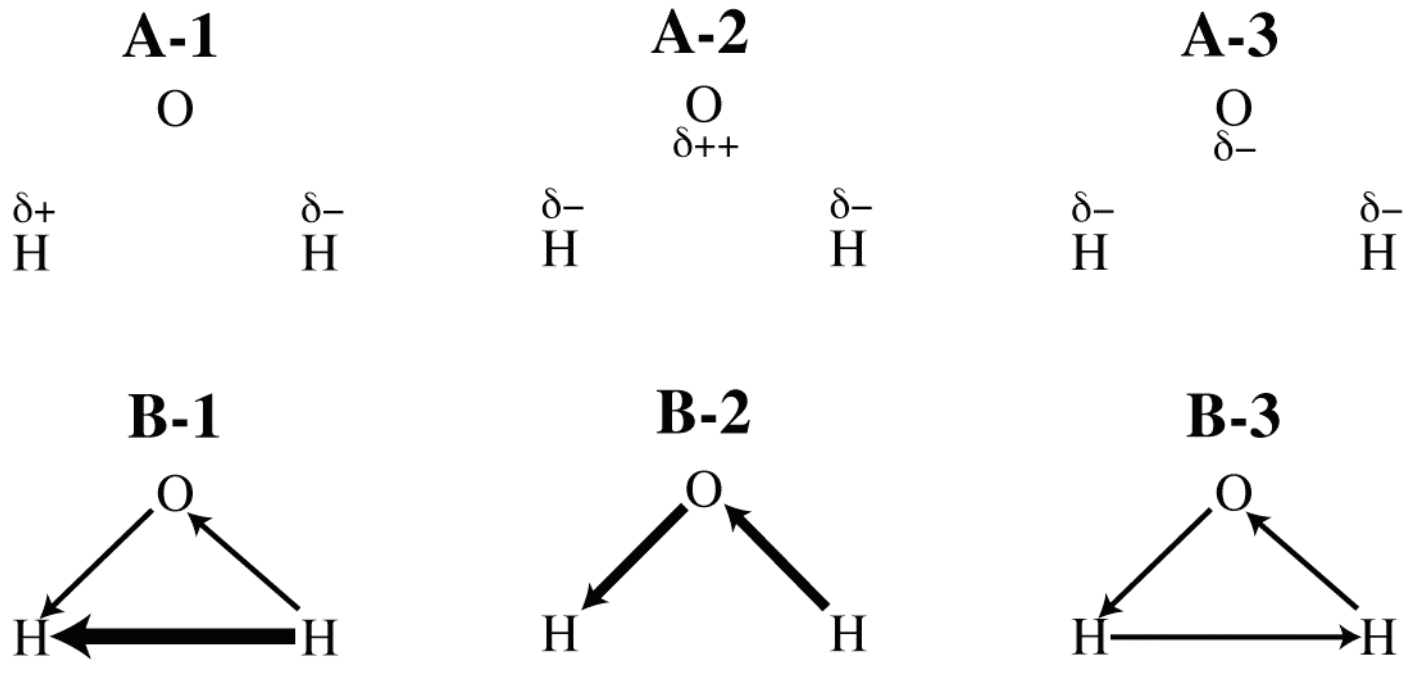

---

Let us rewrite the working equations explicitly in terms of a matrix-vector product by introducing a single pair index λ(*i*,*j*) which runs over unique atom pairs:

$$\lambda(i,j) = \tfrac{1}{2}\max(i,j)\left(\max(i,j)-1\right)+\min(i,j) \tag{3.1}$$

The variables $\{p_\lambda\}$ that minimize the QTPIE energy function of Eq. (2.8) are then solutions to the following linear system of equations

$$\begin{aligned}
\frac{\partial E}{\partial p_\lambda} &= \sum_\nu A_{\lambda\nu} p_\nu - V_\lambda = 0 \\
V_{\lambda(i,j)} &= f_{ij} - f_{ji} \\
A_{\lambda(i,j)\nu(k,l)} &= \frac{1}{2}\frac{\partial^2 E}{\partial p_\lambda \partial p_\nu} = \tfrac{1}{2}\left(J_{ik}+J_{jl}-J_{jk}-J_{il}\right)
\end{aligned} \tag{3.2}$$

where $V$ is a vector of pairwise electronegativity differences, which can be interpreted as potential differences between atomic pairs. The real and symmetric bond hardness matrix **A** thus represents a linear map from charge transfer variables into pairwise voltages differences.

For systems with $N > 2$ atoms, the bond hardness matrix **A** is rank deficient. As an illustration, we diagonalize it for a single water molecule in its equilibrium ground-state geometry. The details of the atomic hardnesses and orbitals defining the elements of **A** may be found in Chapter 2. However, the results shown are general in that they do not depend on these details. The eigenvectors of the bond hardness matrix may be thought of as normal modes for charge flow in the system. These bond-space eigenvectors are visualized in Figure 3.1 along with the atom-space eigenvectors of the Coulomb matrix **J**, which is the atomic hardness matrix for QEq. Even for this small triatomic molecule, the bond hardness matrix has a non-trivial kernel or nullspace. In this case, it is spanned by the vector $u^{\perp} = \frac{1}{\sqrt{3}}\left(-1,1,-1\right)$ that describes cyclic charge transport. The effect of the kernel is given by the scalar product

$$0 = u^{\perp} \cdot p = \sum_{\lambda} u_{\lambda}^{\perp} p_{\lambda} = \tfrac{1}{\sqrt{3}}\left(-p_{21} + p_{31} - p_{32}\right) = \tfrac{1}{\sqrt{3}}\left(p_{12} + p_{23} + p_{31}\right) \tag{3.3}$$

showing that this combination of the charge-transfer variables cannot contribute any net potential difference to the system. This is closely related to Kirchhoff's voltage law, namely that there is no change in the electrostatic potential when a charge is transported about a closed loop. This law reflects the conservative nature of the electrostatic potential embodied by the bond hardness matrix.

## 3.3. The physical significance of rank deficiency

The rank deficiency of the bond hardness matrix for $N > 2$ atoms is not an unfortunate numerical accident, but rather is an unavoidable consequence of electrostatics combined with the representation in charge-transfer variables. The second term in Eq. (2.10) can be rewritten purely in terms of atomic charges as

$$\tfrac{1}{2}\sum_{i=1}^{N}\sum_{j=1}^{N}\sum_{k=1}^{N}\sum_{l=1}^{N} p_{ki} p_{lj} J_{ij} = \tfrac{1}{2}\sum_{i=1}^{N}\sum_{j=1}^{N} q_i q_j J_{ij} \tag{3.4}$$

This relationship can be rewritten using matrix notation as $\frac{1}{2}\vec{p}^T \mathbf{A}\vec{p} = \frac{1}{2}\vec{q}^T \mathbf{J}\vec{q}$, where the bond hardness matrix **A** is a linear mapping between charge transfer variables while the hardness matrix **J** is a linear mapping between atomic variables. Eq. (3.4) clearly shows that there exists a linear transformation **T** that maps from **p** to **q** by acting on the left, i.e. $\vec{q} = \mathbf{T}\vec{p}$. The corresponding adjoint transformation, $\mathbf{T}^T$, then maps from **q** to **p** by acting on the right, i.e. $\vec{q}^T = \vec{p}^T \mathbf{T}^T$. Note that **T** is a real transformation, and hence its adjoint is equal to its transpose. This allows us to show by associativity that

$$\vec{p}^T \mathbf{A}\vec{p} = \vec{q}^T \mathbf{J}\vec{q} = \left(\vec{p}^T \mathbf{T}^T\right)\mathbf{J}\left(\mathbf{T}\vec{p}\right) = \vec{p}^T\left(\mathbf{T}^T \mathbf{J}\mathbf{T}\right)\vec{p} \tag{3.5}$$

This shows that **A** and **J** are related by a linear transformation, i.e. $\mathbf{A} = \mathbf{T}^T \mathbf{J}\mathbf{T}$, and that the ranks of **A** and **J** are equal since the transformation is an information-preserving projection from atom space into bond space. The positive definiteness of the Coulomb interaction guarantees that the rank of **J** is *N*, as long as there is no linear dependence among the atomic sites. The problem therefore has rank $N-1$ since there is one constraint of electrical neutrality. As shown in the next section, the rank of the matrix **T** is $N-1$. Eq. (3.5) then implies that **A** must also have rank $N-1$, and the constraint of neutrality is accounted for implicitly by the skew-symmetry of the charge transfer variables. Therefore, there are only $N-1$ physically significant degrees of freedom regardless of representation in charges or charge-transfer variables.

We can also interpret the rank of the bond hardness matrix **A** as a consequence of Kirchhoff's voltage law. We introduce a graph $G=(V,E)$ as a convenient bookkeeping construct, with each vertex $v \in V$ corresponding to an atom or its charge, and each edge $e \in E$ corresponding to a unique charge-transfer variable $p_\lambda$. An arbitrary set of charge–transfer variables and its corresponding charges can then be mapped onto a corresponding graph *G*. The relevant physics is expressed by Kirchhoff's voltage law, which states that the change in potential as charge is transported about a closed loop vanishes. Therefore, every set of charge-transfer variables that map onto a graph containing a closed loop is linearly dependent. Hence linearly independent sets of charge–transfer variables must correspond to graphs *G* that do not contain cycles. At the same time, charges are allowed to flow between any pair of atoms in our model, unlike other models that enforce *a priori* constraints on pairwise charge flow.[7-19] Hence by definition, the physically interesting sets of charge–transfer variables must have graphs *G* that are spanning trees. An elementary result of graph theory[20] immediately yields that trees that connect all *N* vertices of *G* have $N-1$ edges, since adding any more edges would introduce a cycle. Hence in order for QTPIE to be consistent with the conservative nature of the electrostatic potential, only $N-1$ charge–transfer variables can be linearly independent.

In summary, the linear dependency of the full set of charge-transfer variables is reflected in the rank-deficiency of the bond hardness matrix **A**, which is defined in Eq. (3.2). In the next section, we prove that the matrix **T** has rank N – 1, and in the following section, we provide a formal proof using the theory of matroids,[21, 22] showing the existence of an isomorphism between the combinatorial properties of matrices and the combinatorial properties encapsulated in suitably defined graphs. The reader who is not interested in the formal proofs may skip directly to Section 3.6 with no loss of continuity.

### 3.4. The rank of the mapping from bond space to atom space

In this Section, an explicit construction by Prof. Dirk Hundermark is provided that proves that the rank of the matrix **T** which transforms from bond space to atom space is $N - 1$, where $N$ is the number of atoms. We can define the linear map **T** as

$$\mathbf{T} : \mathbb{R}^{N(N-1)/2} \to \mathbb{R}^{N-1} \cong \left\{ \mathbf{q} \in \mathbb{R}^{N} \mid \sum_{i=1}^{N} q_i = 0 \right\} \tag{A1}$$

$$\mathbf{q} = \mathbf{Tp} \tag{A2}$$

It is sufficient to show that **T** is surjective, a.k.a. onto. Physically, this is equivalent to showing that for every possible charge configuration **q** with total charge zero, there is at least one set of charge transfer variables that gives rise to that charge configuration.

From the charge continuity relation Eq. (2.6) and the antisymmetry of the charge transfer variables, we have

$$\begin{aligned} q_1 &= p_{21} + p_{31} + \dots + p_{N1} \\ q_2 &= p_{12} + p_{32} + \dots + p_{N2} \\ &= -p_{21} + p_{32} + \dots + p_{N2} \\ q_3 &= p_{13} + p_{23} + p_{43} + \dots + p_{N3} \\ &= -p_{31} - p_{32} + p_{43} \dots + p_{N3} \\ &\vdots \\ q_N &= p_{1N} + p_{2N} + \dots + p_{N-1,N} \\ &= -p_{N1} - p_{N2} - \dots - p_{N,N-1} \end{aligned} \tag{A3}$$

Then we have an algorithm for constructing a set of charge transfer variables compatible with an arbitrary charge distribution.

**Algorithm.** Given the charge variables $\mathbf{q} = \left(q_1, q_2, \dots, q_N\right)$,

1. Set $p_{21} = q_1$ and $p_{k1} = 0$ for all $k > 2$. Then
$$q_2 = -p_{21} + p_{32} + \ldots + p_{N2} = -q_1 + 0 + \ldots + 0 \quad \text{(A4)}$$

2. Set $p_{32} = q_1 + q_2$ and $p_{k1} = 0$ for all $k > 3$. Then
$$q_3 = -p_{31} - p_{32} + p_{43} \dots + p_{N3} = -0 - \left(q_1 + q_2\right) + 0 + \dots + 0 \quad \text{(A5)}$$

3. Continue similarly, i.e. set $p_{n,n-1} = \sum_{i=1}^{n} q_i$ and $p_{kn} = 0$ for all $k > n+1$.

This therefore gives a recipe providing one set of charge transfer variables that gives rise to that charge configuration, and is always possible for any arbitrary (overall neutral) charge configuration. Thus **T** is onto, i.e.

$$\operatorname{range} \mathbf{T} = \mathbb{R}^{N-1} \quad \text{(A6)}$$

and therefore

$$\operatorname{rank} \mathbf{T} \equiv \dim \operatorname{range} \mathbf{T} = \dim \mathbb{R}^{N-1} = N - 1 \quad \text{(A7)}$$

A different proof of this fact is provided on p. 102 of Ref. 23.

### 3.5. The relationship between the linear dependencies of charge transfer variables and the rank of the bond hardness matrix

In this section we provide another formal proof that the bond hardness matrix **A** must have rank $N-1$ as stated above, which was justified from an intuitive counting argument of the degrees of the freedom in the problem. We also explore its implications for the linear dependencies of charge transfer variables. The proof is most elegantly stated in the language of matroid theory.[24] We use the notation $|X|$ to denote the cardinality of a set *X* and furthermore assume familiarity with basic concepts of set theory and graph theory. We omit proofs of established results which may be found in any standard text on matroid theory.

**Definition 3.4.1**. A matroid $M$ is an ordered pair $(E, \mathfrak{I})$ where $\mathfrak{I}$ is the set of subsets of $E$ such that $\mathfrak{I}$ contains the empty set, i.e. $\{\ \} \in \mathfrak{I}$, all subsets of elements of $\mathfrak{I}$ are themselves elements of $\mathfrak{I}$, i.e. for all $I \in \mathfrak{I}$ and $I' \subset I$, then $I' \in \mathfrak{I}$, and $\mathfrak{I}$ obeys the independence augmentation axiom, i.e. for all $I_1, I_2 \in \mathfrak{I}$ such that $|I_1| < |I_2|$, $e \in I_2 - I_1$ such that $I_1 \cup \{e\} \in \mathfrak{I}$. $E$ is called the ground set of $M$ and an element of $\mathfrak{I}$ is called an independent set.

**Definition 3.4.2**. Two matroids $M1$ and $M2$ are isomorphic, denoted $M_1 \cong M_2$, if there exists a bijection $f : E(M_1) \to E(M_2)$ between the base sets of each matroid, and any subset $X \subseteq E(M_1)$ in $M_1$ is independent if and only if its image $f(X)$ is also independent in $M_2$.

**Lemma 3.4.1**. Let **A** be a real square matrix of dimension $\frac{1}{2}N(N-1)$ with columns $a_1, \ldots, a_{N(N-1)/2}$. Then there exists a matroid $M[\mathbf{A}]$ called the vector matroid induced by **A** with columns $E = \left\{ a_v : v = 1, \ldots, \frac{1}{2}N(N-1) \right\}$ forming the ground set and independent sets as linearly independent subsets of $E$. **A** is called a $\mathbb{R}$-representation of $M[\mathbf{A}]$.

**Lemma 3.4.2**. Let $G = (V(G), E(G))$ be a graph with vertices $v_i \in V(G)$ and edges $e_{ij} \in E(G)$ connecting $v_i$ and $v_j$. Then there exists a matroid $M(G)$ called the cycle matroid with ground set equal to the edge set, i.e. $E(M(G)) = E(G)$ and independent sets corresponding to acyclic subsets of $E(G)$. We note that $G = K_N$, the graph of $N$ vertices connected by all possible unique edges, describes the connectivity of our QTPIE charge model. We do not assume any a priori connectivity information and hence our model must in principle consider all possible pairwise charge transfers. This is formalized in the following lemma.

**Lemma 3.4.3**. Consider the linear algebra problem $\sum_{v} A_{v\lambda} p_v = V_\lambda$ where **A** is a real square matrix of dimension $\frac{1}{2} N (N-1)$. Then there exist a bijection between $E(M[\mathbf{A}])$ and $\{p_v\}$, a bijection between $E(M[\mathbf{A}])$ and $\{V_v\}$, and a bijection between $E(M[\mathbf{A}])$ and $E(K_N)$.

**Proof**. The first two are trivial, since any index of the columns of **A** also indexes the corresponding row entry of $p_v$. From the rules of matrix-vector multiplication, the identity mapping

$$f: E(M[\mathbf{A}]) \to \{p_v\}, f(a_v) = p_v \tag{3.6}$$

is a trivial bijection. Furthermore since **A** is symmetric, every column of **A** is identical to the transpose of its corresponding row, and so the rules of matrix multiplication also show that the identity mapping

$$g: E(M[\mathbf{A}]) \to \{V_v\}, f(a_v) = V_v \tag{3.7}$$

is another trivial bijection. The third bijection is the following identity mapping

$$h: E(M[\mathbf{A}]) \to E(M(K_N)) = E(K_N), h(a_v) = e_{ij} \tag{3.8}$$

where $e_{ij}$ is the edge connecting the vertices $v_i$ and $v_j$ and $v$ is related to *i* and *j* by

$$v(i,j) = \tfrac{1}{2}\max(i,j)(\max(i,j)-1) + \min(i,j) \tag{3.9}$$

Q.E.D.

An immediate consequence of the preceding lemma is that the each edge $e_{ij} \in E(K_N)$ in the graph $K_N$ can be associated with two weights $p_v$ and $V_v$.

We now prove that Kirchhoff's voltage law determines the various properties of **A** that we had claimed earlier. To do so we first prove this main theorem.

**Theorem 3.4.1**. Let **A** be the matrix defined in the preceding lemma and furthermore let the set $\{V_v\}$ obey the holonomic constraints that for all dependent subsets $X \subseteq E(K_N)$, $\sum_{v \in g(h^{-1}(X))} v = 0$. Then $M[\mathbf{A}] \cong M(K_N)$.

**Proof**. The map $h$ defined in Lemma 3.4.3 lemma provides the necessary bijection to demonstrate isomorphism. Now consider $X \subseteq E(M[\mathbf{A}])$. We now want to show that $h(X) \subseteq E(M(K_N))$ is independent in $M(K_N)$ if and only if $X \subseteq E(M[\mathbf{A}])$ is independent in $M[\mathbf{A}]$.

First suppose that $X \subseteq E(M[\mathbf{A}])$ is a dependent set. Then its elements must be linearly dependent, i.e. there exists real coefficients $\{c_\mu : c_\mu \in \mathbb{R} \setminus \{0\}\}$ such that $\sum_{\mu=1}^{|X|} c_\mu x_\mu = \mathbf{0}$ for $x_\mu \in X$. This implies that $\mathbf{0} = \sum_{\mu=1}^{|X|} c_\mu V \cdot x_\mu = \sum_{\mu=1}^{|X|} c_\mu p_\mu$. Since $\{p_\mu\}$ obey detailed balance, there must exist a charge transfer variable $r = -p_v$ such that $c_v r = \sum_{\mu \neq v} c_\mu p_\mu$. This is only possible if there is more than one path connecting the vertices $v_i$ and $v_j$ where one of these paths is provided by the edge $e = h(f^{-1}(p_v))$ and at least one path defined by $h(f^{-1}(X \setminus \{e\}))$ where $v = v(i,j)$ as defined in the preceding lemma. Hence $h(X)$ contains at least one cycle and therefore $h(X)$ is dependent in $M(K_N)$. Taking the contrapositive completes proof of the backward statement.

Now suppose that $Y \subseteq E(K_N)$ is a dependent set, i.e. is a cycle. Then the constraints on $\{V_v\}$ immediately give $\mathbf{0} = \sum_{\mu=1}^{|Y|} V_\mu = \sum_{\mu=1}^{|Y|} x_\mu \cdot p$, showing that the elements of $h^{-1}(Y) \subseteq E(M[\mathbf{A}])$ are linearly dependent. Taking the contrapositive completes proof of the forward statement.

The isomorphism established in the preceding theorem is a very powerful one, for it allows a collaboration of concepts in linear algebra with analogous notions in graph theory. One such instance is in generalizing the notion of basis as follows:

**Definition 3.4.3**. The set $\mathrm{B}$ is a set of bases of a matroid $M$ if and only if $\mathrm{B}$ is not empty, and $\mathrm{B}$ satisfies the basis exchange axiom, i.e. for $B_1, B_2 \in \mathrm{B}$ and $x \in B_1 - B_2$ then there is an element such that $(B_1 - \{x\}) \cup \{y\} \in \mathrm{B}$.

It immediately follows that each element of $\mathrm{B}$ is a maximally independent sets that generalizes the concept of a complete basis that spans the range of a matrix **A**, and that each element of $\mathrm{B}$ has the same cardinality. The generalization of this to graphs is as follows:

**Lemma 3.4.4**. Let $G = (V(G), E(G))$ a graph with *k* components. Then the bases of the corresponding cycle matroid $M(G)$ are the edge sets of spanning forests of *G*, each of cardinality $|V(G)| - k$.

The rank of a matroid is defined as the cardinality of any of its basis sets. The implications for our matrix **A** immediately follow:

**Corollary**. The matrix **A** has rank $N-1$.

**Proof.** The complete graph $K_N$ is connected, and therefore the matroid $M\left(K_N\right)$ has rank $\left|V\left(K_N\right)\right|-1=N-1$. Since $M\left(K_N\right)\cong M\left[\mathbf{A}\right]$, $M\left[\mathbf{A}\right]$ must have the same rank as $M\left(K_N\right)$. Hence $M\left[\mathbf{A}\right]$, and $\mathbf{A}$ itself in turn, must have rank $N-1$.

### 3.6. Numerical solution of the rank-deficient system in Eq. (3.2)

The singular and indefinite nature of the bond hardness matrix **A** necessitates a careful choice of numerical algorithm to solve the QTPIE equations. Singular value decomposition (SVD) has previously been used in the context of electronegativity equalization methods,[2, 25] but is computationally very costly. We now describe a faster algorithm employing complete orthogonal decomposition[26] (COD) which identifies and projects out the nullspace;[3] this is formally equivalent to the method used to find the minimum-norm least-squares solution for underdetermined equations. The key transformation that allows the nullspace of a matrix to be identified numerically is the rank-revealing QR factorization,[3] which is an orthogonal factorization that employs column pivoting to separate the range of a matrix from its kernel. Rank-revealing QR decomposes a square matrix **A** of dimension $M$ and rank $\rho$ into the matrix product

$$\mathbf{A}=\mathbf{Q}\begin{pmatrix}\tilde{\mathbf{R}} & \tilde{\mathbf{S}}\\ \mathbf{0} & \mathbf{0}\end{pmatrix}\mathbf{P}^{-1} \tag{3.10}$$

where **Q** is an orthogonal $M\times M$ matrix, $\tilde{\mathbf{R}}$ is an upper triangular $\rho\times\rho$ matrix, $\tilde{\mathbf{S}}$ is a rectangular $\rho\times\left(M-\rho\right)$ matrix, and **P** is a permutation matrix describing the sequence of pivots used to compute the factorization. Furthermore, it is possible to construct a complete orthogonal decomposition from Eq. (3.10) of the form

$$\mathbf{Q}^T\mathbf{A}(\mathbf{PQ}) = \begin{pmatrix} \tilde{\mathbf{R}} & \tilde{\mathbf{S}} \\ \mathbf{0} & \mathbf{0} \end{pmatrix}\mathbf{P}^{-1}(\mathbf{PQ}) = \begin{pmatrix} \tilde{\mathbf{T}} & \mathbf{0} \\ \mathbf{0} & \mathbf{0} \end{pmatrix} \tag{3.11}$$

Since **A** is symmetric, $\mathbf{Q}^T$, and therefore its permutation $\mathbf{PQ}^T$, can act from the right to zero out $\tilde{\mathbf{S}}$, projecting all useful information about the range of **A** into the square, full-rank submatrix **T** of dimension $\rho\times\rho$. The COD given in Eq. (3.11) is sufficient to construct an algorithm to solve the linear problem $\mathbf{A}\vec{p} = \vec{V}$, which can now be written as

$$\mathbf{Q}^T\mathbf{A}\vec{p} = \begin{pmatrix} \tilde{\mathbf{T}} & \mathbf{0} \\ \mathbf{0} & \mathbf{0} \end{pmatrix}\left(\mathbf{Q}^T\mathbf{P}^{-1}\right)\vec{p} = \mathbf{Q}^T\vec{V} \tag{3.12}$$

This equation shows explicitly that only the rows of **Q** that span the range of $\rho$ contribute to the norm of the solution. It is therefore useful to define a partition of $\mathbf{Q} = \begin{pmatrix}\mathbf{U} & \mathbf{Z}\end{pmatrix}^T$, where **U** is the rectangular matrix **U** of dimension $\rho\times M$ which is formed by the first $\rho$ rows of **Q**. We can therefore calculate the minimum-norm solution $\vec{p}_0$ to Eq. (2.8) using

$$\tilde{\mathbf{T}}\left(\mathbf{U}^T\mathbf{P}\vec{p}_0\right) = \mathbf{U}^T\vec{V} \tag{3.13}$$

which can be solved using conventional techniques such as Cholesky factorization for $\mathbf{U}^T\vec{p}_0$; left multiplication by **PU** completes the algorithm.

We note that had $\tilde{\mathbf{T}}$ been diagonalized, we would have solved the problem using SVD; the computational savings of using this COD algorithm arises precisely from our ability to solve the equations without a complete diagonalization. This results in a reduction in asymptotic complexity from $O\left(M^3\right)$ in SVD to $O\left(\rho M^2\right)$ in COD.[3]

### 3.7. Transformations between bond and atom space

When applied to the QTPIE model given in Eq. (3.2), which has dimension $M = \frac{1}{2}N(N-1) = O(N^2)$ and rank $\rho = N - 1 = O(N)$, the COD algorithm scales as $O(N^5)$ while SVD scales as $O(N^6)$. This therefore represents significant savings in computational costs. However, both algorithms remain costly as the problem is formulated in the space of charge-transfer variables. Perhaps surprisingly, it is possible to derive an explicit transformation from the set of charge-transfer variables to the set of atomic charge variables, thus enabling the reformulation of Eq. (3.2) in a space of significantly smaller dimensionality.

Again, graph theory provides a framework for understanding why. The transformation of variables arises from a dual relationship between the vertex set $V$ and edge set $E$ of a complete graph $G = (V, E)$ of order $N$, i.e. the graph with edges connecting every possible pair of vertices $v \in V$. $G$ then reflects the underlying topology of the QTPIE system in that every atom (represented by vertices) is connected to every other atom.

We now use the convenient notation $e = \overrightarrow{v_j v_i}$ for an edge $e \in E$ that connects two vertices $v_i, v_j \in V$. This can be interpreted as a bookkeeping device for the charge variable $p_{ji}$, which quantifies the amount of charge transferred from atom $j$ to atom $i$. The atoms themselves accounted for by their respective vertices. Recall that the charge-transfer variables are related to the atomic charges by the continuity relation of Eq. (2.6). Using the graph-theoretic notions above, we can consider the atomic charges as a vector $\vec{q} = (q_1, \cdots, q_n) = (q_v)$, $v \in V$ in a vector space $V_q(\mathbb{R}^n)$, which we call the atom space.

Similarly, the charge-transfer variables can be used to define a vector $\vec{p} = \left(p_{21}, \cdots, p_{n,n-1}\right) = \left(p_e\right), e \in E$ in its corresponding vector space $V_p\left(\mathbb{R}^{\frac{1}{2}n(n-1)}\right)$, which we call the bond space. For example, Figure 3.2 shows diagrams that visualize the atom and bond spaces for a single water molecule.

**Figure 3.2.** A schematic diagram of (a) atoms and atom pairs in a water molecule, with atoms enumerated in superscripts; (b) charge-transfer variables in bond space; and (c) charges in atom space.

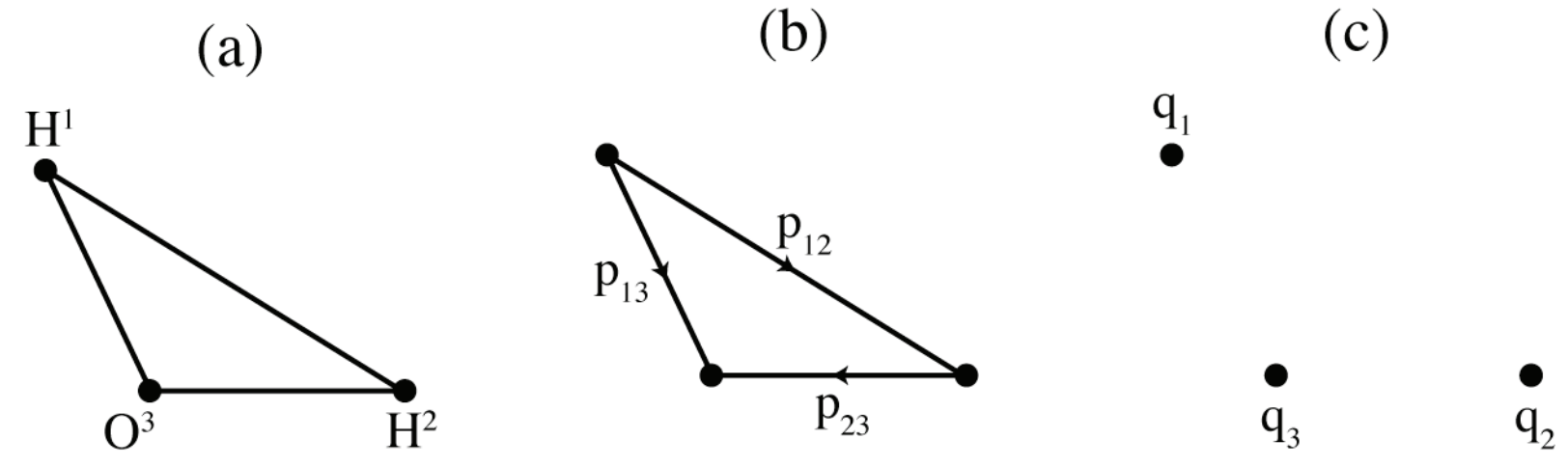

We can now express the relationship between these two sets of variables in terms of a linear transformation $\mathbf{T}$ such that

$$\mathbf{T}: V_p \to V_q, \quad \mathbf{T}\vec{p} = \vec{q} \tag{3.14}$$

When represented by a matrix, $\mathbf{T}$ is identical to the incidence matrix[20] of the underlying directed graph *G*, i.e. $\mathbf{T}$ is the mapping $E(G) \to V(G)$ from edges to vertices of the graph *G*, with elements $T_{ve}$ equal to 1 if the edge *e* connects *v* and points toward *v*, -1 if the edge *e* connects *v* and points away from *v*, and 0 if the edge *e* does not connect the vertex *v* to another vertex.

For QTPIE, QEq and similar models, the lack of topological constraints on the flow of charge means that the underlying graph *G* must be the complete graph of order *N*, which is the graph where each vertex is connected to every other vertex. Furthermore, the antisymmetry of the charge-transfer variables specifies a particular orientation of the graph. We therefore obtain the incidence matrix for all such models as:

$$T_{ve} = \delta_{va} - \delta_{vb},\ e = \overrightarrow{ba} \tag{3.15}$$

where $\delta$ is the usual Kronecker delta. It is easy to verify that Eqs. (3.14) and (3.15) recover the continuity relationship between the charge transfer variables and atomic charges as given in Eq. (2.6). We now wish to compute a transformation matrix $\mathbf{T}'$ that serves as a suitable transformation in the reverse sense, i.e.

$$\mathbf{T}' : V_q \to V_p, \quad \mathbf{T}'\vec{q} = \vec{p} \tag{3.16}$$

Since $V_q$ and $V_p$ in general have different dimensions, **T** is a rectangular matrix. Thus $\mathbf{T}'$ in general cannot be a true inverse, but must be the pseudoinverse, or generalized inverse, that satisfies the Moore-Penrose conditions.[3, 23] Indeed, the rank of T can be easily proven to be of rank *N* - 1[23]. However, we have seen that the forward transformation encoded in the incidence matrix **T** is information preserving, so that there is no information that can be represented in charge transfer variables but not in atomic charges. Therefore the inverse transformation that we seek is not only the pseudoinverse, but the pseudoinverse gives us an inverse transformation that is also information preserving as well.

It is straightforward to verify that the elements of $\mathbf{T}'$ are simply

$$\left(\mathbf{T}'\right)_{ev} = \frac{\delta_{va} - \delta_{vb}}{N},\ e = \overrightarrow{ba} \tag{3.17}$$

so that the inverse relation between the charge and charge-transfer variables is simply

$$p_e = \sum_{v \in V} \mathbf{T}'_{ev} q_v = \sum_{v \in V} \frac{\delta_{va} - \delta_{vb}}{N} q_v = \frac{q_a - q_b}{N} \quad \forall e = \overrightarrow{ba} \in E, \quad b, a \in V \tag{3.18}$$

This relation has a completely unexpected simplicity that allows the original QTPIE energy function of Eq. (2.8) to be rewritten as

$$E = \sum_i q_i \sum_j \frac{\left(\chi_i - \chi_j\right) f_{ij}}{N} + \tfrac{1}{2} \sum_{ij} q_i q_j J_{ij} \tag{3.19}$$

This is our main result, which gives a much more compact set of working equations as we can now solve the QTPIE model in exactly the same way as the QEq model. This reformulated problem is mathematically equivalent to Eq. (3.2) in that the predicted charge distributions are identical. However, the reformulated problem given in Eq. (3.19) is much more amenable to solution with conventional linear algebra algorithms as we have analytically projected out the nullspace in the construction of $\mathbf{T}'$. In the next section, we provide a formal proof of Eq. (3.17).

This reformulation in Eq. (3.19) of the QTPIE model is more than just a mathematical convenience, as it also furnishes some insight into why the model works as well as it does. Our previous expression for the pairwise electronegativity[2] is $\chi_j f_{ij} = \chi_j k_{ij} S_{ij}$ where $\chi_j$ is the electronegativity of atom *j*, $k_{ij}$ is a charge-independent constant factor, and $S_{ij}$ is the overlap integral between atoms *i* and *j*. By substituting this expression into Eq. (3.19), we obtain

$$E = \sum_i q_i \sum_j k_{ij} \frac{\left(\chi_i - \chi_j\right) S_{ij}}{N} + \tfrac{1}{2} \sum_{ij} q_i q_j J_{ij} \tag{3.20}$$

Interestingly, the effect of introducing the bond pairwise electronegativity is to renormalize the atomic electronegativities by an amount that depends on the electronegativities of all other atoms in the system. We previously introduced the overlap integrals as strongly distance-dependent attenuation factors that would allow the charge model to exhibit the correct asymptotic behavior at dissociation limits.[1, 2] These overlap integrals now appear in the atom-space formulation as weighting factors for the averaging of electronegativity differences, and allows the definition of effective atomic electronegativities $v_i = \sum_j k_{ij} \left( \chi_i - \chi_j \right) S_{ij} / N$ that, through the overlap integrals, is sensitive to the molecular environment around each atom. Then after introducing the chemical potential $\mu$ to enforce the charge conservation constraint $\sum_i q_i = 0$, the QTPIE model in atom space reduces to solving the linear system

$$\begin{pmatrix} \mathbf{J} & \mathbf{1} \\ \mathbf{1}^T & 0 \end{pmatrix} \begin{pmatrix} \mathbf{q} \\ \mu \end{pmatrix} = \begin{pmatrix} -\mathbf{v} \\ 0 \end{pmatrix} \tag{3.21}$$

The matrix above is real, symmetric and full-rank, but indefinite, thus necessitating some care in the choice of the numerical algorithms used to solve it.

One final detail to consider is the charge-independent factor $k_{ij}$, which was introduced as part of the pairwise electronegativity in QTPIE.[2] While we had initially set $k_{ij}$ to constant values for the small molecules reported in Chapter 2 and our earlier work,[2] the energy function re-expressed in atomic variables as in Eq. (3.20) makes it clear that $k_{ij}$ must scale as $N$ in order to guarantee the correct size-extensivity of the atomic electronegativities. However, this still does not allow us to completely define $k_{ij}$, except to note that it must not depend on the charge: it is still possible for it to depend parametrically on external factors such as the molecular geometry. Considering that QTPIE was created as a refinement of QEq, it is reasonable to specify $k_{ij}$ in such a way that the predicted charge distribution of QTPIE reduces to that of QEq in some way. Two reasonable choices then present themselves. First, we can specify $k_{ij}$ such that QTPIE will reduce to QEq for all possible diatomic systems at some fixed bond length, e.g. at equilibrium bond lengths $R_{ij}^0$. Then $k_{ij}$ must have the form

$$k_{ij} = \frac{N}{S_{ij}\left(R_{ij}^0\right)} \tag{3.22}$$

where $S_{ij}\left(R_{ij}^0\right)$ is the overlap integral between the basis functions for atoms *i* and *j* when their centers are separated by a distance of $R_{ij}^0$. This leads to the effective atomic electronegativity

$$v_i = \sum_j \left(\chi_i - \chi_j\right) \frac{S_{ij}}{S_{ij}\left(R_{ij}^0\right)} \tag{3.23}$$

It is straightforward to show that this choice makes QTPIE agree with QEq for diatomic systems at the bond length $R_{ij}^0$. Alternatively, we can choose

$$k_{ij} = \frac{N}{\sum_{j'} S_{ij'}} \tag{3.24}$$

This second choice, which is independent of the index *j*, leads to the effective atomic electronegativity

$$v_i = \frac{\sum_j \left(\chi_i - \chi_j\right) S_{ij}}{\sum_{j'} S_{ij'}} \tag{3.25}$$

In this latter choice of $k_{ij}$, the effective atomic electronegativities have a particularly appealing form as it turns out to be the averaged electronegativity differences relative to every other atom and weighted by the corresponding overlap integrals.

We have found empirically that both choices for $k_{ij}$ show very similar behavior for equilibrium geometries and have similar rates of approach to dissociation limits, and hence that choosing either prescription does not significantly alter the qualitative nature of the charge distribution. None of the results reported in this Chapter depend significantly on resolving this ambiguity. We now proceed to prove Eq. (3.17), i.e. that the pseudoinverse of the incidence matrix **T** is proportional to its transpose. Again, the reader who is not interested in the formal details may skip forward to Section 3.9 without loss of continuity.

### 3.8. The pseudoinverse of the incidence matrix of the complete graph

Eq. (3.17) states that the pseudoinverse of the incidence matrix of the complete graph $K_N$ is proportional to its transpose, with a constant prefactor of 1 / *N*. We now prove this.

**Definition 3.7.1.** The incidence matrix **T** = **T** (*G*) of an oriented graph *G* is defined as the matrix **T** with elements

$$T_{ij} = \begin{cases} +1 & \text{if edge } e_j \text{ enters vertex } v_i \\ -1 & \text{if edge } e_j \text{ leaves vertex } v_i \\ 0 & \text{else} \end{cases} \tag{3.26}$$

Thus **T** is a linear mapping from the set of edges to the set of vertices.

**Lemma 3.7.1.** The column sum of each column of **T** is equal to 0, i.e.

$$\sum_j T_{ij} = 0 \tag{3.27}$$

**Proof**. Every edge $e_i$ connects exactly one vertex $v_{j'}$ in a positive sense and one vertex $v_{j''}$ in a negative sense. In other words, there exist a $j'$ and $j''$ such that $T_{ij'} = +1$ and $T_{ij''} = -1$. For all other values of *j*, $T_{ij} = 0$. Thus the sum evaluates to

$$\sum_j T_{ij} = 0 + \ldots + 0 + 1 + 0 + \ldots + 0 + (-1) + 0 + \ldots = 0 \tag{3.28}$$

**Definition 3.7.2**. The degree matrix $\mathbf{D} = \mathbf{D}(G)$ of an oriented graph $G$ is defined as the matrix with elements

$$D_{ij} = \deg v_i \, \delta_{ij} \tag{3.29}$$

where $\delta$ is the Kronecker delta and deg $v_i$ is the degree of vertex $v_i$, i.e. how many other vertices it is connected to. Note that $\mathbf{D}$ is a diagonal matrix with dimensions equal to the number of vertices in $G$.

**Definition 3.7.3.** The adjacency matrix $\mathbf{C} = \mathbf{C}(G)$ of an oriented graph $G$ is defined as the matrix with elements $C_{ij}$ equal to the number of edges connecting vertices $v_i$ and $v_j$. Note that $\mathbf{C}$ is a matrix with dimensions equal to the number of vertices in $G$.

**Definition 3.7.4.** The Kirchhoff matrix or Laplacian matrix $\mathbf{\Delta} = \mathbf{\Delta}(G)$ of an oriented graph is defined as

$$\mathbf{\Delta} = \mathbf{D} - \mathbf{C} \tag{3.30}$$

**Lemma 3.7.2.** The Laplacian matrix of a simple directed graph $G$ is the outer product of the incidence matrix with itself, i.e.

$$\mathbf{\Delta} = \mathbf{T}\mathbf{T}^T \tag{3.31}$$

**Proof.** By explicit calculation,

$$
\begin{aligned}
&\left(\mathbf{TT}^T\right)_{ij} \\
&= \sum_k T_{ik} T_{jk} \\
&= \sum_k \begin{cases} +1 & \text{if edge } e_k \text{ enters vertex } v_i \\ -1 & \text{if edge } e_k \text{ leaves vertex } v_i \\ 0 & \text{else} \end{cases} \times \begin{cases} +1 & \text{if edge } e_k \text{ enters vertex } v_j \\ -1 & \text{if edge } e_k \text{ leaves vertex } v_j \\ 0 & \text{else} \end{cases} \\
&= \sum_k \begin{cases} +1 & \text{if edge } e_k \text{ enters vertex } v_i \text{ and edge } e_k \text{ enters vertex } v_j \\ +1 & \text{if edge } e_k \text{ leaves vertex } v_i \text{ and edge } e_k \text{ leaves vertex } v_j \\ -1 & \text{if edge } e_k \text{ enters vertex } v_i \text{ and edge } e_k \text{ leaves vertex } v_j \\ -1 & \text{if edge } e_k \text{ leaves vertex } v_i \text{ and edge } e_k \text{ enters vertex } v_j \\ 0 & \text{else} \end{cases}
\end{aligned}
\tag{3.32}
$$

The cases where $T_{ik}T_{jk} = +1$ can only occur when $i = j$, as an edge can neither enter two different vertices nor leave two different vertices. For the cases where $T_{ik}T_{jk} = -1$, consider the diagonal and off-diagonal subcases separately. The diagonal subcase corresponds to a loop, i.e. a directed edge starting and ending on the same vertex, which would evaluate to +1 + ( –1) = 0. Hence on the diagonal, where $i = j$, have by definition of degree

$$
\begin{aligned}
\left(\mathbf{TT}^T\right)_{ii} &= \sum_k \begin{cases} +1 & \text{edge } e_k \text{ connects some vertex to vertex } v_i \\ 0 & \text{else} \end{cases} \\
&= \deg v_i
\end{aligned}
\tag{3.33}
$$

and on the off diagonal $i \neq j$, have

$$
\begin{aligned}
\left(\mathbf{TT}^T\right)_{ij} &= \sum_k \begin{cases} -1 & \text{edge } e_k \text{ connects vertex } v_i \text{ to vertex } v_j \\ 0 & \text{else} \end{cases} \\
&= \begin{cases} -1 & \text{vertex } v_i \text{ is connected to vertex } v_j \\ 0 & \text{else} \end{cases}
\end{aligned}
\tag{3.34}
$$

as the each pair of vertices is connected by at most one edge $e_k$ for a simple graph.

Adding the two cases shows that $\mathbf{TT}^T = \mathbf{D} + (-1)\mathbf{C} = \boldsymbol{\Delta}$, which is the desired result.

**Lemma 3.7.3.** For the complete graph $G = K_N$,

$$\boldsymbol{\Delta} = N\mathbf{I} - \mathbf{11}^T \tag{3.35}$$

where **I** is the identity matrix and **1** is a column vector of ones.

**Proof.** Each node in $K_N$ has degree $N - 1$, as it is connected to every other node by exactly one edge. Thus the adjacency matrix has entry 1 on every off-diagonal and 0 on the diagonal. Simple arithmetic thus yields the desired result, noting that $\mathbf{11}^T$ produces a square matrix with each entry equal to one.

We note that a generalization of this result to arbitrary graphs has been provided by Ijiri.[27]

**Theorem 3.7.1.** For the complete graph $G = K_N$, the Moore-Penrose pseudoinverse of the incidence matrix is proportional to its transpose, i.e.

$$\mathbf{T}'(K_N) = \frac{1}{N}\mathbf{T}^T(K_N) \tag{3.36}$$

**Proof.** The Moore-Penrose pseudoinverse, if it exists, is unique. Hence it is sufficient to show that $\mathbf{T}^T / N$ obeys the Moore-Penrose conditions, i.e.

$$\begin{aligned} \mathbf{T} &= \mathbf{TT}'\mathbf{T} \\ \mathbf{T}' &= \mathbf{T}'\mathbf{TT}' \\ (\mathbf{TT}')^* &= \mathbf{TT}' \\ (\mathbf{T}'\mathbf{T})^* &= \mathbf{T}'\mathbf{T} \end{aligned} \tag{3.37}$$

Since **T** is real, showing that $\mathbf{T}' = \mathbf{T}^T / N$ is equivalent to demonstrating the following conditions hold:

$$
\begin{aligned}
N\mathbf{T} &= \mathbf{T}\mathbf{T}^T\mathbf{T} \\
N\mathbf{T}^T &= \mathbf{T}^T\mathbf{T}\mathbf{T}^T \\
\left(\mathbf{T}\mathbf{T}^T\right)^T &= \mathbf{T}\mathbf{T}^T \\
\left(\mathbf{T}^T\mathbf{T}\right)^T &= \mathbf{T}^T\mathbf{T}
\end{aligned}
\tag{3.38}
$$

The last two conditions hold trivially. To prove the other two, note that Lemma 3.7.2 implies that the only remaining condition to prove is

$$N\mathbf{T} = \mathbf{\Delta}\mathbf{T} \tag{3.39}$$

as taking the transpose recovers the other condition, noting that the Laplacian matrix is symmetric and thus $\mathbf{\Delta}^T = \mathbf{\Delta}$ Using Lemma 3.7.3, the right hand side evaluates to

$$\mathbf{\Delta}\mathbf{T} = \left(N\mathbf{I} - \mathbf{1}\mathbf{1}^T\right)\mathbf{T} = N\mathbf{T} - \mathbf{1}\left(\mathbf{1}^T\mathbf{T}\right) \tag{3.40}$$

Finally, note that Lemma 3.7.1 implies that $\mathbf{1}^T\mathbf{T} = 0$ and hence the desired result follows.

The analysis of this section turns out to be isomorphic to the studies of the algebra of dc circuits by Bott and Duffin,[28, 29] who first introduced the notion of generalized network inverse that is essentially identical to the notion of pseudoinverse discussed in this chapter.[23] The implications of this Bott-Duffin inverse have been explored to develop the notions of circuit duality and its implications for circuit theory.[30-32]

## 3.9. Results and discussion

The earlier graph-theoretic analysis shows that there is an algebraic isomorphism between models formulated using either atomic charges in atom space or charge-transfer variables in bond space. QTPIE was simpler to formulate in bond space as this allowed us to construct electronegativities that are explicitly pairwise dependent. However, Eq. (3.21) has significantly lower computational complexity owing to the smaller size of the linear system in atom space and thus has a significant numerical advantage over the corresponding bond-space formulation of Eq. (3.2). The pair of transformations given by Eqs. (2.6) and Eq. (3.18) thus allows us to have the best of both spaces.

We now turn to the pragmatic issue of solving Eqs. (3.2) and (3.21). Figure 3.3 shows how the execution time of the various implementations of the QTPIE model scales with system size for linear water chains of increasing length (one representative water chain is shown in the inset). We do not exploit sparsity in any way for these tests, using dense matrix multiplication and factorization routines throughout. Thus, these results should be considered as upper bounds on the computational costs of the various solution methods.

We solved the model in the bond space using both the SVD and COD approaches detailed above. Singular value decomposition was carried out using the DGELSS routine from the LAPACK linear algebra library.[26] The COD method was implemented using routines from the LAPACK and BLAS libraries. The algorithm is similar to the LAPACK routine DGELSY, but without the time-consuming step of numerical rank determination since for our problems the rank of these matrices are known exactly. Both SVD and COD methods scale as $O(N^6)$ for the range of system sizes investigated here. However, the COD method is faster by roughly an order of magnitude. In practice, we also find that COD tended to be numerically unstable without preconditioning; however, a simple diagonal (Jacobi) preconditioner was sufficient to observe convergence.

In contrast to the overcomplete bond-space problem, the reformulated atom-space model of Eq. (3.21) can be solved much more efficiently due to the intrinsically smaller matrix. A direct solution using the DGESV routine from LAPACK showed an asymptotic complexity of $O(N^{2.6})$ while an implementation of the iterative generalized minimal residuals (GMRES) algorithm[5] using dense matrix multiplications exhibited an asymptotic complexity of $O(N^{1.8})$. The charges computed using both atom-space methods are identical essentially to within machine precision, while the charges computed using the bond-space methods agree to three decimal places, reflecting the greater intrinsic numerical instability of the bond-space problem. Thus as expected, our model is much more practical to solve when reformulated in atom-space charge variables compared to its original formulation in bond-space charge-transfer variables. A complete implementation of the QTPIE model using the direct and iterative algorithms for the atom-space formulation is provided in Appendix B.

The transformations of Eqs. (2.6) and (3.18) illustrate the existence of an underlying topological duality between models formulated in atom space, such as QEq, and models formulated in bond space, such as QTPIE. With these transformations, any bond-space model can be related to an equivalent atom-space model that predicts the same charge distribution, and vice versa. The reformulation of atom-space models in bond space is trivial. For the reverse case, consider the most general charge model in bond space that has a quadratic energy function:

$$E = \sum_{i,j=1}^{N} L_{ij} p_{ji} + \sum_{i,j,k,l=1}^{N} M_{ijkl} p_{ji} p_{lk} \tag{3.41}$$

**Figure 3.3.** Execution time of QTPIE for coplanar, linear chains of water molecules (such as the one shown in the inset) for four methods of solving the system of equations: bond-space singular value decomposition (SVD) (red dotted line), bond-space complete orthogonal decomposition (COD) (blue dashed line), atom-space direct matrix solver (green dash-dotted line), and atom-space iterative solution using the generalized minimal residuals (GMRES) algorithm (black solid line). All calculations were run on a single core of an AMD Opteron 175 CPU with 2.2 GHz clock rate.

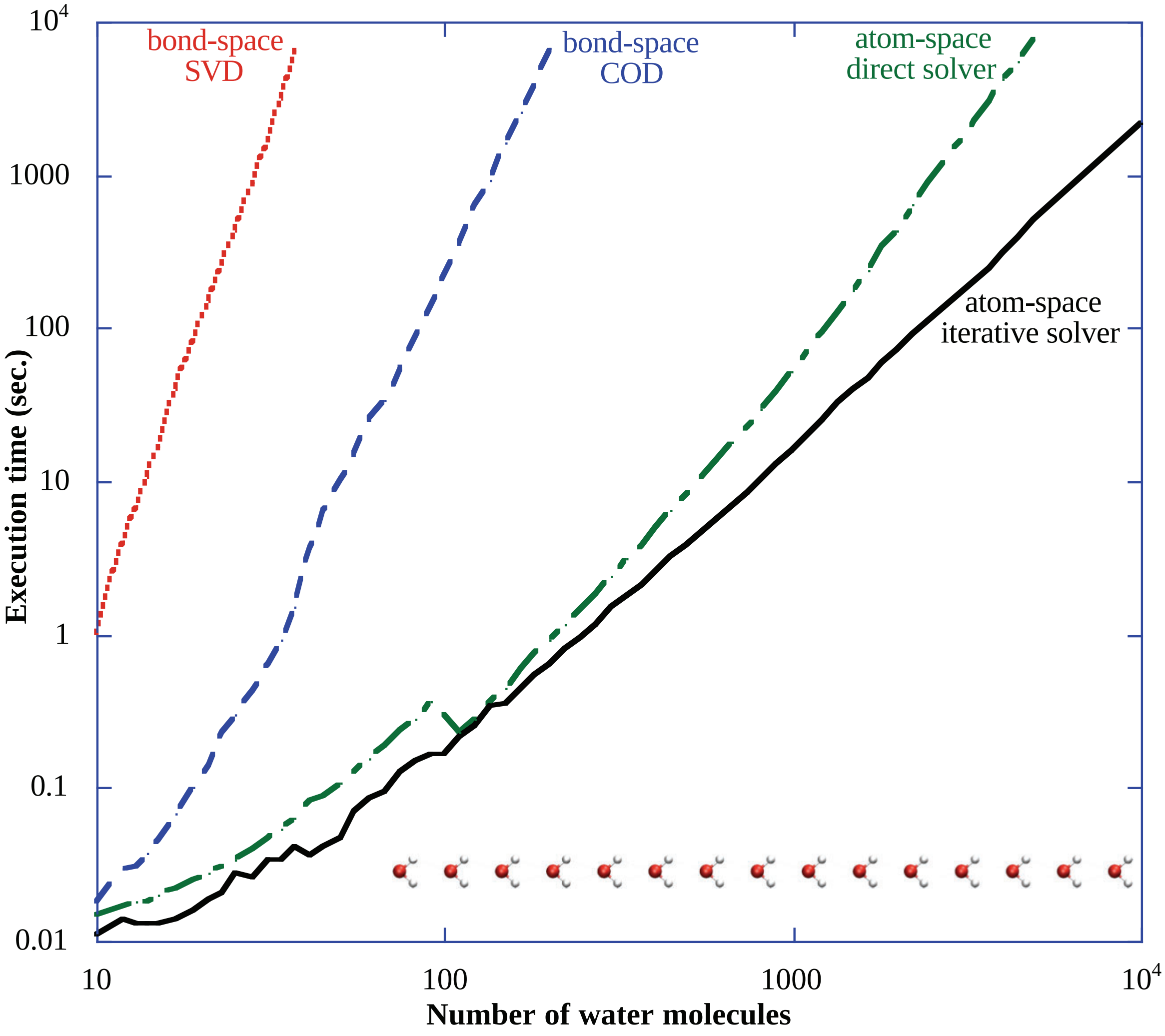

It is straightforward to show from Eq. (3.18) that this bond-space model is exactly equivalent to the analogous model in atom space

$$E = \sum_{i=1}^{N} X_i q_i + \sum_{i,j=1}^{N} Y_{ij} q_i q_j \tag{3.42}$$

where

$$X_i = \sum_{j=1}^{N} \left( L_{ij} - L_{ji} \right) / N \tag{3.43}$$

$$Y_{ij} = \sum_{k,l=1}^{N} \left( M_{ikjl} - M_{ilkj} - M_{kijl} + M_{kilj} \right) / N^2 \tag{3.44}$$

Thus any quadratic charge model in bond space can be rewritten exactly as an equivalent quadratic charge model in atom space, which can be solved more efficiently. These equations can be used to generalize any diatomic model to arbitrary polyatomic systems, including models that include bond hardnesses[33-35] which have until now have not been successfully generalized to multiple atoms without remaining in bond space, such as the atom-bond electronegativity equalization method (ABEEM) [36-40] or the atom-atom charge transfer (AACT) model.[41, 42] In particular, this analysis highlights the severe linear dependency problems in the ABEEM model, as that model employs both charge and charge transfer variables, whereas we have already shown that either set of variables is sufficient to encapsulate all the relevant information about the charge distribution. Thus even the ABEEM model can be reformulated exactly as an atom-space model with renormalized parameters. This analysis can also be easily extended to much more general charge models containing terms of arbitrary order in the charge-transfer variables and atomic charges respectively.

Our analysis also provides a straightforward prescription for deriving mappings analogous to Eqs. (2.6) and (3.18) for fluctuating-charge models with *a priori* topological constraints on charge flow.[7, 9-15, 18, 19, 25, 42-47] In such models, the mapping from atom space to bond space is still represented by the incidence matrix of the graph encoding the topological constraints. Although the reverse mapping $\mathbf{T}'$ may not be as simple as that given in Eq. (3.18), it can nevertheless be computed using any method for calculating the pseudoinverse. Interestingly, there exists a specialized algorithm for calculating the pseudoinverse of arbitrary incidence matrices.[27] At any rate, the transformation need only be calculated once for any given model — it is only necessary to recompute $\mathbf{T}'$ when the incidence matrix changes. Furthermore, as long as both mappings have rank $N - 1$, the bijection between bond-space models and atom-space models will still hold.

**Figure 3.4.** Diagram showing the energy of two fluctuating-charge models, QEq (blue dashed line) and QTPIE (black solid line) for a generic diatomic system as a function of the atomic charge for a typical equilibrium geometry, and at dissociation.

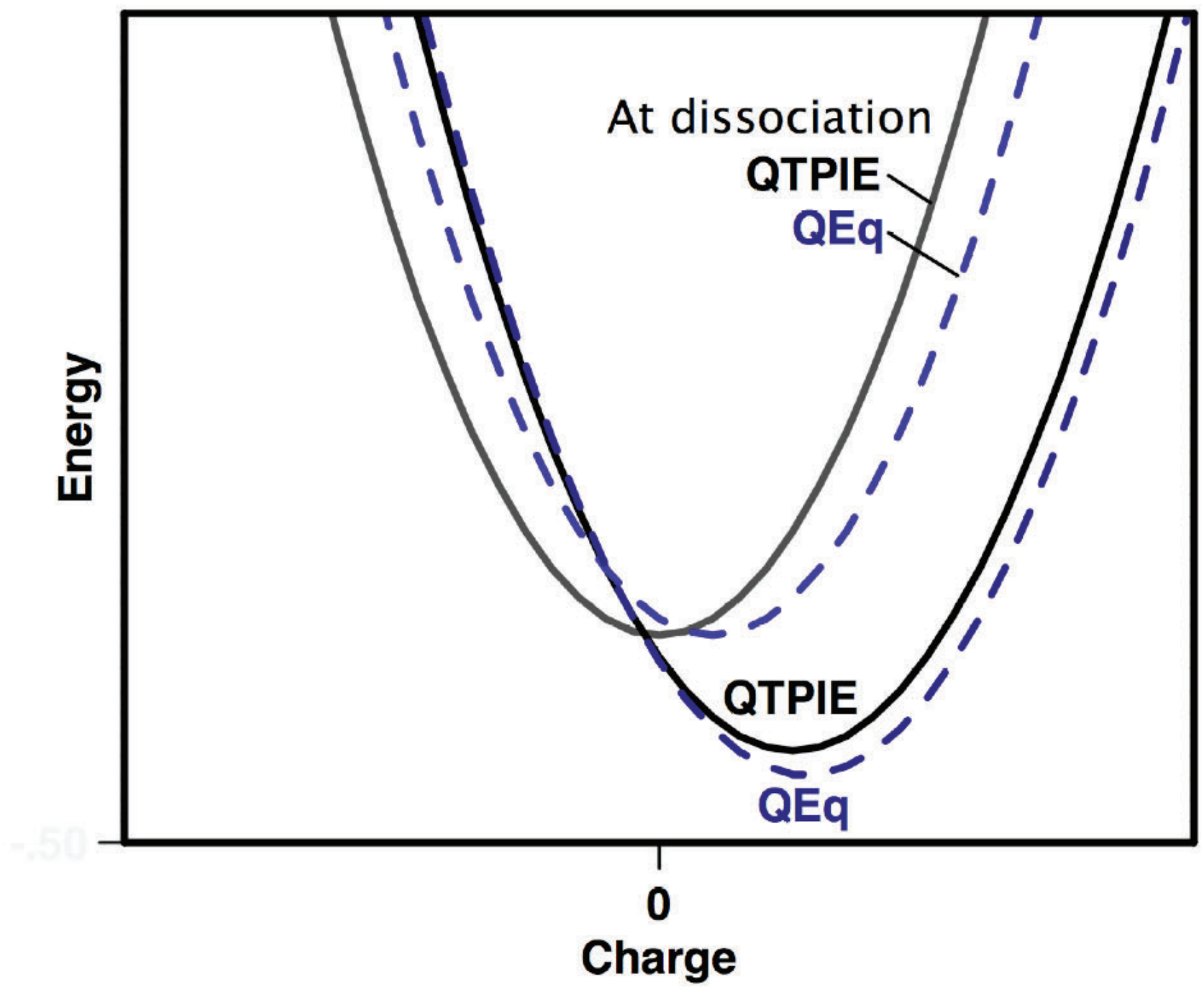

## 3.10. Reexamining QTPIE's dissociation behavior in atom space

The atom-space reformulation of QTPIE in Eq. (3.20) gives us additional insight as to why the QTPIE model is able to show the correct asymptotic behavior at dissociation. The behavior of fluctuating-charge models is intimately related to the properties of density functional theory.[48] It is well known that in the infinitely separated, noninteracting limit, exact density functional theory requires that the energy become nondifferentiable at integer particle numbers.[1, 49-54] This derivative discontinuity requires fluctuating-charge models to have energy functions that become piecewise linear and therefore predict integer charges at dissociation limits.[1] This therefore requires the electronegativities to become discontinuous and the chemical hardnesses to exhibit delta-function-like singularities at integer particle numbers.[33, 34, 55, 56] However, previous work in the Martínez group[1, 54] has shown that it is not possible to enforce such behavior with a quadratic energy function without recourse to ensemble densities. In lieu of this, we note that a dimensional analysis of Eq. (3.19) shows that the predicted charge distribution has dimensions of electronegativity divided by hardness. In QTPIE, the electronegativities are modified to vanish at the dissociation limit, in contrast to having hardnesses that become infinite in this limit, which is the behavior obtained from explicit solution of the electronic Schrödinger equation.[1, 33, 54] Either prescription would give us vanishing charge transfer at dissociation, which is sufficient for the purposes of calculating the charge distribution.

In order to further understand the relationship between these two seemingly distinct ways to enforce the dissociation limit, we now examine how the energy in Eq. (3.20) varies as a function of atomic charge for a neutral diatomic molecule after analytically enforcing the charge constraint $q_1 + q_2 = 0$. Figure 3.4 shows the energy functions for both QEq and QTPIE for a generic diatomic system in an equilibrium geometry, as well as their behaviors at the dissociation limit. The minimum of the energy parabola for QTPIE clearly moves toward zero charge at the dissociation limit, whereas this is clearly not the case for QEq. We can understand this behavior analytically: by completing the square and discarding an irrelevant charge-independent constant, the QTPIE energy as a function of the atomic charge is a perfect parabola of the form

$$E = \tfrac{1}{2}\left(J_{11} - 2J_{12} + J_{22}\right)\left(q_1 - q_1^0\right)^2 \tag{3.45}$$

where $q_1^0 = S_{12}\left(\chi_1 - \chi_2\right)\left(k_{12} + k_{21}\right) / \left(J_{11} - 2J_{12} + J_{22}\right)$. In contrast, the QEq energy has a similar parabolic form, but with minimum $q_1^0 = \left(\chi_1 - \chi_2\right) / \left(J_{11} - 2J_{12} + J_{22}\right)$ instead. At the dissociation limit, both $q_1^0$ and $J_{12}$ vanish in QTPIE and the model clearly predicts the expected charge distribution $q_1 = q_2 = 0$; however, $q_1^0$ does not vanish at infinity in the QEq model. We therefore see that the expected asymptotic behavior in QTPIE, which is enforced by the attenuation of the pairwise electronegativities, causes the energy minimum to shift to zero at the dissociation limit.

We now compare the behavior of QTPIE with that of the charge-constrained minimal basis valence bond (CC-VB2)[1] model for the same diatomic system, which has a well-understood foundation in *ab initio* theory. In simplified notation, the CC-VB2 model has a continuous and piecewise differentiable energy function of the form

$$E\left(q\right) = \begin{cases} E_0 + q\chi^+ + q^2\eta^+, q > 0 \\ E_0 + q\chi^- + q^2\eta^-, q < 0 \end{cases} \tag{3.46}$$

where $E_0$ is a constant, $\chi^{\pm}$ is the pairwise electronegativity and $\eta^{\pm}$ is the pairwise chemical hardness on the positive or negative branch as denoted in the superscript. The exact relationship between these parameters and quantities arising from valence bond theory have been discussed in an earlier work;[1] however, it is sufficient for the purposes of this current discussion to know that the energy of CC-VB2 exhibits a local (if nondifferentiable) maximum at $q = 0$ for small internuclear separations, and as the dissociation limit is approached, $q = 0$ becomes a local (nondifferentiable) minimum. As illustrated in Fig. 1 of Ref. 1, there exists an intermediate regime where the energy is approximately flat in the neighborhood of $q = 0$. In this regime, $\chi^{\pm} \approx 0$ and $\eta^{\pm} > 0$ and so the energy is piecewise quadratic and minimized by $q \approx 0$. Thus the energy in this regime can be approximated very well by a single analytic parabola with a minimum at $q = 0$, as is the case for QTPIE. The main qualitative difference between QTPIE and the CC-VB2 model is that while the energy function for the latter becomes piecewise linear at the dissociation limit, the energy of the former remains quadratic. This preceding analysis allows us to conclude that while the quadratic approximation inherent in QTPIE results in the inability to model the correct piecewise linear behavior in the dissociation limit, it nevertheless affords a reasonable description of the charge transfer up to the aforementioned intermediate regime. As the approximation of retaining the quadratic character does not change the predicted charge distributions in both regimes, this is therefore essentially equivalent to the exact behavior from a pragmatic point of view.

## 3.11. Classical electrical circuit representations of fluctuating-charge models

In order to gain further physical insight into fluctuating-charge models, we investigate the physical systems that are described by the same working equations. It turns out that classical dc circuits of capacitors and batteries can be described with the same energy functions as fluctuating-charge models. In other words, molecular systems are described by purely classical dc electrical circuits in fluctuating-charge models.

Fluctuating-charge models assume thermodynamic equilibrium, and therefore the resulting charge distributions they predict must be stable to fluctuations in time. Thus, they can only describe dc circuits in the absence of any net current flow. Therefore, electronegativity equilibration models can only correspond to circuits that contain capacitors and batteries, i.e. dc sources of electromotive forces, since these are the only elementary electrical circuit components that exhibit nontrivial behavior in the absence of any net electrical current flow.

To illustrate the concepts that we will use later, we will consider first the very simple circuit of Figure 3.5, consisting of a single battery with electromotive force (colloquially termed 'voltage') *V* connected to a lone capacitor of capacitance *C*.

---

**Figure 3.5.** A minimal circuit with one capacitor C and one battery V. Each atomic site in QEq can be represented as such a minimal circuit element.

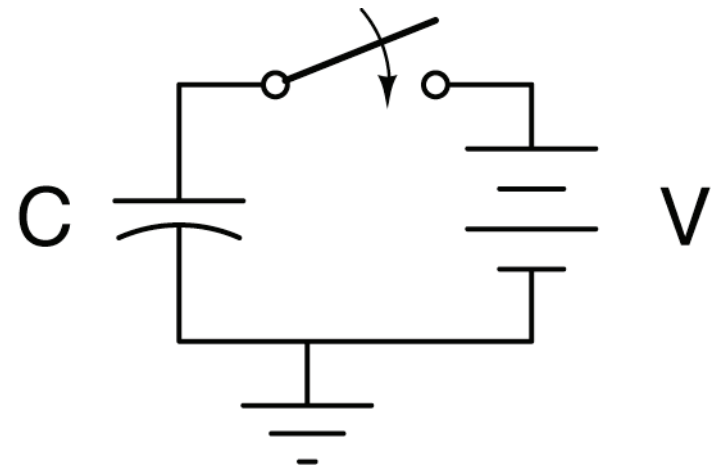

We now calculate the charge $q$ that eventually accumulates on the capacitor once the circuit is closed using a Hamiltonian approach. The energy function for the circuit in Figure 3.5 is

$$E(q) = -Vq + \tfrac{1}{2}C^{-1}q^2 \tag{3.47}$$

where the first term is the work done by the system to charge up the capacitor and is subtracted from the total energy. For this circuit to be in equilibrium the total energy must be minimized; elementary calculus then derives the well-known constitutive relation $q = CV$. The enforcement of the total charge on a system can be implemented straightforwardly by introducing a Lagrange multiplier $\mu$ that corresponds to the chemical potential. This effectively shifts the bias of the ground to a non-zero voltage $\mu$.

The preceding discussion can be used in principle to relate any fluctuating-charge model with the existence of batteries and capacitors respectively. We note that others have explored similar ideas in defining connections between fluctuating charge models and electrical circuit theory, but using resistors instead.[10, 13] This also extends earlier observations that the hardness is inversely related to charge capacitance. [57-64] Existing circuit duality identities permit the transformation of our capacitor-battery circuits into current-resistor circuits; however, we believe that the capacitor-battery circuit model is physically more reasonable since in the absence of magnetic fields, it is not possible for classical physical systems to sustain quasi-steady currents at equilibrium.

We now show that the QEq model corresponds to the circuit in Figure 3.6, created by making $N$ copies of the minimal circuit in Figure 3.5 and connecting them all together with a common ground with bias $\mu$. The dashed lines along the wires denote multiple copies of the minimal circuit omitted from the diagram, while the additional dotted lines between the capacitors represent additional terms arising from mutual interactions between the charge distributions of each capacitor.

The energy function corresponding to the circuit in Figure 3.6 is

$$E\left(q_1,\cdots,q_i,\cdots,q_n\right)=\sum_{i=1}^{N}\left(-V_i q_i+\tfrac{1}{2}C_i^{-1}q_i^2\right)+\sum_{i<j}\left(C^{-1}\right)_{ij}q_i q_j \tag{3.48}$$

where the extra terms are parameterized in terms of coefficients of induction[65] $\left(C^{-1}\right)_{ij}$ that represent the mutual Coulomb interactions of the charges built up on every capacitor. Again, we introduce a bias $\mu$ to the ground voltage as a Lagrange multiplier to enforce the constraint on the total charge $Q=\sum_{i=1}^{N}q_i$. In comparison, the QEq model[66] for a $N$-atom system has the form:

$$E\left(q_1,\cdots,q_n\right)=\sum_{i=1}^{N}\left(\chi_i q_i+\tfrac{1}{2}\eta_i q_i^2\right)+\sum_{i<j}J_{ij}q_i q_j \tag{3.49}$$

---

**Figure 3.6.** The circuit diagram corresponding to the QEq charge model, consisting of *n* minimal circuits (atoms) connected by a common ground.

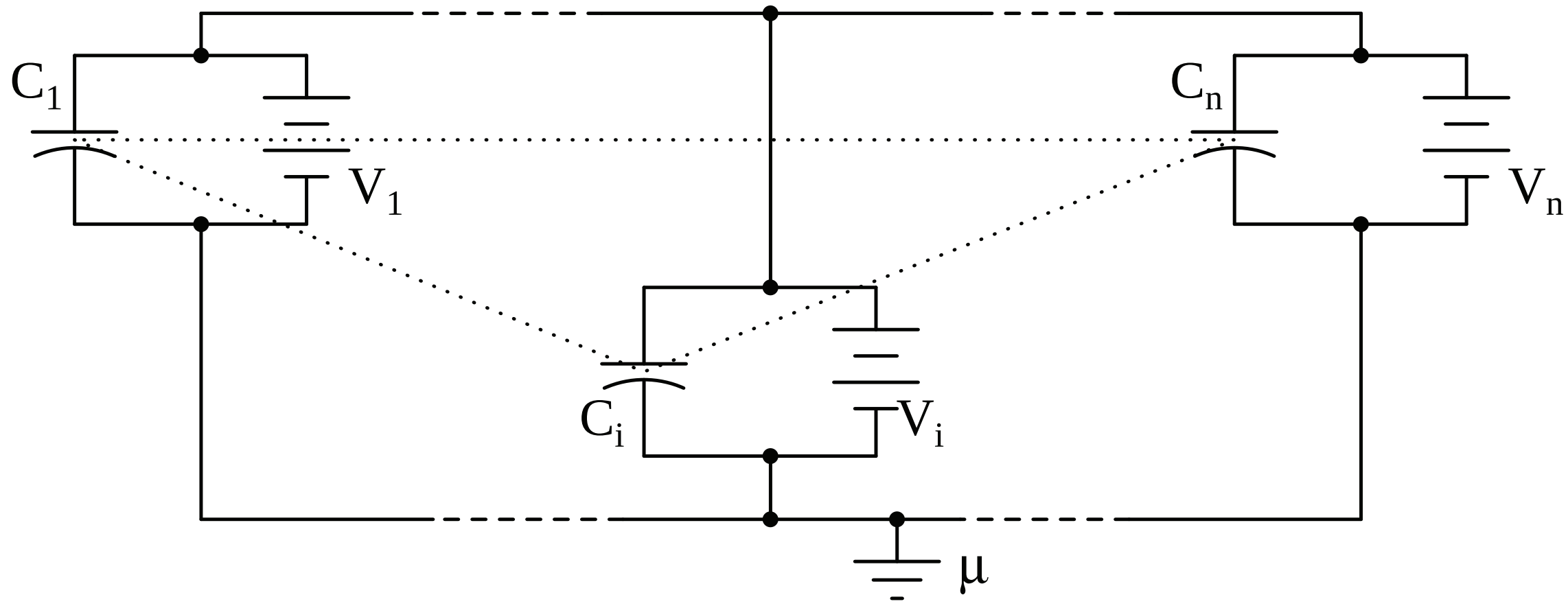

The QEq parameters map perfectly onto the parameters describing a capacitor-battery pair: the electronegativities $\chi_i = -V_i$ are directly related to internal electromotive forces, the chemical hardnesses $\eta_i = C_i^{-1}$ are identical to elastances or inverse capacitances, and the screened Coulomb interactions $J_{ij} = \left(C^{-1}\right)_{ij}$ are equivalent to coefficients of inductance[65]. Furthermore, these relations are dimensionally consistent. Therefore, with only a minor relabeling of the relevant quantities, the QEq model is equivalent to the electrical circuit in Figure 3.6.

From a similar argument we can construct a circuit diagram corresponding to the QTPIE model, shown in Figure 3.7.[2] (Wires not meeting at a dot junction are not connected.) In the same way we constructed the QTPIE model from QEq, we obtain this circuit diagram in two steps. First, the transformation of variables is equivalent to replacing all batteries in Figure 3.6 with equivalent batteries connected along all possible pairs of capacitors. Second, to obtain the QTPIE model, the only essential modification of the QEq model was to allow the electronegativities to be pairwise dependent on the distance between pairs of atoms. Hence, the corresponding circuit elements must be variable voltage dc sources straddling each pair of capacitors.

As discussed earlier, the relationship between bond-space and atom-space fluctuating charge models is intimately related to the notion of circuit duality, which have been studied extensively for dc circuits.[23, 28-32] We will now demonstrate this explicitly using a diatomic system as an illustration. The corresponding circuits in bond space and atom space are given in Figure 3.8. The bond-space equation is given simply by

$$C^{-1} q = V \tag{3.50}$$

and the atom-space equation is

**Figure 3.7.** The circuit diagram corresponding to the QTPIE charge model.

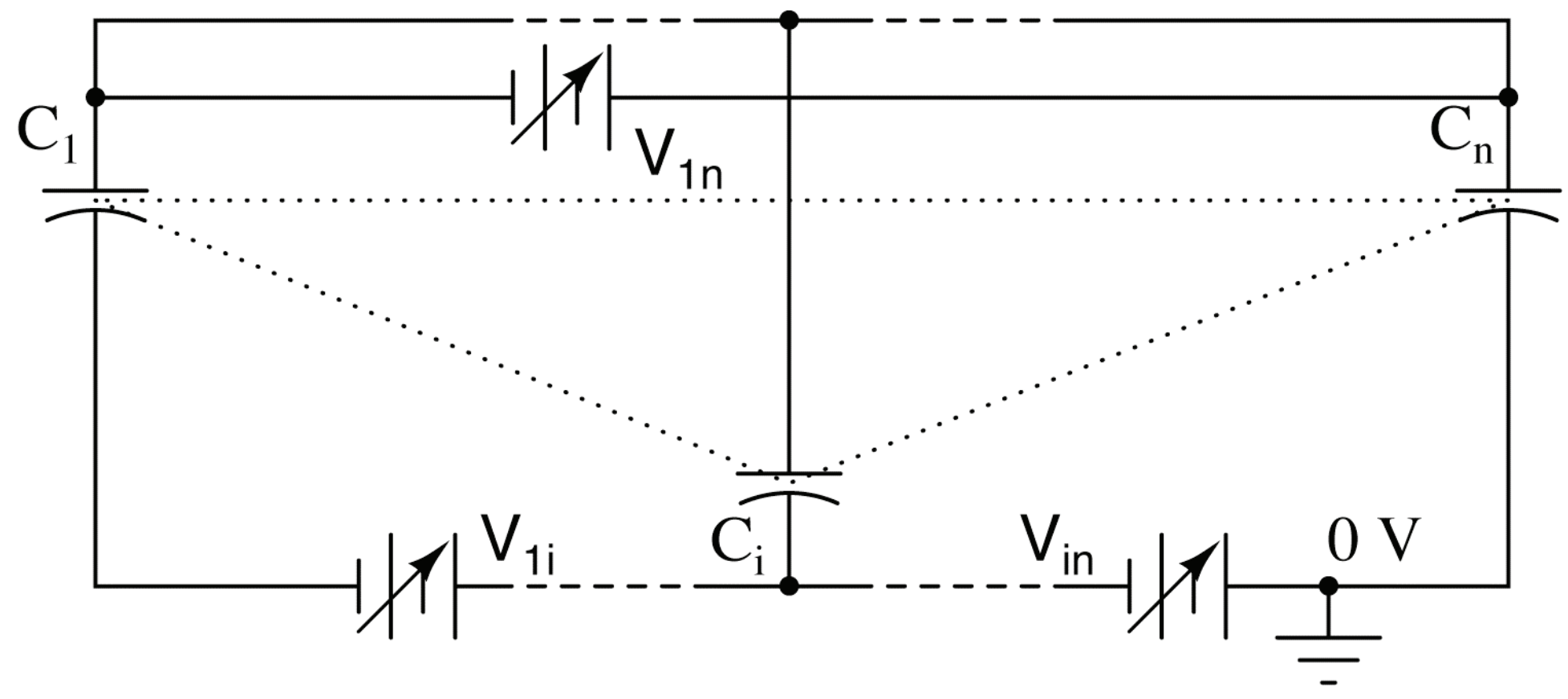

**Figure 3.8.** An illustration of the circuit diagrams for a bond-space fluctuating-charge model (left) and an atom-space fluctuating-charge model (right) for a diatomic system.

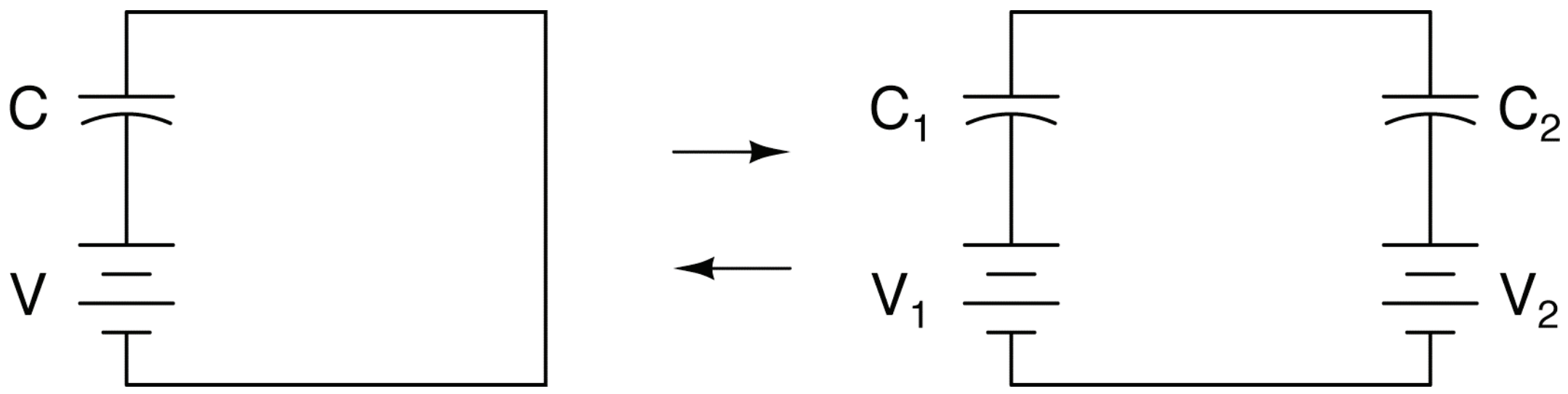

$$\begin{pmatrix} C_1^{-1} & J \\ J & C_2^{-1} \end{pmatrix} \begin{pmatrix} q_1 \\ q_2 \end{pmatrix} = \begin{pmatrix} V_1 \\ V_2 \end{pmatrix} \tag{3.51}$$

where charge conservation requires $q_1 = -q_2$. Topological considerations give the incidence matrix for this system as $\begin{pmatrix} 1 & -1 \end{pmatrix}^T$, which has pseudoinverse $\frac{1}{2}\begin{pmatrix} 1 & -1 \end{pmatrix}$. To transform Eq. (3.50) into a form of the type given in Eq. (3.51), apply the mapping of Eq. (3.18) so that

$$\begin{pmatrix} 1 \\ -1 \end{pmatrix} \left( C^{-1} \right) \frac{1}{2} \begin{pmatrix} 1 & -1 \end{pmatrix} \begin{pmatrix} 1 \\ -1 \end{pmatrix} (q) = \begin{pmatrix} 1 \\ -1 \end{pmatrix} V \tag{3.52}$$

which simplifies to

$$\tfrac{1}{2}\begin{pmatrix} C^{-1} & -C^{-1} \\ -C^{-1} & C^{-1} \end{pmatrix}\begin{pmatrix} q \\ -q \end{pmatrix} = \begin{pmatrix} V \\ -V \end{pmatrix} \tag{3.53}$$

This tells us that the dual circuit with $C_1^{-1} = C_2^{-1} = -J = \tfrac{1}{2}C^{-1}$ and $V_1 = -V_2 = V$ predicts a charge distribution $q_1 = -q_2 = q$ that is compatible with the bond-space charge transfer under the mapping of Eq. (3.18).

Conversely, to transform Eq. (3.51) into a form similar to Eq. (3.50), apply the mapping of Eq. (3.9) to obtain

$$\tfrac{1}{2}\begin{pmatrix} 1 & -1 \end{pmatrix}\begin{pmatrix} C_1^{-1} & J \\ J & C_2^{-1} \end{pmatrix}\begin{pmatrix} 1 \\ -1 \end{pmatrix}\tfrac{1}{2}\begin{pmatrix} 1 & -1 \end{pmatrix}\begin{pmatrix} q_1 \\ q_2 \end{pmatrix} = \tfrac{1}{2}\begin{pmatrix} 1 & -1 \end{pmatrix}\begin{pmatrix} V_1 \\ V_2 \end{pmatrix} \tag{3.54}$$

Taking into account charge neutrality, this simplifies to

$$\left(C_1^{-1} - 2J + C_2^{-1}\right)q_1 = \left(V_1 - V_2\right) \tag{3.55}$$

In this example, the presence of Kirchhoff's law in the mapping of Eq. (3.9) allowed us to derive the well-known combination rule for capacitance $C^{-1} = C_1^{-1} + C_2^{-1}$ (by neglecting the Coulomb coupling $J$ as is common in classical circuit analysis), and the combination rule for voltages $V = V_1 - V_2$, noting that $V_2$ is the voltage of a battery oriented in opposite way relative to $V_1$. In addition, the charge distribution $q_1 = -q_2 = q$ is indeed what we had expected.

While is tempting to associate the charge transfer variables with electrical currents, this is dimensionally inconsistent since they have dimensions of charge, not current. However, a consistent interpretation of these variables is that they are integrated traces of transient currents as the system equilibrates. The variables defined in this manner retain the property of detailed balance, yet are compatible with the concept of equilibrium since there are no net current flows. Although external potentials induce current flow, the buildup of charge in the capacitors decreases the potential difference driving such currents, eventually establishing equilibrium when the potentials are equalized. By identifying such potentials as electronegativities, we therefore see that in the QTPIE model, electronegativity equalization[67, 68] comes from the formation of countercurrents induced from polarization effects, i.e. charge buildup in atomic capacitors.

We conclude this discussion by noting that since the concepts native to fluctuating-charge models can be related to exactly analogous concepts in classical electrical circuits with no equilibrium current flow. This shows that fluctuating-charge models are essentially classical in nature, with all quantum effects subsumed into the parameterization of the electronegativities and hardnesses. It is tempting to speculate that superior fluctuating-charge models could be constructed by similar analogies with quantum circuits, especially when considering quantum effects such as capacitance quantization. In addition, the equilibrium nature of this analysis can be viewed as a manifestation of the Born-Oppenheimer approximation of assuming that the electronic problem is always completely equilibrated relative to the nuclear dynamics. This suggests that more complicated electrical circuits, particularly those containing time-dependent components such as inductors, may allow the incorporation of non-adiabatic effects.

### 3.12. Conclusions

Our previously introduced QTPIE model is a fluctuating-charge model that exhibits correct asymptotic behavior for dissociating molecular systems. Formulating our new model in terms of charge-transfer variables allows us to construct explicitly distance-dependent pairwise electronegativities. However, the linear system of Eq. (3.2) which determines the bond space variables that minimize the QTPIE energy exhibits significant linear dependencies which complicate its numerical solution. We have discovered that the rank deficiency in our QTPIE model, and in bond-space fluctuating-charge models in general, can be attributed to the conservative nature of the laws of electrostatics, thus showing that the rank deficiency has a genuine physical basis, and is not merely a numerical inconvenience. With this knowledge, we constructed a numerical algorithm based on complete orthogonal decomposition that had better asymptotic complexity than singular value decomposition; this allowed an order of magnitude reduction in the time needed to solve Eq. (3.2). However, the computational complexity of this algorithm was still considerably higher than that for solving atom-space models.

We then showed that each fluctuating-charge model defined in bond space is equivalent *via* the mappings of Eqs. (2.6) and (3.18) to a related model of the form given in Eq. (3.20) formulated in atom space that predicts exactly the same charge distribution. Therefore, it is possible to formulate fluctuating-charge models with pairwise electronegativities that nonetheless retain the same asymptotic computational complexity as conventional atom-space models. In the process, we have discovered a framework which unifies fluctuating–charge models with and without topological constraints. In particular, we have shown that the underlying graphical structure of a topologically unconstrained fluctuating-charge model is that of a complete directed graph; thus fluctuating-charge models can be considered a special case of a larger class of graph charge models.

Finally, the QEq and QTPIE fluctuating-charge models can be described using the classical theory of electrical circuits, but with Coulomb interactions playing a significant role on the atomic scale that could otherwise be neglected in the description of macroscopic circuits. The circuit interpretation helps us establish some intuition for the duality mappings that are encapsulated in Eqs. (2.6) and (3.18).

# Chapter 4. The calculation of electrostatic properties in fluctuating-charge models

## 4.1. Introduction

In this Chapter, we investigate how to use fluctuating-charge models to calculate electrostatic properties such as dipole moments and polarizabilities. These seemingly routine calculations turn out to be surprisingly fraught with subtleties that have led to significant confusion in the literature. This confusion is due to incomplete understanding of how the constraint of charge conservation plays an integral role in electronegativity equalization, in addition to the well-understood driving effects of imbalances in atomic electronegativities. The constraint of charge conservation can be treated numerically without significant difficulty; however, an analytic formula for the charge distribution allows us to understand the issues much more clearly. Such an analytic solution has previously been published for the ES+ model of Streitz and Mintmire[1] and elsewhere.[2, 3] However, none of these authors had discussed the physical significance of the solutions that were obtained.

It turns out that charge conservation is critical for providing the correct spatial transformation properties of dipole moments and polarizabilities. Unfortunately, there is significant confusion over the calculation of such electrostatic properties, and the enforcement of their correct translational symmetries, in the literature. Many published formulae for dipole moments and polarizabilities do not exhibit the correct translational symmetries without special choices of coordinate origin, and it is often necessary to select the origin carefully to avoid spurious coordinate dependence of these electrostatic properties.[4-10] It is difficult to reconcile such choices with the translational and rotational symmetries required by classical electrostatics, which require that polarizabilities and dipole moments (for neutral systems, in the latter case) be translationally invariant.[11, 12] As it turns out, the correct treatment of the terms arising from charge conservation solves this problem naturally.

## 4.2. Analytic solution of fluctuating-charge models

We now show how the method of Lagrange multipliers and Gaussian elimination produce an analytic solution for the charge distribution that contains two different terms, clearly separating the contribution of charge conservation from that of charge–charge interactions and chemical hardness. This allows us to identify the roles of these separate terms in electrostatic observables such as dipole moments and polarizabilities.

Recall from Section 1.6 that a fluctuating-charge model is solved by a charge distribution that minimizes an energy expression, which in most modern models is quadratic in the charges and takes the form

$$E(\mathbf{q}) = \sum_{i=1}^{N} q_i \chi_i + \frac{1}{2}\sum_{i,j=1}^{N} q_i q_j J_{ij} = \mathbf{q}^T \boldsymbol{\chi} + \frac{1}{2}\mathbf{q}^T \mathbf{J}\mathbf{q} \tag{4.1}$$

where $q_i$ is the charge on atom *i*, $\chi_i$ is the intrinsic Mulliken electronegativity[13-15] of atom *i*, $J_{ii}$ is the chemical hardness[16] of atom *i*, and $J_{ij}$ are screened Coulomb interactions, the details of which vary from model to model.[17] We also introduce the boldface convention for vectors and matrices acting in the space of atomic charge variables, i.e. **q** and $\boldsymbol{\chi}$ are *N*–vectors and **J** is a $N \times N$ real and symmetric matrix. We have previously discussed in Chapter 3 that there is an exact analogy between fluctuating–charge models and classical electric circuits, where atoms can be interpreted as serial pairs of batteries of voltage $\chi_i$ and capacitors of capacitance $J_{ii}^{-1}$, which are coupled with coefficients of inductance $J_{ij}^{-1}$. This expression is formally equivalent to a Taylor series expansion of the energy with respect to the atomic charges to quadratic order.[18] Thus assuming that the hardness matrix **J** is invertible, and after discarding an irrelevant constant term, the energy can be expressed as the quadratic form

$$E(\mathbf{q}) = \frac{1}{2}\left(\mathbf{q} + \mathbf{J}^{-1}\chi\right)^T \mathbf{J}\left(\mathbf{q} + \mathbf{J}^{-1}\chi\right) \tag{4.2}$$

At first blush, it is tempting to note that this quadratic form is minimized by the solution

$$\mathbf{q} = \mathbf{q}_u = -\mathbf{J}^{-1}\chi \tag{4.3}$$

and hence assert that $\mathbf{q}_u$ is the solution to the fluctuating-charge model. However, this solution does not account for charge conservation, which is essential for electronegativity equalization.[19-23] (Thus, we have used the subscript $u$ to denote an unconstrained solution.) This physical conservation law imposes a constraint on the total charge of the system

$$\sum_{i=1}^{N} q_i = \mathbf{q}^T \mathbf{1} = Q \tag{4.4}$$

where $Q$ is the total charge and **1** is a column $N$–vector with all entries equal to unity. We use the method of Lagrange multipliers to reformulate the original problem as an unconstrained minimization:[24] by introducing the Lagrange multiplier $\mu$, which has a physical interpretation as the chemical potential of charge, we construct and minimize the Lagrange function

$$F(\mathbf{q},\mu) = E(\mathbf{q}) - \mu\left(\mathbf{q}^T \mathbf{1} - Q\right) = \mathbf{q}^T\left(\chi - \mu\mathbf{1}\right) + \frac{1}{2}\mathbf{q}^T \mathbf{J}\mathbf{q} + \mu Q \tag{4.5}$$

Minimizing this Lagrange function leads to a linear system which can be written in the $2 \times 2$ block matrix form

$$\begin{pmatrix} \mathbf{J} & \mathbf{1} \\ \mathbf{1}^T & 0 \end{pmatrix} \begin{pmatrix} \mathbf{q} \\ \mu \end{pmatrix} = \begin{pmatrix} -\boldsymbol{\chi} \\ Q \end{pmatrix} \tag{4.6}$$

where again we note that **J** is a $N \times N$ matrix; **1**, **q** and $\boldsymbol{\chi}$ are column *N*–vectors, and 0, $\mu$ and *Q* are scalars. Thus we have converted a constrained optimization in *N* unknowns to a linear system of *N* + 1 unknowns. This type of linear system is known in the numerical analysis literature as a saddle point problem, and many computationally efficient methods have been developed for solving such problems numerically.[25] (This terminology should not be confused with calculations to find saddle point geometries in electronic structure theory.)

We now use Gaussian elimination to derive an analytic solution. As before, we assume that the hardness matrix **J** is invertible, pre–multiply the first row by $-\mathbf{1}^T \mathbf{J}^{-1}$ and add the resulting equation to the second row. After some rearrangement, we obtain the solution

$$\begin{pmatrix} \mathbf{q} \\ \mu \end{pmatrix} = \begin{pmatrix} -\eta^{-1}\left(\boldsymbol{\chi} + \mu\mathbf{1}\right) \\ -\dfrac{Q + \mathbf{1}^T \mathbf{J}^{-1}\boldsymbol{\chi}}{\mathbf{1}^T \mathbf{J}^{-1}\mathbf{1}} \end{pmatrix} \tag{4.7}$$

from which it is immediately clear that the solution **q** differs from the unconstrained solution $\mathbf{q}_u$ in Eq. (4.3) by an additional term $-\mu\mathbf{J}^{-1}\mathbf{1}$ that arises directly from the charge conservation constraint of Eq. (4.4). Thus $-\mathbf{J}^{-1}\boldsymbol{\chi}$ represents the driving effect of electronegativity differences while the other term $-\mu\mathbf{J}^{-1}\mathbf{1}$ captures the restrictions imposed by charge conservation. We refer to these terms as the electronegativity–driven term and the charge–conserving term respectively. Note that the naïve, unconstrained charge distribution $\mathbf{q}_u$ contains only the electronegativity–driven term.

### 4.3. Formulae for dipole moments and polarizabilities

We now study how electronegativity–driven term and the charge–conserving term affect the calculation of electrostatic properties such as multipole moments and polarizabilities. The dipole moment can be obtained immediately from the definition

$$d_\lambda = \sum_{i=1}^{N} q_i R_{i\lambda} = \mathbf{q}^T \mathbf{R}_\lambda \tag{4.8}$$

where the Greek index $\lambda$ denotes a spatial component and $R_{i\lambda}$ is the $\lambda^{\text{th}}$ spatial component of the position of atom *i*. To obtain the dipole polarizability, we use the method of finite fields and employ the usual dipole coupling prescription to construct the energy in the presence of an external electrostatic field $\vec{\varepsilon} = \left(\varepsilon_\lambda\right)_\lambda$ as

$$E\left(\mathbf{q};\varepsilon_\lambda\right) = E\left(\mathbf{q}\right) - \mathbf{q}^T \mathbf{R}_\lambda \varepsilon_\lambda = \mathbf{q}^T\left(\boldsymbol{\chi} - \mathbf{R}_\lambda \varepsilon_\lambda\right) + \frac{1}{2}\mathbf{q}^T \mathbf{J}\mathbf{q} \tag{4.9}$$

where we have used the Einstein implicit summation convention for repeated Greek indices. The external field simply perturbs the atomic electronegativities by an amount $\mathbf{R}_\lambda \varepsilon_\lambda$, which is the potential produced by the external field. Therefore, we can replace $\boldsymbol{\chi}$ by $\boldsymbol{\chi} - \mathbf{R}_\lambda \varepsilon_\lambda$ in Eq. (4.7) to obtain the new charge distribution as

$$\begin{pmatrix} \mathbf{q}\left(\varepsilon_\lambda\right) \\ \mu\left(\varepsilon_\lambda\right) \end{pmatrix} = \begin{pmatrix} -\mathbf{J}^{-1}\left(\boldsymbol{\chi} - \mathbf{R}_\lambda \varepsilon_\lambda + \mu\left(\varepsilon_\lambda\right)\mathbf{1}\right) \\ -\dfrac{Q + \mathbf{1}^T \mathbf{J}^{-1}\left(\boldsymbol{\chi} - \mathbf{R}_\lambda \varepsilon_\lambda\right)}{\mathbf{1}^T \mathbf{J}^{-1}\mathbf{1}} \end{pmatrix} \tag{4.10}$$

which corresponds to an energy of

$$E_0(\varepsilon_\lambda) = -\frac{1}{2}(\boldsymbol{\chi} - \mathbf{R}_\lambda \varepsilon_\lambda)^T \mathbf{J}^{-1} (\boldsymbol{\chi} - \mathbf{R}_\lambda \varepsilon_\lambda) + \frac{1}{2} \frac{\left(Q + \mathbf{1}^T \mathbf{J}^{-1}(\boldsymbol{\chi} - \mathbf{R}_\lambda \varepsilon_\lambda)\right)^2}{\mathbf{1}^T \mathbf{J}^{-1} \mathbf{1}} \tag{4.11}$$

We can verify that the dipole moment given by Eq. (4.8) is the expected derivative of $E_0$ with respect to the external field,

$$\begin{aligned} d_\lambda = \left.\frac{\partial E_0}{\partial \varepsilon_\lambda}\right|_{\varepsilon_\lambda = 0} &= \left[ -\mathbf{R}_\lambda^T \mathbf{J}^{-1} \left( \boldsymbol{\chi} - \vec{\mathbf{R}} \cdot \vec{\varepsilon} \right) - \mu(\vec{\varepsilon}) \left( \mathbf{1}^T \mathbf{J}^{-1} \mathbf{R}_\lambda \right) \right]_{\varepsilon_\lambda = 0} \\ &= -\mathbf{R}_\lambda^T \mathbf{J}^{-1} \boldsymbol{\chi} - \mu \mathbf{R}_\lambda^T \mathbf{J}^{-1} \mathbf{1} \end{aligned} \tag{4.12}$$

The dipole polarizability is the next derivative,

$$\alpha_{\lambda\rho} = \left.\frac{\partial d_\rho}{\partial \varepsilon_\lambda}\right|_{\varepsilon_\lambda = 0} = \left.\frac{\partial E_0}{\partial \varepsilon_\lambda \partial \varepsilon_\rho}\right|_{\varepsilon_\lambda = 0} = -\mathbf{R}_\rho^T \mathbf{J}^{-1} \mathbf{R}_\lambda - \frac{\left(\mathbf{1}^T \mathbf{J}^{-1} \mathbf{R}_\lambda\right)\left(\mathbf{1}^T \mathbf{J}^{-1} \mathbf{R}_\rho\right)}{\mathbf{1}^T \mathbf{J}^{-1} \mathbf{1}} \tag{4.13}$$

Interestingly, the polarizability as calculated by Eq. (4.13) depends only on the hardness, and has no explicit dependence on the atomic electronegativities. This explains our earlier observation in Section 2.3 that QEq and QTPIE predict the same polarizabilities. Also, just as the charge distribution of Eq. (4.7) contains two distinct terms, the above formulae for dipole moments and polarizabilities also contain two separate terms that can also be identified as electronegativity–driven and charge–conserving respectively. The former terms correspond to the formulae obtained from calculations based on the unconstrained charge distribution $\mathbf{q}_u$. We shall see in the next section what the significance of the charge–conserving terms are.

## 4.4. Spatial symmetries of the dipole moment and polarizability

The results of the previous sections show that the presence of separate electronegativity–driven and charge–conserving terms in the charge distribution induce analogous separations of terms in the dipole moment and polarizability, and that the

charge–conserving terms disappear when calculating the electrostatic properties based on the unconstrained charge distribution $\mathbf{q}_u$. It turns out that when the charge–conserving terms are omitted, the dipole moment and dipole polarizability have incorrect translational symmetries, which are restored once these charge-conserving terms are retained. Indeed, under the global coordinate translation $\vec{\mathbf{R}} \mapsto \vec{\mathbf{R}} - \vec{s}\mathbf{1}$, the translational symmetries required by classical electrostatics[11, 12] require that the dipole moment transform as

$$d_\lambda \mapsto d_\lambda + s_\lambda Q \tag{4.14}$$

and the dipole polarizability must be invariant under this transformation. Instead, we see that the naïve dipole moment transforms as

$$d_{\lambda,u} = -\mathbf{R}_\lambda^T \mathbf{J}^{-1} \boldsymbol{\chi} \mapsto -\mathbf{R}_\lambda^T \mathbf{J}^{-1} \boldsymbol{\chi} - s_\lambda \mathbf{1}^T \mathbf{J}^{-1} \boldsymbol{\chi} \tag{4.15}$$

and the polarizability transforms as

$$\alpha_{\lambda\rho,u} = -\mathbf{R}_\rho^T \mathbf{J}^{-1} \mathbf{R}_\lambda \mapsto -\mathbf{R}_\rho^T \mathbf{J}^{-1} \mathbf{R}_\lambda - s_\rho \mathbf{1}^T \mathbf{J}^{-1} \mathbf{R}_\lambda - s_\lambda \mathbf{R}_\rho^T \mathbf{J}^{-1} \mathbf{1} - s_\rho s_\lambda \mathbf{1}^T \mathbf{J}^{-1} \mathbf{1} \tag{4.16}$$

where the subscript $u$ denotes, as before, the quantities derived from the unconstrained charge distribution. The use of these formulae in the literature have always been accompanied by an avid discussion of the need to select carefully a coordinate origin, typically by modifying the formulae in a way that is tantamount to placing the first atom at the origin.[4-9, 26] However, such an arbitrary specification is not compatible with the laws of electrostatics, as discussed above. In contrast, the formulae of Eqs. (4.12) and (4.13) show the correct translational symmetries, as the dipole moment transforms as

$$d_\lambda \mapsto d_\lambda + s_\lambda \mathbf{1}^T \mathbf{q} = d_\lambda + s_\lambda Q \tag{4.17}$$

and the dipole polarizability remains invariant. Importantly, the required physical symmetries are obtained naturally, and without any specific choice of coordinate origin.

Interestingly, the dipole moment and polarizability obey the correct rotational symmetries whether or not the charge constraint terms are present. Under the global coordinate rotation $\vec{\mathbf{R}}_\lambda \mapsto U_{\lambda\rho}\vec{\mathbf{R}}_\rho$ as described by some rotation matrix $U \in SO(3)$, the dipole moment transforms as

$$d_\lambda \mapsto -U_{\lambda\rho}\mathbf{R}_\rho^T \mathbf{J}^{-1}(\chi + \mu\mathbf{1}) = U_{\lambda\rho} d_\rho \tag{4.18}$$

and the dipole polarizability transforms as

$$\alpha_{\lambda\rho} = -U_{\rho\sigma}\mathbf{R}_\sigma^T \mathbf{J}^{-1} U_{\lambda\tau}\mathbf{R}_\tau - \frac{\left(\mathbf{1}^T \mathbf{J}^{-1} U_{\rho\sigma}\mathbf{R}_\lambda\right)\left(\mathbf{1}^T \mathbf{J}^{-1} U_{\lambda\tau}\mathbf{R}_\tau\right)}{\mathbf{1}^T \mathbf{J}^{-1}\mathbf{1}} = U_{\rho\sigma}U_{\lambda\tau}\alpha_{\sigma\tau} \tag{4.19}$$

which are exactly the transformational properties required of first– and second–rank tensors respectively.[12] The results of these calculations do not change when the charge constraint terms are discarded.

### 4.5. The ambiguity of field couplings in QTPIE

The atom-bond duality relationship detailed in Chapter 3 produces a curious ambiguity when coupling the electrostatic field to a fluctuating-charge model. The usual dipole coupling prescription in atom space is equivalent to a perturbation of the atomic electronegativities by an amount equal to the value of the potential due to the field at each atom, as evidenced by the regrouping of terms in Eq. (4.9). It is natural to ask if this observation holds also in the bond space variables of Chapter 2. These two descriptions of the field coupling, however, are *not* equivalent when applied to the QTPIE energy function of Eq. (2.8). The former choice, which is analogous to adding the usual dipole coupling term to the energy, is to couple the field by adding in the potential difference due to the field for each pair of atoms, i.e.

$$E^{I}\left(\mathbf{p};\varepsilon\right)=\sum_{ij}\chi_i f_{ji}p_{ji}+\sum_{ij}\frac{1}{2}\varepsilon_\lambda\left(R_{i\lambda}-R_{j\lambda}\right)p_{ji}+\frac{1}{2}\sum_{ijkl}p_{ki}p_{lj}J_{ij} \tag{4.20}$$

The factor of ½ is introduced to avoid double-counting. Applying the bond-atom transformation Eq. (3.18) allows us to rewrite the field coupling term as

$$\sum_{ij}\frac{1}{2}\varepsilon_\lambda\left(R_{i\lambda}-R_{j\lambda}\right)p_{ji}=\sum_{i}\varepsilon_\lambda\left(R_{i\lambda}-\sum_{j}\frac{R_{j\lambda}}{N}\right)q_i=\sum_{i}\varepsilon_\lambda R_{i\lambda}q_i-\sum_{j}\frac{\varepsilon_\lambda R_{j\lambda}}{N}Q \tag{4.21}$$

The last term vanishes since the formulation in charge transfer variables implicitly assumed overall charge neutrality, due to the skew-symmetry of the charge transfer variables as discussed in Chapter 2. This choice of coupling then reduces to the regular field coupling prescription of the earlier sections.

In contrast, the latter choice of field coupling, namely by perturbing the atomic electronegativities, has the following analogue in bond space

$$\begin{aligned}E^{II}\left(\mathbf{p};\varepsilon\right)&=\sum_{ij}\left(\chi_i-\varepsilon_\lambda R_{i\lambda}\right)f_{ji}p_{ji}+\frac{1}{2}\sum_{ijkl}p_{ki}p_{lj}J_{ij}\\&=\sum_{ij}\chi_i f_{ji}p_{ji}-\sum_{ij}\varepsilon_\lambda R_{i\lambda}f_{ji}p_{ji}+\frac{1}{2}\sum_{ijkl}p_{ki}p_{lj}J_{ij}\end{aligned} \tag{4.22}$$

Applying the bond-atom transformation Eq. (3.18) allows us to rewrite the field coupling term of $E^{II}$ as

$$\begin{aligned}\sum_{ij}\varepsilon_\lambda R_{i\lambda}f_{ji}p_{ji}&=\sum_{ij}\varepsilon_\lambda R_{i\lambda}f_{ji}\frac{q_i-q_j}{N}=\sum_{ij}\varepsilon_\lambda R_{i\lambda}f_{ji}\frac{q_i}{N}-\sum_{ij}\varepsilon_\lambda R_{j\lambda}f_{ij}\frac{q_i}{N}\\&=\sum_{i}\varepsilon_\lambda\left(\frac{\sum_{j'}f_{j'i}}{N}\right)\left(R_{i\lambda}-\frac{\sum_{j}R_{j\lambda}f_{ij}}{\sum_{j''}f_{j''i}}\right)q_i\end{aligned} \tag{4.23}$$

where for the second equality we have relabeled the dummy indices on the second summation term. Applying the definition $f_{ij} = NS_{ij} / \sum_{j'} S_{ij'}$ simplifies the resulting expression to

$$\sum_{ij} \varepsilon_\lambda R_{i\lambda} f_{ji} p_{ji} = \sum_i \varepsilon_\lambda \left( R_{i\lambda} - \frac{\sum_j R_{j\lambda} \sum_{j'} S_{ij'}}{\sum_{j''} S_{ij''}} \right) q_i \tag{4.24}$$

Thus the effect of perturbing the electronegativities is *not* equivalent to the usual dipole coupling prescription, but rather produces a slightly different coupling formula that includes a nonlocal component as well. The reason for this disparity is because the QTPIE model distinguishes between two types of electronegativities. As discussed in Chapter 3, the intrinsic or bare atomic electronegativities $\chi_i$ are different from the effective atomic electronegativities $v_i$ that consist of weighted averages of all other electronegativities. The nonlocality of the field coupling in Eq. (4.24) turns out to be analogous to the nonlocal nature of the effective atomic electronegativities. The nonlocality of the latter can be seen explicitly by rewriting the effective atomic electronegativities of Eq. (3.25) as

$$v_i = \frac{\sum_j \left(\chi_i - \chi_j\right) S_{ij}\left(\left|\vec{R}_i - \vec{R}_j\right|\right)}{\sum_{j'} S_{ij'}\left(\left|\vec{R}_i - \vec{R}_{j'}\right|\right)} = \chi_i - \sum_j \frac{S_{ij}\left(\left|\vec{R}_i - \vec{R}_j\right|\right)}{\sum_{j'} S_{ij'}\left(\left|\vec{R}_i - \vec{R}_{j'}\right|\right)} \chi_j \tag{4.25}$$

where we have shown the coordinate dependence explicitly to highlight the nonlocal nature of the effective atomic electronegativities. Using this field coupling prescription, the analogous calculations of the dipole moment and polarizability yield

$$
\begin{aligned}
d_\nu^{II} = \sum_{i,i',j,j'=1}^{N} & \tau_i \tau_{i'} S_{ij} S_{i'j'} \left( R_{i\nu} - R_{j\nu} \right) \left( \sigma \left( \mathbf{1}^T \mathbf{J}^{-1} \right)_i \left( \mathbf{1}^T \mathbf{J}^{-1} \right)_{i'} - \left( \mathbf{J}^{-1} \right)_{ii'} \right) \\
& \times \left( \chi_{i'} - \chi_{j'} - \sum_\lambda \left( R_{i'\lambda} - R_{j'\lambda} \right) E^\lambda \right)
\end{aligned} \tag{4.26}
$$

$$
\alpha_{\nu\lambda}^{II} = \sum_{i,i',j,j'=1}^{N} \tau_i \tau_{i'} S_{ij} S_{i'j'} \left( R_{i\nu} - R_{j\nu} \right) \left( \sigma \left( \mathbf{1}^T \mathbf{J}^{-1} \right)_i \left( \mathbf{1}^T \mathbf{J}^{-1} \right)_{i'} - \left( \mathbf{J}^{-1} \right)_{ii'} \right) \left( R_{i'\lambda} - R_{j'\lambda} \right) \tag{4.27}
$$

where for brevity we introduce the notation $\tau_i^{-1} = \sum_{k=1}^{N} S_{ik}$ and $\sigma^{-1} = \mathbf{1}^T \mathbf{J}^{-1} \mathbf{1}$. Note that Eqs. (4.26) and (4.27) still retain the correct translational symmetries that were discussed in the preceding section.

The coupling prescription $E^I$ is equivalent to perturbing the effective electronegativities $v_i$, while $E^{II}$ perturbs the intrinsic electronegativities $\chi_i$. In the next section, we investigate the size extensivity of the dipole moment and polarizability as calculate by these coupling prescriptions. Surprisingly, the usual coupling prescription $E^I$ turns out not to be size extensive, while that of $E^{II}$ does.

## 4.6. The size extensivity of dipole moments and polarizabilities

Polarizabilities are size extensive, namely that they scale linearly with system size in the asymptotic limit of infinitely large systems. In this section, we investigate the size extensivity of the dipole moments and polarizabilities as calculated under the coupling prescriptions $E^I$ and $E^{II}$. Consider a system with $n$ identical copies of a subsystem comprised of $m$ atoms, with each copy separated by a distance $\Delta_\nu$ that is much larger than the spatial extent of one subsystem. We use the overbar to denote quantities related to a single subsystem. The nuclear coordinates of the entire system can then be written in terms of the subsystem positions as

$$\mathbf{R}_\nu = \begin{pmatrix} \bar{\mathbf{R}}_\nu \\ \bar{\mathbf{R}}_\nu + \Delta_\nu \bar{\mathbf{1}} \\ \vdots \\ \bar{\mathbf{R}}_\nu + \Delta_\nu (n-1)\bar{\mathbf{1}} \end{pmatrix} = \begin{pmatrix} \bar{\mathbf{R}}_\nu \\ \bar{\mathbf{R}}_\nu \\ \vdots \\ \bar{\mathbf{R}}_\nu \end{pmatrix} + \Delta_\nu \begin{pmatrix} \bar{\mathbf{0}} \\ \bar{\mathbf{1}} \\ \vdots \\ (n-1)\bar{\mathbf{1}} \end{pmatrix} \tag{4.28}$$

In addition, the intrinsic electronegativities can be written as $\boldsymbol{\chi} = (\bar{\boldsymbol{\chi}}, ..., \bar{\boldsymbol{\chi}})^T$. In the limit of infinite subsystem separation, i.e. $|\Delta| \to \infty$, the subsystems decouple and the hardness matrix **J** becomes approximately block diagonal, with inverse

$$\mathbf{J}^{-1} = \begin{pmatrix} \bar{\mathbf{J}}^{-1} & \mathbf{0} & \cdots & \mathbf{0} \\ \mathbf{0} & \bar{\mathbf{J}}^{-1} & \ddots & \vdots \\ \vdots & \ddots & \ddots & \mathbf{0} \\ \mathbf{0} & \cdots & \mathbf{0} & \bar{\mathbf{J}}^{-1} \end{pmatrix} + O\left(|\Delta|^{-1}\right) \tag{4.29}$$

In this limit, the total dipole moment and polarizability for the usual dipole coupling prescription become

$$d_\nu = n\bar{d}_\nu + \tfrac{1}{2}(n-1)(n-2)\bar{Q}\Delta_\nu + O\left(\Delta_\nu / |\Delta|\right) \tag{4.30}$$

$$\alpha_{\nu\lambda} = n\bar{\alpha}_{\nu\lambda} - \frac{(n-1)(n-2)(n^2-3n-6)}{12n}\Delta_\nu\Delta_\lambda\bar{\sigma} + O\left(\frac{\Delta_\nu\Delta_\lambda}{|\Delta|}\right) \tag{4.31}$$

where the subsystem dipole moment and polarizability are defined analogously to those of the entire system, i.e.

$$\bar{d}_\nu = \bar{\mathbf{R}}_\nu^T \bar{\mathbf{J}}^{-1}\left(\bar{\mathbf{v}} - \sum_\lambda \bar{\mathbf{R}}_\lambda^T \mathrm{E}^\lambda\right) - \bar{\sigma}\left(\bar{\mathbf{1}}^T \bar{\mathbf{J}}^{-1}\bar{\mathbf{R}}_\nu\right)\left(\bar{\mathbf{1}}^T \bar{\mathbf{J}}^{-1}\bar{\mathbf{v}} + \bar{Q}\right) \tag{4.32}$$

$$\bar{\alpha}_{\nu\lambda} = -\bar{\mathbf{R}}_\nu^T \bar{\mathbf{J}}^{-1}\bar{\mathbf{R}}_\lambda + \bar{\sigma}\left(\bar{\mathbf{1}}^T \bar{\mathbf{J}}^{-1}\bar{\mathbf{R}}_\nu\right)\left(\bar{\mathbf{1}}^T \bar{\mathbf{J}}^{-1}\bar{\mathbf{R}}_\lambda\right) \tag{4.33}$$

where $\bar{\sigma}^{-1} = \bar{\mathbf{1}}^T \bar{\mathbf{J}}^{-1}\bar{\mathbf{1}}$ and $\bar{Q} = Q/n$ is the total charge of each identical subsystem. The second term in Eq. (4.30) represents the summed contributions of *m* point charges, each of charge $\bar{Q}$ and placed at coordinates $\bar{\mathbf{0}}, \bar{\mathbf{1}}, ..., (n-1)\bar{\mathbf{1}}$ respectively. When $Q = 0$, the dipole moment in Eq. (4.30) becomes size-extensive. However, the second term in the polarizability expression Eq. (4.31) grows cubically with $n$, which is physically incorrect.

When we apply the subsystem decomposition of Eqs. (4.28) and (4.29) to Eqs. (4.26) and (4.27), the overlap matrix element decays exponentially quickly with interatomic distance and thus attenuates inter-subsystem interactions; the effective atomic electronegativities become

$$\bar{v}_i = \sum_{j=1}^{m} S_{ij} \left( \chi_i - \chi_j \right) / \left( \sum_{k=1}^{m} S_{ik} \right) + O\left( e^{-|\Delta|} \right) \tag{4.34}$$

Then the dipole moment and polarizability show the correct size extensivity

$$d_v = n\bar{d}_v + O\left( e^{-|\Delta|} \right) \tag{4.35}$$

$$\alpha_{v\lambda} = n\bar{\alpha}_{v\lambda} + O\left( n^3 e^{-|\Delta|} \right) \tag{4.36}$$

Therefore, the overlap factors give rise to size-extensivity. Importantly, this does not come at the price of forbidding intermolecular charge transfer *a priori*, unlike previously proposed topological solutions to the size-extensivity problem.[26-33] In the next section, we apply the field coupling $E^{II}$ to a simple water model and show that it is possible to obtain reasonable results with it.

## 4.7. Application to a liquid water model

As a simple application of our QTPIE model, we study a series of simple water systems. As is well known, the dipole moment of a single molecule of water is 1.85 D in the gas phase[34] but increases to 2.95±0.20 D in the liquid phase[35] due to cooperative polarization between the water molecules in condensed phases. The reproduction of such cooperative behavior is a useful test of polarizable water models. Here, we study whether the QTPIE model is able to reproduce the onset between gas-like behavior to bulk-like behavior in planar water chains. To better study the size extensivity, we use idealized geometries instead of optimized geometries for each chain. The oxygen atoms are collinear and spaced 2.870 Å apart; the hydrogen atoms are all coplanar with transverse separations of 1.514 Å and with O–H bond lengths of 1.000 Å. The O-O internuclear distances of 2.87 Å is chosen to be the O-O internuclear separation in the ground state geometry of the water dimer. The water molecules are chosen to be coplanar and aligned along their dipole moments. While such intermolecular geometries are physically unlikely to be observed, they are useful for studying the transition from gas-like to bulk-like behavior in an essentially one-dimensional system. As a further test of our charge models, we choose to parameterize the models using data only from monomer and dimer geometries, and see if these models satisfactorily reproduce the dipole moments in longer water chains. This would be a sensitive indicator of the quality of the intermolecular electrostatic interactions.

To eliminate systematic error arising from improper parameterization, we reparameterized both the QEq and QTPIE models to be applied specifically to three-site water models. 1,230 monomer geometries were generated by systematically varying the internal coordinates and bond lengths, and 890 dimer geometries were generated from fictitious high temperature molecular dynamics runs at 30,000 K with a systematic variation in the Lennard–Jones attraction parameters to sample a wide variety of inter–monomer distances. For each geometry, *ab initio* dipole moments were calculated with density-fitted local second-order Møller-Plesset perturbation theory[36] using the augmented Dunning correlation-consistent valence triple-zeta basis set (DF-LMP2/aug-cc-pVTZ).[37] We then optimized the weighted root mean square deviation between the model's predictions and the ab initio calculation using the derivative-free simplex algorithm,[24] with weights given by Boltzmann factors at a temperature of 10,000 K. This temperature has no physical significance and is merely chosen to generate convenient weights to penalize the contribution of geometries of higher energies that were produced in the systematic exploration of configuration space—some geometries were as high as *ca.* 0.4 Hartrees above the minimum energy configurations and for all practical purposes lie in energetically inaccessible, and hence irrelevant, regions of the relevant potential energy surfaces. The resulting parameters are compared with the original QEq parameters in Table 4.1.

**Table 4.1.** Parameters for the QTPIE and QEq models for a three-site water model.

| Parameter (eV) | QTPIE | QEq (original)[a] | QEq (reparameterized) |
|---|---|---|---|
| H electronegativity | 5.366 | 4.528 | 3.678 |
| H hardness | 11.774 | 13.890 | 18.448 |
| O electronegativity | 7.651 | 8.741 | 9.591 |
| O hardness | 13.115 | 13.364 | 17.448 |

[a]From Ref. 38.

As a test of the water models obtained by this procedure, we use the parameters obtained from monomer and dimer data to calculate the dipole moments and polarizabilities of longer one–dimensional water chains. Figure 4.1 shows the dipole moments calculated from QEq and QTPIE, together with dipole moments with high quality *ab initio* calculations at the DF-LMP2/aug-cc-pVTZ level of theory. Figures 4.2—4.4 show similar plots for the components of the polarizability. In addition, we compare the results to the AMOEBA model available in the TINKER molecular dynamics package, which is a polarizable multipole model parameterized to the same level of *ab initio* theory.[39]

The *ab initio* data show that dipole moment per molecule increases rapidly as a function of the chain length, and beyond approximately five water molecules gradually saturates toward a limiting value of 2.50 D per molecule. As expected, the AMOEBA model reproduces the ab initio data very well. By comparison, the QTPIE model is also able to reproduce the trends exhibited by the ab initio data and the AMOEBA model, which is especially encouraging when taking into account the much simpler description of electrostatics in QTPIE as compared to AMOEBA. Surprisingly, we see that the QEq model, using the original parameters, show a decrease in the dipole moment per molecule with increasing chain length. This behavior is absent in the reparameterized model, but instead saturates to a value of 2.25 D per molecule, which is significantly lower than for the QTPIE and AMOEBA models.

The polarizability results in Figures 4.2—4.4 are more interesting. The transverse polarizability shown in Figure 4.2, being the component parallel to the H—H axes, is well described by both QEq and QTPIE. However, the longitudinal polarizability shown in Figure 4.3 shows that QEq drastically overestimates the polarizability along the O—O axis. Reparameterization did not ameliorate this superlinear scaling to any significant degree. This result is in agreement with the analysis of previous sections. In contrast, the $E^{II}$ coupling for the QTPIE model allows the recovery of size extensivity, with the longitudinal polarizability saturating to a value of 1.69 Å$^3$ per molecule, a result that is surprisingly close to the *ab initio* data which show saturation to 1.65 Å$^3$ per molecule. Finally, we note that the out of plane component of the polarizability vanishes for QEq and QTPIE, as shown in Figure 4.4. While clearly a disappointing result, this result is not unexpected, as it is not possible to polarize a planar system out of plane because there are no charge sites off the plane to receive or donate charge. This is a known problem of fluctuating–charge models,[29] which for similar reasons are also unable to describe the polarization of single atoms.[30]

Finally, we reiterate that these results were obtained in the presence of significant intermolecular charge transfer, as shown in Figure 4.5. This is in stark contrast to previous studies, where charge transfer had to be curtailed topologically in order to guarantee the correct size extensivity.[5, 26, 40, 41] We note that both QEq and QTPIE predict charge transfer from the hydrogen bond donating end of the water chain to the hydrogen bond accepting end, a result which is in qualitative agreement with chemical intuition as well as Mulliken population analysis of the *ab initio* wavefunctions. The discrepancy in absolute values is not significant as Mulliken population analysis, and any charge analysis scheme in general, cannot be unambiguously defined for atoms in molecules.[42-47]

The results suggest that QTPIE affords a qualitatively superior description of intermolecular electrostatic interactions over QEq, as even reparameterizing QEq could not produce the bulk–like dipole moments to the same level of accuracy. In contrast, the results of the QTPIE model are comparable with those of the significantly more costly AMOEBA water model, which has 14 parameters specifically for electrostatic interactions, as well as nonlinear, higher–order multipole interaction equations (up to the quadrupole—quadrupole level) to solve for.[39] In contrast, the QTPIE model requires only four parameters and solving a linear system of equations for charge—charge interactions only. Thus, the three-site water model based on QTPIE is able to reproduce satisfactorily the cooperative polarization behavior in these planar water chains with just four independent parameters, and therefore shows great promise for providing a comparable level of accuracy with more computationally costly and more highly parameterized models.

**Figure 4.1.** Dipole moments per molecule for a sequence of planar water chains, with consecutive O-O internuclear separations of 2.87 Å and O-H bond lengths of 1.00 Å and internal angle 105º, as calculated by QTPIE (black solid line), DF-LMP2/aug-cc-pVTZ (blue broken line), AMOEBA (green short-dashed line), QEq (brown dashed line), and reparameterized QEq (purple dash-dotted line). The parameters used for the QTPIE and QEq models are given in Table 4.1.

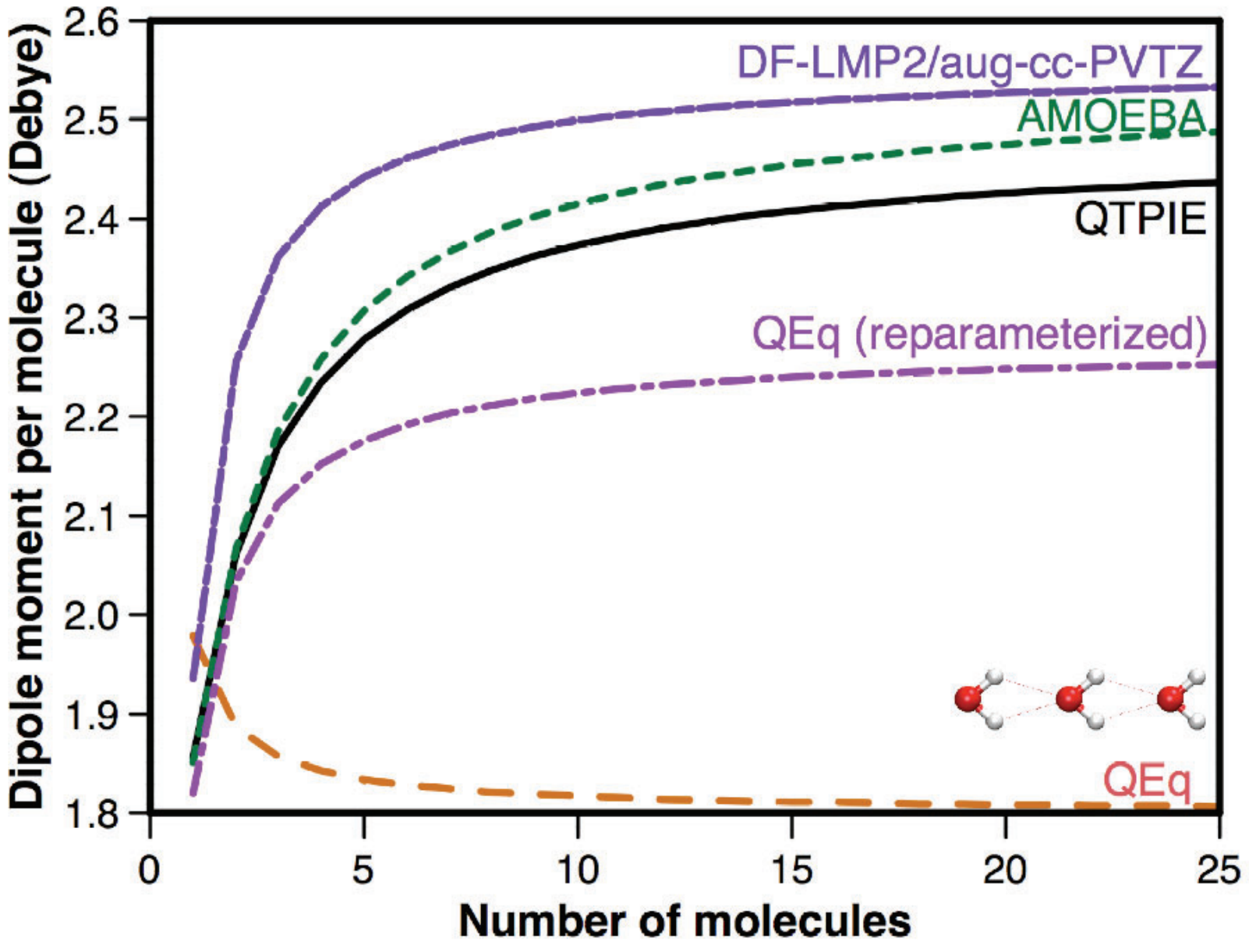

**Figure 4.2.** Transverse polarizability per molecule for a sequence of planar water chains, with consecutive O-O internuclear separations of 2.87 Å and O-H bond lengths of 1.00 Å and internal angle 105º, as calculated by QTPIE (black solid line), DF-LMP2/aug-cc-pVTZ (blue broken line), AMOEBA (green short-dashed line), QEq (brown dashed line), and reparameterized QEq (purple dash-dotted line). The parameters used for the QTPIE and QEq models are given in Table 4.1. The polarization response occurs parallel to the H–H axes.

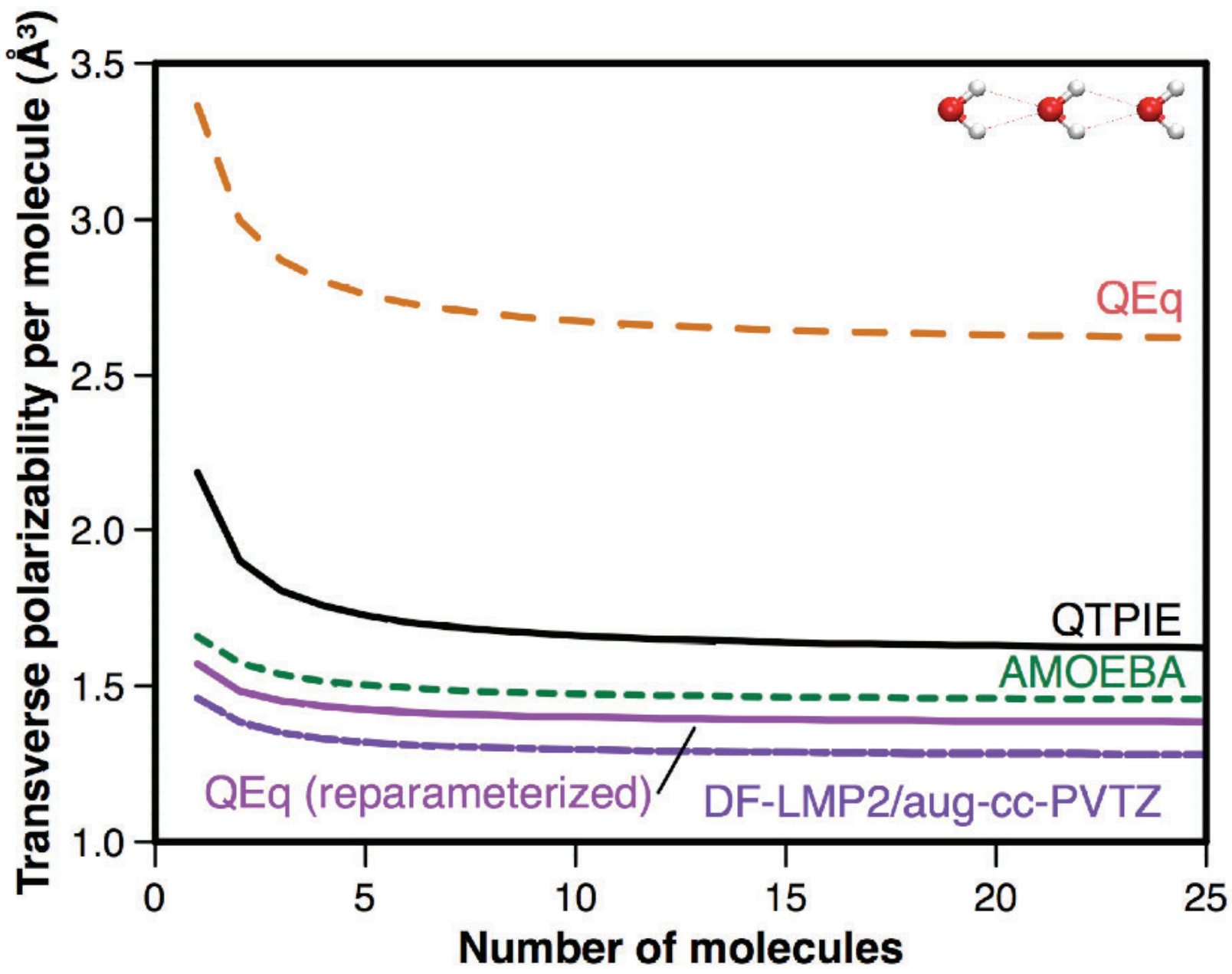

**Figure 4.3.** Longitudinal polarizability per molecule for a sequence of planar water chains, with consecutive O-O internuclear separations of 2.87 Å and O-H bond lengths of 1.00 Å and internal angle 105º, as calculated by QTPIE (black solid line), DF-LMP2/aug-cc-pVTZ (blue broken line), AMOEBA (green short-dashed line), QEq (brown dashed line), and reparameterized QEq (purple dash-dotted line). The parameters used for the QTPIE and QEq models are given in Table 4.1. The polarization response occurs along the shared O–O axis.

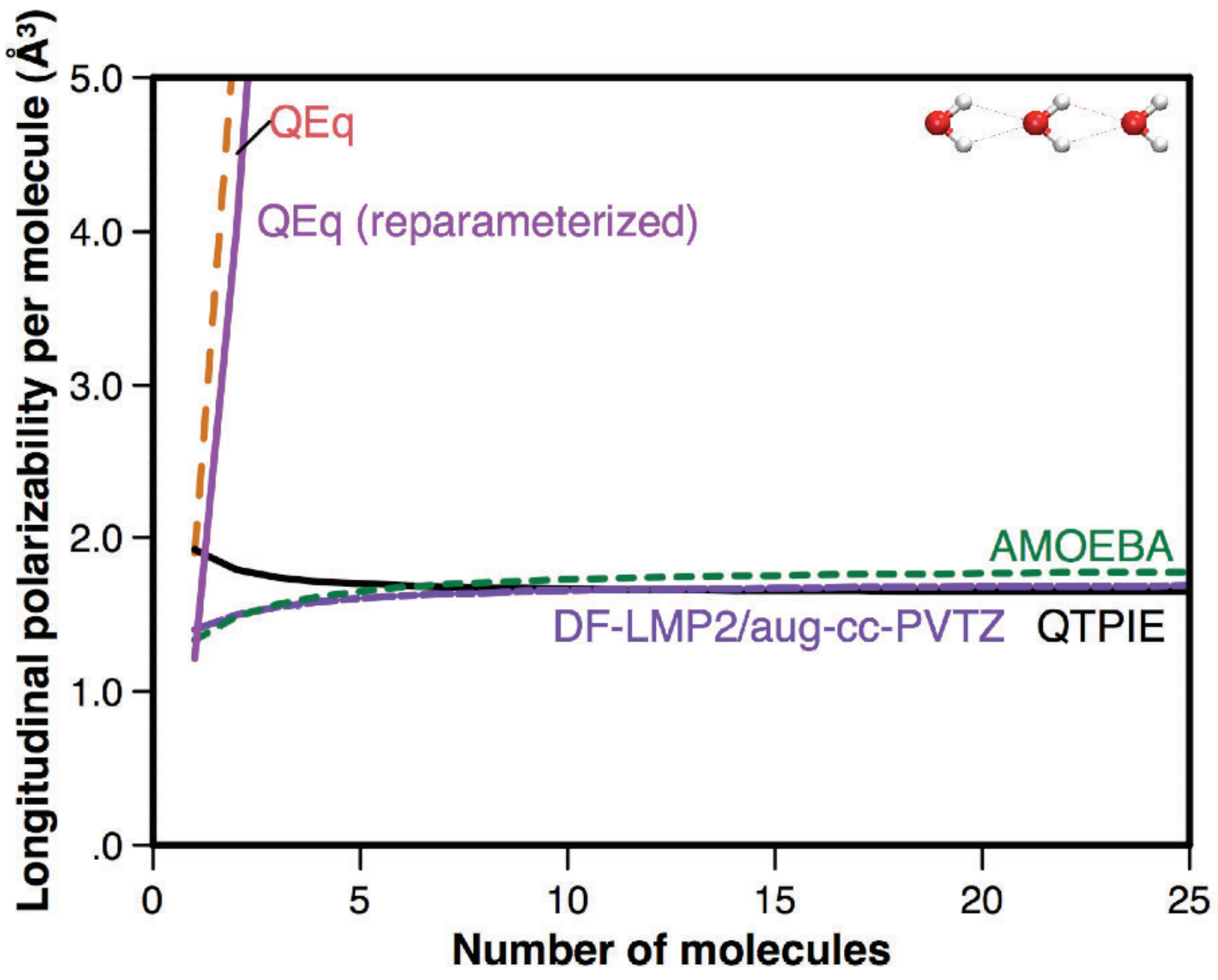

**Figure 4.4.** Out of plane polarizability per molecule for a sequence of planar water chains, with consecutive O-O internuclear separations of 2.87 Å and O-H bond lengths of 1.00 Å and internal angle 105º, as calculated by QTPIE (black solid line), DF-LMP2/aug-cc-pVTZ (blue broken line), AMOEBA (green short-dashed line), QEq (brown dashed line), and reparameterized QEq (purple dash-dotted line). The parameters used for the QTPIE and QEq models are given in Table 4.1.

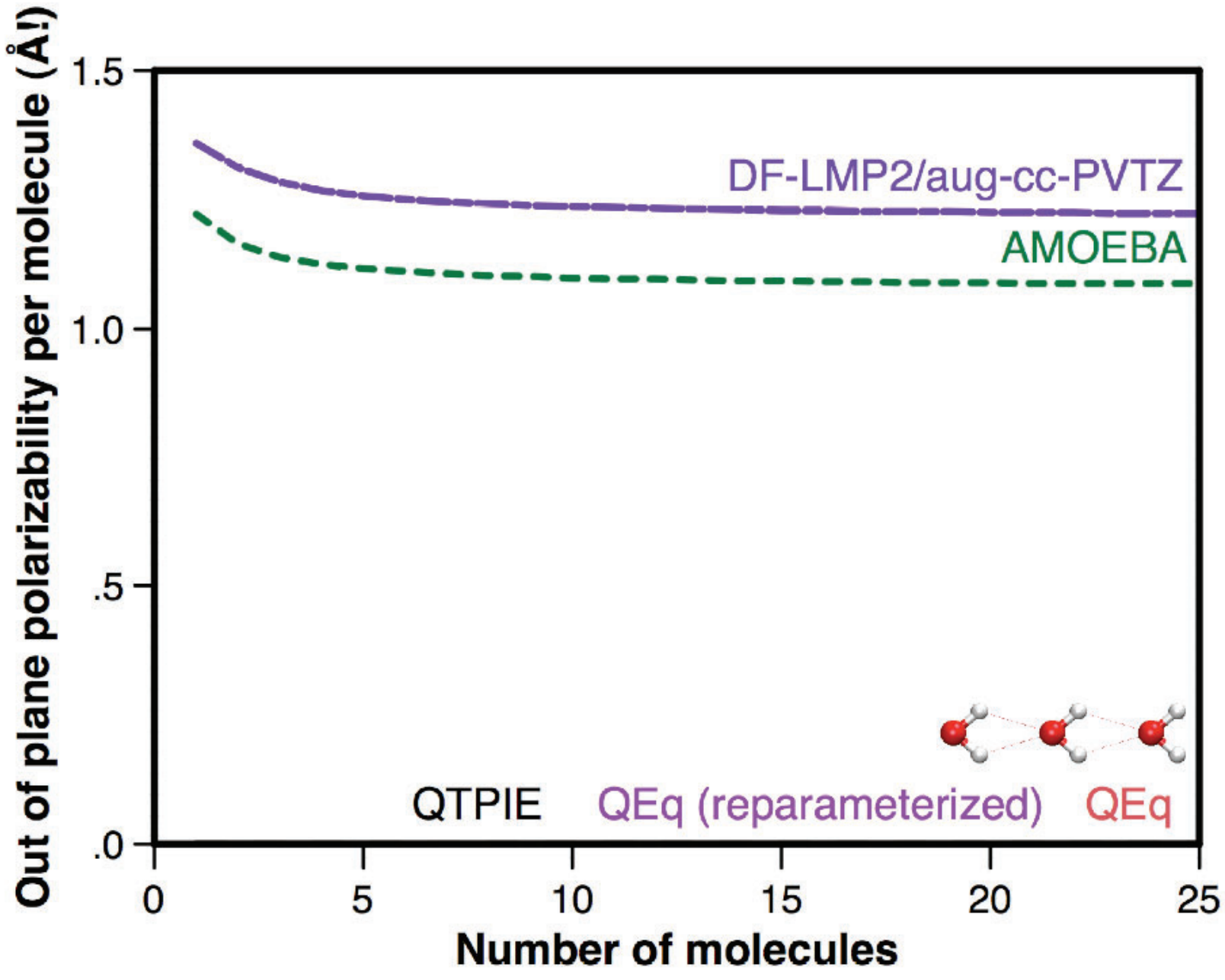

**Figure 4.5.** Charge on each molecule in a planar chain of 15 water molecules, with consecutive O-O internuclear separations of 2.87 Å and O-H bond lengths of 1.00 Å and internal angle 105º, as calculated by QTPIE (black solid line), QEq (red dashed line), reparameterized QEq (purple dash–dotted line), and Mulliken analysis of the DF-LMP2/aug-cc-pVTZ wavefunction (green broken line). The parameters used for the QTPIE and QEq models are given in Table 4.1.

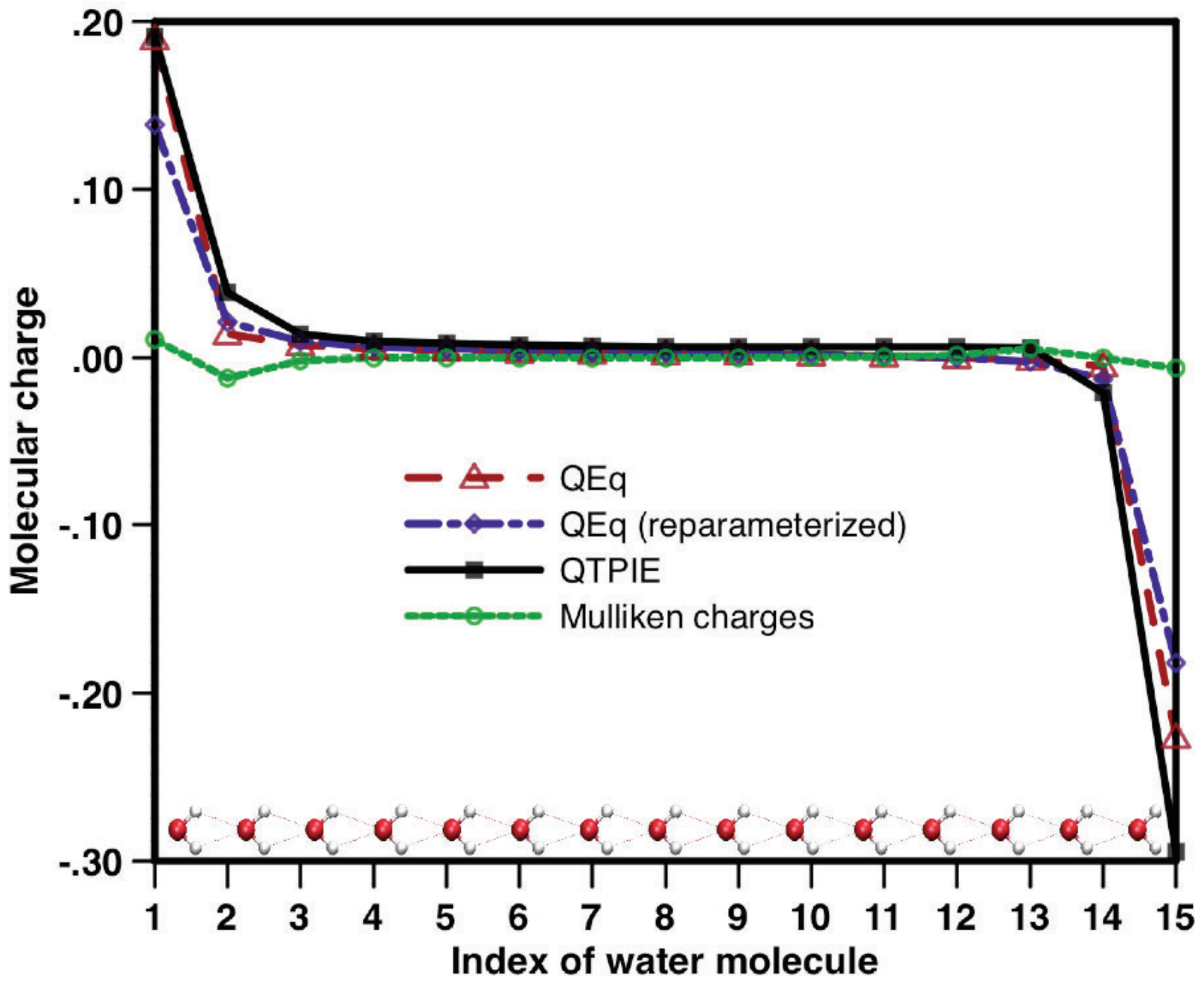

# Chapter 5. Conclusions and Outlook

## 5.1. Summary and conclusions

In this thesis, we have explored and proposed solutions to two of most well-known problems with fluctuating-charge models. In Chapter 2, we have studied why fluctuating-charge models predict fractional charge separation for dissociated systems, and have proposed the charge transfer with polarization current equilibration (QTPIE) model that exhibits the correct attenuation of charge transfer in this asymptotic limit. We accomplished this by introducing geometry dependent bond electronegativities. However, this came at the cost of making a change of variables—where we originally had one variable per atom describing the charge residing on that atom, we needed to change to a representation in charge-transfer variables, so that the linear terms could have coefficients that were bond electronegativities. This made the QTPIE model very costly from a computational standpoint.

Using our initial implementation of the QTPIE model in bond space, we discovered empirically that the linear system of equations arising from electronegativity equalization turned out to be rank deficient for systems with more than two atoms. In Chapter 3, we studied this rank deficiency and proved that the rank of this system of equations had to be one less than the number of atoms in the systems, where the difference of one was due to the imposition of the charge conservation constraint. This demonstrated that there was no difference in the information capacity when representing the model in atomic charge variables, which span atom space, as compared to charge transfer variables, which span bond space. This is a surprising result at face value, as there are many more charge transfer variables than atomic charges. However, we found that the conservative nature of the electrostatic potential generated conservation symmetries that gave rise to this equivalence between atom and bond spaces. Furthermore, we were able to determine analytically the exact mappings that allowed bidirectional interconversion between atomic charges and charge transfer variables. These mappings turned out to be intimately related to the incidence matrix of the graph that captured the topological relationship between atom and bond spaces, thus allowing us for the first time to build explicit connections between fluctuating-charge models and topological charge models, as well as understand the topological implications of imposing *ad hoc* restrictions on intermolecular charge transfer. In addition, this allowed us to implement the QTPIE charge model with negligible cost overhead relative to other fluctuating-charge models. An example of this implementation in Fortran 90 is given in Appendix A.

Finally in Chapter 4, we calculated the electrostatic properties predicted by fluctuating-charge models. We discovered a point of confusion in the literature regarding the choice of coordinate origin needed for the correct calculation of polarizabilities, and showed that the imposition of charge conservation in fluctuating-charge models gave rise to additional terms in the dipole moment and polarizability that were crucial to preserving the correct spatial symmetries as demanded by classical electrostatics. Then, we analyzed the size extensivity of the dipole moment and polarizability in considerable detail, and showed that while the dipole moment was correctly size extensive in these calculations, the polarizability turned out to exhibit asymptotically cubic scaling with system size. This contradicts our experience with the onset of bulk polarization behavior in macroscopic systems. We found that the usual dipole coupling prescription for coupling an external electrostatic field to fluctuating-charge models was equivalent to perturbing the atomic electronegativities by the potential corresponding to the external field being introduced. However, the introduction of geometry dependent bond electronegativities in the QTPIE model broke this equivalence, and that coupling the field by perturbing atomic electronegativities allowed the recovery of correct size extensivity for polarizabilities in a way that did not require *ad hoc* restrictions on charge transfer. We have taken advantage of this to develop a new three-site water model that was able to reproduce the correct size extensivity of dipole moments and polarizabilities in model one-dimension water chains.

### 5.2. Unresolved issues with fluctuating-charge models

The work described in this thesis does not resolve all outstanding issues with fluctuating-charge models. The problem of taming the superlinear scaling of molecular polarizabilities remains one of the important unsolved problems with these models. All existing solutions to this problem in the literature come at the price of artificially restricting charge transfer,[1-7] and the problems associated with superlinear scaling return once such constraints are removed. We believe this to be an unsatisfactory solution because it removes one of the primary advantages of fluctuating-charge models, namely its ability to treat both polarization and charge transfer phenomena in the same unified theoretical framework. We have observed that the QTPIE model gives rise to some ambiguity in the field couplings used to calculate electrostatic properties, and we do not understand the significance of these ambiguity. Although we have been able to use this ambiguity to discover a solution to the size extensivity problem for liquid water in Chapter 4, this proposal is completely inadequate for the correct description of semiconducting or metallic systems. That the QTPIE model is inadequate for metallic systems can be easily seen from the introduction of overlap functions to model the geometry dependence of bond electronegativities; this results in the forcible attenuation of long-distance intermolecular charge transfer, even for metallic systems where this is the correct behavior. Thus just as QEq and other fluctuating-charge models introduce the uncontrolled approximation of metallicity, the QTPIE model can be seen as forcibly introducing the approximation that everything behaves as an insulator. Further work is needed to understand how the geometry dependence of bond electronegativities introduced in Chapter 2 must be suitably modified in order to interpolate correctly between insulating, semiconducting and metallic systems.

In addition, one of the biggest remaining problems of fluctuating-charge models is the inability to describe polarization out of plane for planar systems.[8] As discussed earlier, polarization is modeled in such models by moving charge in the direction of polarization. However, charge is only allowed to flow between charge sites. As there are no charge sites available out of plane for planar systems, it is not possible to polarize planar systems out of plane. The introduction of dummy atoms to create such charge sites is a possible solution, and has been explored in the construction of four-site water models like TIP4P-FQ.[4, 9, 10] However, this workaround introduces additional parameters into the molecular model and must be applied on a case by case basis – it is unclear how to extend this systematically . Another possibility is to introduce charge sites with *p*-type angular momentum into the fluctuating-charge model, as in the York and Yang model.[11] Such extensions of fluctuating-charge models are formally equivalent to constructing hybrid inducible dipole – fluctuating charge models,[3] and can be readily extended to higher order multipoles if necessary.[12] Yet another possibility is to allow the charge sites to drift away from the center of the nucleus, as allowed in Dinur's hybrid Drude oscillator – fluctuating-charge model.[13] At present, all these solutions come as the cost of additional parameters and working variables, and the possibility of further simplification and retention of the minimal parameterization, perhaps by discovering relationships between the parameters of hybrid models, remains to be studied.

## 5.3. Understanding the theoretical foundations of fluctuating-charge models

It is now clear that fluctuating-charge models bear very close resemblance to density functional theory.[1, 14-18] The electronegativities and hardnesses that play such a fundamental role in fluctuating-charge models have been explored in great detail in density functional theory.[19-26] As density functional theory works with the charge distribution as a continuous function over real three-dimensional space and fluctuating-charge models deal with the charge distribution as a discrete collection of point charges, it is clear that fluctuating-charge models must be, on some level, coarse-grained versions of density functional theory where the molecular charge density has been partitioned into atomic chunks, which are then approximated by point charges with various shape factors. This highlights not only the theoretical origins of fluctuating-charge models in density functional theory, but also highlights clearly the equally important question of how the notion of atoms in molecules can be suitably defined.[19] This has vitally important consequences for understanding the reference states for which parameters such as electronegativities and hardnesses are derived from.[11, 16, 27-33]

The derivation of fluctuating-charge models from higher level semiempirical or *ab initio* theories is extremely appealing not only for theoretical reasons, but would also allow for a much more comprehensive understanding of the physical content of the electronegativities and hardnesses that parameterize such models.[15, 28, 30, 34-36] Unfortunately, the large amount of work put into parameterization efforts strongly suggest that it is better to treat the electronegativities and hardnesses as purely empirical fitting parameters rather than insisting on their calculation from existing theoretical connections,[2, 4, 10, 13, 37-48] which strongly suggests that our physical understanding of these parameters is still incomplete. It is clear that the issue of atoms in molecules is one of the key unresolved aspects of this problem. It is worth recalling that the distinction between isolated atoms and atoms in molecules dates back to the birth of quantitative electronegativity scales, as even Mulliken's seminal paper on electronegativities takes great care to stress that electronegativities "must, however, in general, be calculated not in the ordinary way, but for suitable 'valence states' of the positive and negative ion."[49] However, Mulliken's use[49, 50] of van Vleck valence states[51] has proven to be troublesome when extended beyond the first two periods of the periodic table.[52-57] In addition, van Vleck's valence states belie an inherent assumption of using minimal basis sets, and cannot be extended straightforwardly to the complete basis set limit. This necessitates a deeper study of the problem of defining atoms in molecules. Wavefunction-based charge analysis schemes such as Coulson and Mulliken population analyses are notoriously dependent on the size of basis sets, and are therefore not particularly well-suited to the task at hand. At present, the most popular theoretical framework for studying atoms in molecules is the AIM topological index analysis of Bader;[58, 59] however, Bader's theory cannot be satisfactorily applied to the derivation of fluctuating-charge models from density functional theory. It is widely known that Bader's AIM analysis yields atomic charges that severely overestimate dipole moments and hence cannot be used to produce charge distributions that accurately represent molecular electrostatic properties. In addition, the use of zero-flux boundaries to define partitionings is fundamentally incompatible with the notion of entities engaging in charge transfer that forms a fundamental component of the electronegativity equalization principle that is so central to fluctuating-charge models.[21, 29, 33, 36, 60-62] For these reasons, partitionings based on Hirshfeld's stockholder principle[63-65] or that of the partition theory of Cohen and coworkers[66-69] are much more appealing theoretically. However, some work has suggested that Hirshfeld partitionings do not produce reasonable charges,[70] and partition theory has to date only been applied to very simple molecules.[68, 69] The success of these methods, while promising, remains to be seen.

In addition to the definition of atoms in molecules, relating fluctuating-charge models to wavefunction-based theories or density functional theory must also address the observation of Perdew, Parr, Levy and Balduz (PPLB),[71] whereby the energy functional was shown to have discontinuous derivatives that have very strong implications for the behavior of density functionals. It is now increasingly widely accepted that the resolution of the derivative discontinuity problem lies in considering mixed states and density ensembles to handle changes in particle number.[19] Much work on the generalization toward ensemble density functional theory[72] shows considerable progress in understanding the consequences of such derivative discontinuities,[73] and in particular for long-range charge transfer,[74] and the principle of electronegativity equalization.[75, 76] The modeling of Fukui functions[77] in fluctuating-charge models could in principle help in alleviating these issue.[22, 24, 78, 79] However, the breakdown of the description afforded by small grand canonical ensembles[62, 80] suggests caution when attempting to use such ensemble generalizations of density functional theory to derive fluctuating-charge models. This not only has serious implications for relating fluctuating-charge models to density functional theory, but also has great significance for reconciliation with wavefunction-based theories. While Mulliken's original proposal of electronegativities was made in the context of a small valence-bond configuration interaction space,[49, 50] subsequent work by Morales and Martínez has shown that quantum mechanical studies in very small state spaces can lead to results that are difficult to interpret physically, such as the onset of complex temperatures or ensembles with negative probabilities.[80, 81] Furthermore, the bond-space duality relation of fluctuating-charge models introduced in Chapter 3, which necessarily imposes total charge neutrality by definition of the charge-transfer variables, suggests that a canonical ensemble formalism could have more well-behaved properties than a grand canonical ensemble one. Indeed, some intriguing results have shown that the duality between external potential and charge density in density functional theory can be extended to a canonical ensemble framework in a way that has proven to be extremely difficult for the grand canonical ensemble.[82] The implications for such work for fluctuating-charge models have yet to be investigated.

# Appendix A. Source code for a Fortran 90 implementation of QTPIE in atom space

## A.1. Makefile

```
#--- OPTION 1: Portland Group Fortran compiler on Linux ---
#FC = pgf90
#Use the following line for very strict checking of code and allowing
debugging
#FFLAGS = -C -g -Kieee -Ktrap=fp,denorm,unf -Mdclchk -Mneginfo -Mbounds
-Mchkstk -Mchkptr -Mchkfpstk  -Minform=inform
#Use the following line for optimized code
#FFLAGS = -fast
#LINKOPT = -lblas
#--- OPTION 2: GNU Fortran compiler on MacOSX 10.5 ---
FC = gfortran
OPTS = -m64 -march=nocona #-fopenmp
#Use the following line for gfortran strict checking
FFLAGS = $(OPTS) -Wall -Wextra -Waliasing -Wsurprising -pedantic \
-C -g3 -ggdb -fbounds-check -dH -fbacktrace -frange-check \
-fimplicit-none -ffpe-trap=invalid,zero,overflow
#Use the following line for optimized code
#FFLAGS = $(OPTS) -g -O3 -ftree-vectorize -ffast-math -malign-double -
ffinite-math-only
#LINKOPT = $(OPTS) -framework Accelerate

#--- OPTION 3: Intel fortran ---
#FC = ifort
#FFLAGS=-C -debug all -fpe0 -ftrapuv -g -traceback -warn all
#FFLAGS=-C -g -fpe0 -debug full -ftrapuv
LINKOPT=-framework Accelerate
#LINKOPT = -L/usr/local/intel/mkl801/lib/32 -lmkl
/usr/local/intel/mkl801/lib/32/libmkl_lapack.a

TINKERDIR = $(HOME)/src/tinker

DOCDIR = ../doc
BINDIR = ../bin

GMRES = ../3rdparty/gmres/dPackgmres.f
SOLVER = cg.o
#SOLVER = gmres.o solver.o
CORE = api_tinker.o io.o atomicunits.o $(SOLVER) factorial.o geometry.o
parameters.o sparse.o sto-int.o gto-int.o qtpie.o properties.o
ALL = $(CORE) matrixutil.o
onexyz: onexyz.o libqtpie.a
     $(FC) -o $(BINDIR)/onexyz $(FFLAGS) onexyz.o libqtpie.a $(LINKOPT)
test: test.o libqtpie.a
     $(FC) -o $(BINDIR)/test $(FFLAGS) test.o libqtpie.a $(LINKOPT)
     $(BINDIR)/test
tinker: libqtpie.a api_tinker.o
     ar r libqtpie.a api_tinker.o
     ranlib libqtpie.a
```

```
     echo QTPIE is ready to be interfaced with TINKER, call libqtpie.a
     make -C $(TINKERDIR)
     make -C $(TINKERDIR) all
     make -C $(TINKERDIR) rename
     make -C $(TINKERDIR) create_links
libqtpie.a: $(CORE)
     ar r libqtpie.a $(CORE)
     ranlib libqtpie.a
doc: $(CORE)
     doxygen

.f.o:
     $(FC) -c $(FFLAGS) $*.f
onexyz.o: libqtpie.a
test.o: libqtpie.a
api_tinker.o: parameters.o
io.o: parameters.o
dqtpie.o: parameters.o
qtpie.o: parameters.o
sto-int.o: parameters.o factorial.o
gmres.o: $(GMRES)
     $(FC) -c $(FFLAGS) $(GMRES) -o gmres.o
parameters.o: atomicunits.o sparse.o
clean:
     rm \#* *~ *.o *~ *.mod *.out *.a *.log ../bin/onexyz ../bin/test
```

## A.2. api_tinker.f

```
!>
!! Subroutine to interface QTPIE with the Tinker MM dynamics package
!! Inputs
!! \param n     : number of "atoms" (charge sites)
!! \param x,y,z : arrays of coordinates
!! \param Atoms : array of atomic numbers
!! Outputs
!! charge: array of atomic charges
!! energy: QTPIE contribution energy
!! grad  : matrix of QTPIE energy gradients indexed by direction, then
site index
!<
      subroutine QTPIEFromTinker(n,x,y,z,Atoms,charge,energy,grad)
        use Parameters
        implicit none
        integer, intent(in) :: n
        double precision, intent(in), dimension(n) :: x, y, z
        integer, intent(in), dimension(n) :: Atoms
        double precision, intent(out), dimension(n) :: charge
        double precision, intent(out) :: energy
        double precision, intent(out), dimension(3,n), optional :: grad
C       internally used variables
        logical :: isParameterized, ParameterFileExists
        integer :: j,l
        type(Molecule), save :: Mol
```

```
C       allocate memory for atoms and coordinate data
c       Do this (and parameterization) ONLY if the number of atoms
change
c       which should happen only ONCE in a MD simulation
        if (Mol%NumAtoms .ne. n) then
           Mol%NumAtoms = n
           call NewAtoms(Mol%Atoms, n)

C          Parameterize atoms by matching atomic numbers
           do j=1,n
              isParameterized = .false.
              do l=1,numParameterizedAtoms
                 if (Atoms(j).eq.ParameterizedAtoms(l)%Z) then
                    Mol%Atoms(j)%Element = ParameterizedAtoms(l)
                    isParameterized = .true.
                 end if
              end do

C             assign basis set
              if (isParameterized) then
                 call AssignsGTOBasis(Mol%Atoms(j))
              else
                 print *, "QTPIE Error: Unknown element, Z=", Atoms(j)
                 stop
              end if
           end do
c          Read parameters from file, if one exists
           inquire(file="parameter.txt", EXIST=ParameterFileExists)
           if (ParameterFileExists) then
               print *, "Loading parameters from parameter.txt"
               call UpdateParameters("parameter.txt", Mol)
           end if
        end if

c       Update positions
        Mol%Atoms(1:n)%Position(1) = x(1:n) * Angstrom
        Mol%Atoms(1:n)%Position(2) = y(1:n) * Angstrom
        Mol%Atoms(1:n)%Position(3) = z(1:n) * Angstrom
        Mol%Atoms(1:n)%Basis%Position(1) = x(1:n) * Angstrom
        Mol%Atoms(1:n)%Basis%Position(2) = y(1:n) * Angstrom
        Mol%Atoms(1:n)%Basis%Position(3) = z(1:n) * Angstrom

C       Call QTPIE
        call DosGTOIntegrals(Mol)
        call QTPIE(Mol)
c        return charges calculated by QTPIE
        charge(1:n) = Mol%Atoms(1:n)%Charge

C       return energy in kcal/mol
        energy = Mol%Energy / kcal_mol

        if (present(grad)) then
          print *, "Calculating gradients by finite difference"
          call DoGradientsByFiniteDifference(Mol)
c        call DoGradientsAnalytically(Mol)
```

```
C          set energy gradients in kcal/mol per Angstrom
           do j=1,3
             do l=1,n
               grad(j,l) = Mol%EGradient(l,j) / (kcal_mol / Angstrom)
             end do
           end do
         end if
C        write log file
c         call WriteLog(Mol, "qtpie.log")
c         call WriteXYZ(Mol, "qtpie.xyz")
c         print *, "QTPIE is done. Back to TINKER."
      end subroutine
```

## A.3. atomicunits.f

```
!>
!! A Fortran module storing conversion factors
!!
!!
!!        Our QTPIE charge model works exclusively in atomic units
!!        The values stored here are conversion factors to convert
!!        from that unit into atomic units
!!        There is no dimensional checking implemented!
!<
      module AtomicUnits
        implicit none
        save
        double precision, parameter :: ONE = 1.0d0
        double precision, parameter :: ZERO = 0.0d0
        double precision, parameter :: eV = 3.67493245d-2 !< electron
volt to Hartree
        double precision, parameter :: kJ_mol = 6.6744644952d-3
!<kilojoule per mole to Hartree
        double precision, parameter :: kcal_mol = 1.5952353d-3
!<kilocalorie per mole to Hartree
        double precision, parameter :: invAngstrom = 455.6335252760d0
!<inverse ?ngstrom to Hartree
        double precision, parameter :: Debye = 0.3934302014076827d0
!<Debye to atomic unit of dipole moment
        double precision, parameter :: Angstrom = 1.0d0/0.529177249d0
!<?ngstrom to bohr
      end module AtomicUnits
```

## A.4. cg.f

```
!>
!!    \param N : an integer specifying the size of the problem
!!    \param A : a real, positive definite NxN matrix
!!    \param b : a real vector with N elements
!!    \param x : (Output) solution to matrix equation
!<
      subroutine solver(N, A, b, x)
      implicit none
      integer, intent(in) :: N
```

```
      double precision, dimension(N,N), intent( in) :: A
      double precision, dimension(N)  , intent( in) :: b
      double precision, dimension(N)  , intent(inout) :: x

      double precision, external :: dnrm2
      external :: dcg
      double precision, external :: ConditionNumber

      integer i

      if (abs(x(1)) .lt. 1.0d-8) x = 0.0d0
!     Check that b is not zero vector, else return 0
      if (dnrm2(N,b,1) .gt. 1.0d-8) then
!     Use conjugate gradients routine
        call dcg(N, A, b, x)
!     Use LAPACK SVD-based solver
!      call lapack_svdsolver(N, A, b, x)
      else
        print *, 'WARNING, b = 0'
        x = 0.0d0
      end if
!      print *, ConditionNumber(N, A)
!      print *, "Condition number = ", ConditionNumber(N, A)
!      print *, "Norm of b = ", dnrm2(N,b,1)

!        print *, "b = "
!      do i=1,N
!        print *, b(i)
!        end do
      end subroutine solver

      subroutine lapack_svdsolver(N, AA, b, x)
      implicit none

      integer, intent(in) :: N
      double precision, dimension(N,N) :: A
      double precision, dimension(N,N), intent( in) :: AA
      double precision, dimension(N)  , intent( in) :: b
      double precision, dimension(N)  , intent(inout) :: x

!     Used for LAPACK solver
      double precision, dimension(N) :: S !< Matrix of singular values
      integer :: Rank, stat, WorkSize
      double precision, dimension(:), allocatable :: WORK
      double precision, parameter :: RCond = 1.0d-8
      external :: dgelss

!     Avoids bug where LAPACK could overwrite matrix
      A = AA

      x = b
C     First find optimal workspace size
      allocate(WORK(1))
      call dgelss(N, N, 1, A, N, x, N, S, RCond, Rank, WORK,
     &     -1, stat)
```

```
      WorkSize = WORK(1)
      deallocate(WORK)

      allocate(WORK(WorkSize))
      call dgelss(N, N, 1, A, N, x, N, S, RCond, Rank, WORK,
     &     WorkSize, stat)
      deallocate(WORK)
      end subroutine

!>
!! Use LAPACK routine to calculate condition number of the NxN matrix A
!<
      double precision function ConditionNumber(N, AA)
      implicit none

      integer, intent(in) :: N
      double precision, dimension(N,N), intent( in) :: AA
      double precision, dimension(N,N) :: A

      double precision, dimension(N) :: c

!     Used for LAPACK solver
      double precision, dimension(N) :: S !< Matrix of singular values
      integer :: Rank, stat, WorkSize
      double precision, dimension(:), allocatable :: WORK
      double precision, parameter :: RCond = 1.0d-8
      external :: dgelss

      integer :: i
      c = 0.0d0

c     Known bug: can mess around with the matrix A for some reason
c     This is a workaround
      A = AA
C     First find optimal workspace size
      allocate(WORK(1))
      call dgelss(N, N, 1, A, N, c, N, S, RCond, Rank, WORK,
     &     -1, stat)
      WorkSize = WORK(1)
      deallocate(WORK)

      allocate(WORK(WorkSize))
      call dgelss(N, N, 1, A, N, c, N, S, RCond, Rank, WORK,
     &     WorkSize, stat)
      deallocate(WORK)

      if (Rank .eq. N) then
         ConditionNumber = S(1)/S(N)
      else
         print *, "Matrix found to be singular"
         ConditionNumber = S(1)/S(Rank)
      end if

      print *, "Singular values:"
      do i =1, N
```

```
        print *, S(i)
      end do

      end function ConditionNumber

!>
!!    Double precision conjugate gradient solver with Jacobi
preconditioner
!!
!!     Solves the matrix problem Ax = b for x
!!    Implemented from Golub and van Loan's stuff
!!    \author Jiahao Chen
!!    \date   2008-01-28
!!    \param N : an integer specifying the size of the problem
!!    \param A : a real, positive definite NxN matrix
!!    \param b : a real vector with N elements
!!    \param x : (Output) solution to matrix equation
!!    On input, contains initial guess
!<
      subroutine dcg(N, A, b, x)
      implicit none
      integer, intent(in) :: N
      double precision, dimension(N,N), intent( in) :: A
      double precision, dimension(N)  , intent( in) :: b
      double precision, dimension(N)  , intent(inout) :: x

      integer :: max_k = 100000 !< maximum number of iterations
      double precision, parameter :: tol = 1.0d-7 !< Convergence
tolerance

      integer :: k !< Iteration loop counter
      external :: Precondition

c     Residual vector, p, q, z
      double precision, dimension(N) :: r, p, q, z
      double precision :: alpha, norm, critical_norm, gamma, gamma0

c     BLAS routines
      double precision, external :: ddot, dnrm2
      external :: dcopy, daxpy, dgemv

c      logical, parameter :: Verbose = .True.
      logical, parameter :: Verbose = .False.

*     Termination criterion norm
      critical_norm = tol * dnrm2(N,b,1)

*     Calculate initial guess x from diagonal part P(A) x = b
!     The secret code to want an initial guess calculated is to pass an
!     initial guess with the first entry equal to floating-point zero.
!     If not, we'll just use the pre-specified initial guess that's
already in x

      if (abs(x(1)) .lt. 1.0d0-10) call Precondition(N,A,b,x)
```

```
*     Calculate residual r = b - Ax
c     r = b (Copy b into r)
      call dcopy(N,b,1,r,1)
c     r = r - Ax
      call dgemv('N',N,N,-1.0d0,A,N,x,1,1.0d0,r,1)

c     Calculate norm
      norm = dnrm2(N,r,1)
      if (Verbose) then
         print *, "Iteration",0,":",norm, norm/critical_norm
      end if

      if (norm.lt.critical_norm) goto 1

      do k=1,max_k
*        Generate preconditioned z from P(A) z = r
         call Precondition(N,A,r,z)
c         z = r
*        Propagate old vectors
         if (k.ne.1) gamma0 = gamma
c        gamma  = r . z
         gamma  = ddot(N,r,1,z,1)

         if (k.ne.1) then
c           p = z + gamma/gamma0 * p
c           With BLAS, first overwrite z,then copy result from z to p
c           z = z + gamma/gamma0 * p
            call daxpy(N,gamma/gamma0,p,1,z,1)
         end if
c        p = z
         call dcopy(N,z,1,p,1)
*        Form matrix-vector product
c        q = A p
         call dgemv('N',N,N,1.0d0,A,N,p,1,0.0d0,q,1)

*        Calculate step size
c        alpha = gamma / p.q
         alpha = gamma / ddot(N,p,1,q,1)

*        Propagate by step size
c        x = x + alpha * p
         call daxpy(N, alpha,p,1,x,1)
c        r = r - alpha * q
         call daxpy(N,-alpha,q,1,r,1)

*        Calculate new norm of residual
         norm = dnrm2(N,r,1)

*        If requested, print convergence information
         if (Verbose) then
            print *, "Iteration",k,":",norm, norm/critical_norm
         end if
*        Check termination criterion

c        Done if || r || < tol || b ||
```

```
         if (norm.lt.critical_norm) goto 1
      end do

c     Oops, reached maximum iterations without convergence
      print *, "dcg: maximum iterations reached."
      print *, "ERROR: Solution is not converged."
      stop
c     Finally, return the answer
 1    if (Verbose) then
         print *, "dcg: solution found with residual", norm
      end if
      end subroutine dcg

!>
!! Calculates the solution x of the approximate preconditioned problem
!! \f[
!! \mathbf{P}(\mathbf{A}) \vec{x} = \vec{b}
!! \f]
!<
      subroutine Precondition(N,A,b,x)
      implicit none
      integer, intent(in) :: N
      double precision, dimension(N,N), intent( in) :: A
      double precision, dimension(N)  , intent( in) :: b
      double precision, dimension(N)  , intent(out) :: x

      integer :: i
c      double precision :: ReciprocalSumOfDiagonals, MatrixElement

c     Use NO preconditioning
c      x = b

c     Use Jacobi preconditioning
      do i=1,N
         if (abs(A(i,i)) .gt. 1.0d-10) then
         x(i) = b(i) / A(i,i)
         else
         print *, "Error: divide by zero!"
         print *, "Error in column", i,":", A(i,i)
         stop
        end if
      end do

!     Use Gauss-Siedel preconditioning
!      do i=N,1,-1
!         x(i) = b(i)
!         do j=i+1,N
!           x(i) = x(i) - x(j) * A(j,i)
!         end do
!         x(i) = x(i) / A(i,i)
!      end do

*     Add in exact solution for last column and last row
!     ( 0 v ) ( x ) = (   y v   )
!     ( v w ) ( y ) = ( x.v + wy)
```

```
!      ReciprocalSumOfDiagonals = 0.0d0
!
!      do i = 1,N-1
!         ReciprocalSumOfDiagonals = ReciprocalSumOfDiagonals
!     &        + 1.0d0/A(i,i)
!      end do
!
!      x(N) = - b(N) / ReciprocalSumOfDiagonals
!
!      do i = 1,N-1
!         MatrixElement = 1.0d0/(ReciprocalSumOfDiagonals * A(i,i))
!         x(i) = x(i) + b(N) * MatrixElement
!         x(N) = x(N) + b(i) * MatrixElement
!      end do

*     Calculate initial guess from approximate inverse
*     W is the inverse of the preconditioning matrix
*     W = approximate inverse of A
c      call ApproximateInverse(A,W,N)
*     Form matrix-vector product x = W b
c      call dgemv('N',N,N,1.0d0,W,N,b,1,0.0d0,x,1)

      end subroutine
!>
!!    Calculates an approximate inverse to a matrix of the form
!!    \f[
!!    \mathbf{M}=\left(\begin{array}{cc}\mathbf{J} & 1\\
!!                   1 & 0\end{array}\right)
!!    \f]
!!
!!    The inverse is calculated by approximating J by its diagonal, in
which
!!    case an exact inverse can be constructed.
!<
      subroutine ApproximateInverse(M, W, N)
      implicit none
      integer, intent(in) :: N !< Size of matrix
      double precision, dimension(N,N), intent(in) :: M !< Matrix to
invert
      double precision, dimension(N,N), intent(out) :: W !< Approximate
inverse matrix

      integer :: i!, j
      double precision :: ReciprocalSum
      double precision :: MatrixElement!, OffDiagonalMatrixElement

      W = 0.0d0

      ReciprocalSum = 0.0d0
      do i = 1,N-1
         ReciprocalSum = ReciprocalSum + 1.0d0/M(i,i)
      end do

      W(N,N) = -1.0d0/ReciprocalSum
      do i = 1,N-1
```

```
         MatrixElement = 1.0d0/(ReciprocalSum * M(i,i))
         W(i,N) = MatrixElement
         W(N,i) = MatrixElement

c This is an approximation to the approximate problem, replacing that
which follows

         W(i,i) = 1.0d0/M(i,i)

c This code computes the exact solution to the approximation, but
exhibits
c slower convergence when used as a preconditioner. Go figure.
c         do j = 1, i-1
c            OffDiagonalMatrixElement = -MatrixElement/M(j,j)
c            W(i,j) = OffDiagonalMatrixElement
c            W(j,i) = OffDiagonalMatrixElement
c         end do
c
c         W(i,i) = 0.0d0
c         do j = 1,N-1
c            W(i,i) = W(i,i) - W(i,j)
c         end do
      end do
      end subroutine
```

### A.5. factorial.f

```
! This factorial.f was automatically generated from factorial.py
!>
!! Stored values of the factorial function
!!
!<
      module Factorial
        implicit none
        save
        integer,parameter :: maxFact = 150 !< Largest factorial
computed
        double precision, parameter :: fact(0:maxFact)=(/
     &  1.0d0 ,
     &  1.0d0 ,
     &  2.0d0 ,
     &  6.0d0 ,
     &  24.0d0 ,
     &  120.0d0 ,
     &  720.0d0 ,
     &  5040.0d0 ,
     &  40320.0d0 ,
     &  362880.0d0 ,
     &  3628800.0d0 ,
     &  39916800.0d0 ,
     &  479001600.0d0 ,
     &  6227020800.0d0 ,
     &  87178291200.0d0 ,
     &  1307674368000.0d0 ,
```

```
     &  20922789888000.0d0 ,
     &  355687428096000.0d0 ,
     &  6402373705728000.0d0 ,
     &  1.21645100408832d+17 ,
     &  2.43290200817664d+18 ,
     &  5.109094217170944d+19 ,
     &  1.1240007277776077d+21 ,
     &  2.5852016738884978d+22 ,
     &  6.2044840173323941d+23 ,
     &  1.5511210043330986d+25 ,
     &  4.0329146112660565d+26 ,
     &  1.0888869450418352d+28 ,
     &  3.0488834461171384d+29 ,
     &  8.8417619937397008d+30 ,
     &  2.6525285981219103d+32 ,
     &  8.2228386541779224d+33 ,
     &  2.6313083693369352d+35 ,
     &  8.6833176188118859d+36 ,
     &  2.9523279903960412d+38 ,
     &  1.0333147966386144d+40 ,
     &  3.7199332678990118d+41 ,
     &  1.3763753091226343d+43 ,
     &  5.2302261746660104d+44 ,
     &  2.0397882081197442d+46 ,
     &  8.1591528324789768d+47 ,
     &  3.3452526613163803d+49 ,
     &  1.4050061177528798d+51 ,
     &  6.0415263063373834d+52 ,
     &  2.6582715747884485d+54 ,
     &  1.1962222086548019d+56 ,
     &  5.5026221598120885d+57 ,
     &  2.5862324151116818d+59 ,
     &  1.2413915592536073d+61 ,
     &  6.0828186403426752d+62 ,
     &  3.0414093201713376d+64 ,
     &  1.5511187532873822d+66 ,
     &  8.0658175170943877d+67 ,
     &  4.2748832840600255d+69 ,
     &  2.3084369733924138d+71 ,
     &  1.2696403353658276d+73 ,
     &  7.1099858780486348d+74 ,
     &  4.0526919504877221d+76 ,
     &  2.3505613312828789d+78 ,
     &  1.3868311854568986d+80 ,
     &  8.3209871127413916d+81 ,
     &  5.0758021387722484d+83 ,
     &  3.1469973260387939d+85 ,
     &  1.9826083154044401d+87 ,
     &  1.2688693218588417d+89 ,
     &  8.2476505920824715d+90 ,
     &  5.4434493907744307d+92 ,
     &  3.6471110918188683d+94 ,
     &  2.4800355424368305d+96 ,
     &  1.711224524281413d+98 ,
     &  1.197857166996989d+100 ,
```

```
&  8.5047858856786218d+101 ,
&  6.1234458376886077d+103 ,
&  4.4701154615126834d+105 ,
&  3.3078854415193856d+107 ,
&  2.4809140811395391d+109 ,
&  1.8854947016660498d+111 ,
&  1.4518309202828584d+113 ,
&  1.1324281178206295d+115 ,
&  8.9461821307829729d+116 ,
&  7.1569457046263779d+118 ,
&  5.7971260207473655d+120 ,
&  4.7536433370128398d+122 ,
&  3.9455239697206569d+124 ,
&  3.314240134565352d+126 ,
&  2.8171041143805494d+128 ,
&  2.4227095383672724d+130 ,
&  2.1077572983795269d+132 ,
&  1.8548264225739836d+134 ,
&  1.6507955160908452d+136 ,
&  1.4857159644817607d+138 ,
&  1.3520015276784023d+140 ,
&  1.24384140546413d+142 ,
&  1.1567725070816409d+144 ,
&  1.0873661566567424d+146 ,
&  1.0329978488239052d+148 ,
&  9.916779348709491d+149 ,
&  9.6192759682482062d+151 ,
&  9.426890448883242d+153 ,
&  9.3326215443944096d+155 ,
&  9.3326215443944102d+157 ,
&  9.4259477598383536d+159 ,
&  9.6144667150351211d+161 ,
&  9.9029007164861754d+163 ,
&  1.0299016745145622d+166 ,
&  1.0813967582402903d+168 ,
&  1.1462805637347078d+170 ,
&  1.2265202031961373d+172 ,
&  1.3246418194518284d+174 ,
&  1.4438595832024928d+176 ,
&  1.5882455415227421d+178 ,
&  1.7629525510902437d+180 ,
&  1.9745068572210728d+182 ,
&  2.2311927486598123d+184 ,
&  2.5435597334721862d+186 ,
&  2.9250936934930141d+188 ,
&  3.3931086844518965d+190 ,
&  3.969937160808719d+192 ,
&  4.6845258497542883d+194 ,
&  5.5745857612076033d+196 ,
&  6.6895029134491239d+198 ,
&  8.09429852527344d+200 ,
&  9.8750442008335976d+202 ,
&  1.2146304367025325d+205 ,
&  1.5061417415111404d+207 ,
&  1.8826771768889254d+209 ,
```

```
     &  2.3721732428800459d+211 ,
     &  3.0126600184576582d+213 ,
     &  3.8562048236258025d+215 ,
     &  4.9745042224772855d+217 ,
     &  6.4668554892204716d+219 ,
     &  8.4715806908788174d+221 ,
     &  1.1182486511960039d+224 ,
     &  1.4872707060906852d+226 ,
     &  1.9929427461615181d+228 ,
     &  2.6904727073180495d+230 ,
     &  3.6590428819525472d+232 ,
     &  5.0128887482749898d+234 ,
     &  6.9177864726194859d+236 ,
     &  9.6157231969410859d+238 ,
     &  1.346201247571752d+241 ,
     &  1.8981437590761701d+243 ,
     &  2.6953641378881614d+245 ,
     &  3.8543707171800706d+247 ,
     &  5.5502938327393013d+249 ,
     &  8.0479260574719866d+251 ,
     &  1.1749972043909099d+254 ,
     &  1.7272458904546376d+256 ,
     &  2.5563239178728637d+258 ,
     &  3.8089226376305671d+260 ,
     &  5.7133839564458505d+262  /) !< array of precomputed factorials
      end module Factorial
```

### A.6. fitgto.f

```
!>
!!    Helps fit a GTO to a STO based on the Coulomb self-repulsion
!!    integral generated from it. Yay!
!<
      program fitgto
      use parameters
      implicit none

      type(Atom) :: SlaterAtom, GaussianAtom
      integer :: i, j
      double precision, external :: sSTOCoulInt, sGTOCoulInt
      integer, parameter :: maxIter = 100
      double precision, parameter :: thresh = 1.0d-12
      double precision :: xnew, xold, change

!     do j = numParameterizedAtoms, numParameterizedAtoms
      do j = 1,1

      SlaterAtom%Element = ParameterizedAtoms(j)
      GaussianAtom%Element = ParameterizedAtoms(j)

      call AssignsSTOBasis(SlaterAtom)
      call AssignsSTO1GBasis(GaussianAtom)

      xold = GaussianAtom%Basis%zeta
```

```
      xnew = GaussianAtom%Basis%zeta * 2

      do i = 1, maxIter
         change = fit(SlaterAtom%Basis%zeta, SlaterAtom%Basis%n, xnew)
/
     *       ( fit(SlaterAtom%Basis%zeta, SlaterAtom%Basis%n, xnew)
     *       - fit(SlaterAtom%Basis%zeta, SlaterAtom%Basis%n, xold) )
     *       * (xnew - xold)

         if (abs(change) .le. thresh) then
            xnew = xnew - change
            goto 2
         end if

         xold = xnew
         xnew = xnew - change

!        If change is too large, damp it by an arbitrary factor
         if (change * 2.0d0 .gt. xold) then
            xnew = xold - 0.25 * change
         end if

c         print *, xold, fit(SlaterAtom%Basis%zeta, SlaterAtom%Basis%n,
c     *        xold), change

      end do

 2    print *, j, xnew,
     * fit(SlaterAtom%Basis%zeta, SlaterAtom%Basis%n, xnew)

c      xold = fit(SlaterAtom%Basis%zeta, SlaterAtom%Basis%n,
c     *        xnew, .True.)
      end do

      contains
!>
!!    Now calculate integral to fit, which is
!!    \f[
!!    \int_0^\infty (J^{GTO}(\alpha) - J^{STO}(\zeta) ) e^{-
\left(\frac{\alpha R}{2}\right)^2} dR
!!    \f]
!!    In practice, truncate when we are far out since integral becomes
tiny
!<
      function fit(zeta, n, alpha, zPrint)
      double precision, intent(in) :: zeta, alpha
      integer, intent(in) :: n
      double precision :: fit
      logical, optional :: zPrint

      double precision :: Gaussian, Slater, Density, Distance

      double precision, parameter :: MaxDistance = 1.0d1
      double precision, parameter :: Step = 1.0d-4
```

```
      integer :: i

      fit = 0.0d0

      do i = 1, MaxDistance / Step
         Distance = i*Step * Angstrom

         Slater = sSTOCoulInt(zeta, zeta, n, n, Distance)
         Gaussian = sGTOCoulInt(alpha, alpha, Distance)

         Density = (Slater - Gaussian) *
     *        exp(- 0.5d0 alpha * Distance)

         fit = fit + Step * Density
         if (present(zPrint) .and. zPrint ) then
            print *, Distance , Slater, Gaussian, Density
         end if

      end do


      end function fit

      end program fitgto
```

## A.7. geometry.f

```
!>
!! Computes pairwise distances from Cartesian coordinates
!!
!! \param Point1, Point2: 3-vectors of double precisions
!! \return Cartesian distance in atomic units
!<
      double precision function Distance(Point1, Point2)
        implicit none
        double precision, dimension(3), intent(in) :: Point1, Point2
        double precision :: x, y, z
        x = Point1(1) - Point2(1)
        y = Point1(2) - Point2(2)
        z = Point1(3) - Point2(3)
        Distance = sqrt(x*x + y*y + z*z)
      end function Distance

!>
!! Computes inverse pairwise distance from Cartesian coordinates
!!
!! \param Point1, Point2: 3-vectors of double precisions
!! \return Inverse Cartesian distance in atomic units
!<
      double precision function InverseDistance(Point1, Point2)
        implicit none
        double precision, dimension(3), intent(in) :: Point1, Point2
        double precision :: x, y, z, rsq
        double precision, external :: InvSqrt
```

```
        x = Point1(1) - Point2(1)
        y = Point1(2) - Point2(2)
        z = Point1(3) - Point2(3)
        rsq = x*x + y*y + z*z
c        InverseDistance = InvSqrt(rsq)
        InverseDistance = rsq**(-0.5d0)
      end function InverseDistance

!>
!! Checks if two points is nearer than some distance
!! This function exists because the sqrt is expensive to calculate!
!! \param Point1, Point2: 3-vectors of double precisions
!! \param Threshold Distance beyond which is considered 'far'
!! \return Cartesian distance between points exceed Threshold, return
True, otherwise false
!<
      logical function isNear(Point1, Point2, Threshold)
        implicit none
        double precision, dimension(3), intent(in) :: Point1, Point2
        double precision, intent(in) :: Threshold
        double precision :: x, y, z

c       First check if any component is too large
        x = abs(Point1(1) - Point2(1))
        if (x .gt. Threshold) goto 1

        y = abs(Point1(2) - Point2(2))
        if (y .gt. Threshold) goto 1

        z = abs(Point1(3) - Point2(3))
        if (z .gt. Threshold) goto 1

c       Second, check if l1-norm is too large
        if ((x + y + z) .gt. Threshold) goto 1

c       Third, check if l2-norm is too large
        if ((x*y + y*y + z*z) .gt. Threshold*Threshold) goto 1

c       If we made it this far, it's not far
        isNear = .True.
        goto 2
 1      isNear = .False.
 2    end function isNear

!>
!! Contains lookup table
!<
      double precision function InvSqrt(x)
        implicit none
        double precision, intent(in) :: x
        integer :: ex
        double precision :: ab
        integer*8 :: frac
        equivalence (ab, frac)
```

```
          double precision, parameter :: Accuracy = 1.0d-5
          double precision, parameter :: Spacing = (2.0d0 * 0.25d0 *
     &       Accuracy)

          logical :: haveLUT = .False.
          integer, parameter :: LUTSize = int(0.75d0 / Spacing)

          double precision, dimension(LUTSize) :: LookUpTable
          save haveLUT, LookUpTable
          integer :: LUTIndex
          double precision :: Value
c          integer*8 :: xrepr
c          equivalence (Value, xrepr)

          Value = x

c       the sign bit = ibits(xrepr, 63, 1)
c       We will assume it's always positive

c       Pull out exponent
c        ex = ibits(xrepr, 52, 11)-1023
          ex = exponent(x)

c       Pull out abcissa
          ab = fraction(x)

          if (mod(ex, 2) .eq. 1) then
             ex = ex + 1
             ab = ab * 0.5d0
          end if

          if (.not. haveLUT) then
             Value = 0.25d0
             do LUTIndex = 1, LUTSize
                LookUpTable(LUTIndex) = 0.5d0 * Value ** (-0.5d0)
                Value = Value + Spacing
             end do
             haveLUT = .True.
          end if
          ex = (1 - ex / 2)
          LUTIndex = (ab - 0.25d0) / Spacing
          Value = LookUpTable(LUTIndex)
          InvSqrt = Set_Exponent(Value, ex)

      end function invSqrt
```

### A.8. gto-int.f

```
!>
!! Assigns a Gaussian-type orbital to the atom
!!
!! \note See research notes dated 2008-03-14
!<
      subroutine AssignsGTOBasis(theAtom)
```

```
        use Parameters
        implicit none
        type(Atom) :: theAtom
        double precision, external :: sGTOFromHardness

!       Assign position
        theAtom%Basis%Position = theAtom%Position

!       Assign Gaussian orbital exponent by scaling
!       The diagonal Gaussian integral is simply sqrt(pi)
        theAtom%Basis%zeta = sGTOFromHardness(theAtom%Element%Hardness)
      end subroutine AssignsGTOBasis

      function sGTOFromHardness(Hardness)
        use parameters
        implicit none
        double precision, intent(in) :: Hardness
        double precision :: sGTOFromHardness
        sGTOFromHardness = 0.5d0 * pi * Hardness**2
      end function sGTOFromHardness
!>
!! Use parameters fitted from QTPIE STO orbitals.
!<
      subroutine AssignFittedGTOBasis(theAtom)
        use Parameters
        implicit none
        type(Atom) :: theAtom
        integer :: i
!       Assign Gaussian orbital exponent
        do i=1,numParameterizedAtoms
           if (theAtom%Element%Z .eq. ParameterizedAtoms(i)%Z) then
              theAtom%Basis%zeta = GaussianExponent(i)
           end if
        end do
      end subroutine AssignFittedGTOBasis

!>
!! Calculates best-fit GTO exponent given best-fit STO exponent
!! \param n: principal quantum number
!! \param zeta: exponent for s-type Slater orbital
!! \return the best-fit exponent for the s-type Gaussian orbital
!! \note See research notes dated 2007-08-31
!! \deprecated
!<
      double precision  function sSTO2sGTO(n, zeta)
        implicit none
        integer, intent(in) :: n
        double precision, intent(in) :: zeta
        double precision, parameter :: conversion(1:7) = (/
     &       0.2709498089, 0.2527430925, 0.2097635701,
     &       0.1760307725, 0.1507985107, 0.1315902101,
     &       0.1165917484 /)
        sSTO2sGTO = conversion(n) * zeta * zeta
      end function sSTO2sGTO
```

```
!>
!! Calculates a best-fit Gaussian-type orbital (STO-1G) to
!! the Slater-type orbital defined from the hardness parameters
!! \param Hardness: chemical hardness in atomic units
!! \param n: principal quantum number
!! \note See research notes dated 2007-08-30
!! \deprecated
!<
      subroutine AssignsSTO1GBasis(theAtom)
        use Parameters
        implicit none
        type(Atom) :: theAtom
        double precision, external :: sSTOCoulInt
        double precision, external :: sSTO2sGTO
        integer :: n
        double precision :: zeta

C       Approximate the exact value of the constant of proportionality
C       by its value at a very small distance epsilon
C       since the exact R = 0 case has not be programmed into
STOIntegrals
        double precision :: epsilon = 1.0d-6

C       Assign position
        theAtom%Basis%Position = theAtom%Position

C       Assign principal quantum number
        n = pqn(theAtom%Element)
        theAtom%Basis%n = n

C       Assign orbital exponent
        zeta = (sSTOCoulInt(1.0d0, 1.0d0, n, n, epsilon)
     &      /theAtom%Element%Hardness)**(-1.0d0/(3.0d0 + 2.0d0*n))

C       Rewrite it with best-fit Gaussian
        theAtom%Basis%zeta = sSTO2sGTO(n, zeta)
      end subroutine AssignsSTO1GBasis

!>
!! Computes Coulomb integral analytically over s-type GTOs
!!
!! Computes the two-center Coulomb integral over Gaussian-type orbitals
!! of s symmetry.
!!
!! \param a: Gaussian exponent of first atom in atomic units (inverse
squared Bohr)
!! \param b: Gaussian exponent of second atom in atomic units (inverse
squared Bohr)
!! \param R: internuclear distance in atomic units (bohr)
!! \return the value of the Coulomb potential energy integral
!! \note Reference: T. Helgaker, P. Jorgensen, J. Olsen, Molecular
Electronic Structure Theory
!!                  Wiley, NY, 2000, Equations (9.7.21) and (9.8.23)
!<
      double precision  function sGTOCoulInt(a, b, R)
```

```
        implicit none
        double precision, intent(in) :: a,b,R
        intrinsic :: erf
        double precision :: p

        p = sqrt(a * b / (a + b))
        sGTOCoulInt = erf(p * R) / R
      end function sGTOCoulInt

!> Computes overlap integral analytically over s-type GTOs
!!
!! Computes the overlap integral over two Gaussian-type orbitals of s
symmetry.
!! \param a: Gaussian exponent of first atom in atomic units (inverse
squared Bohr)
!! \param b: Gaussian exponent of second atom in atomic units (inverse
squared Bohr)
!! \param R: internuclear distance in atomic units (bohr)
!! \note Reference: T. Helgaker, P. Jorgensen, J. Olsen, Molecular
Electronic Structure Theory
!!                   Wiley, NY, 2000, Equation (9.2.41)
!! \note With normalization constants added, calculates
!! \f[
!! S = \left(\frac{4\alpha\beta}{(\alpha + \beta)^2}\right)^\frac{3}{4}
!!     \exp\left(-\frac{\alpha\beta}{\alpha+\beta} R^2 \right)
!! \f]
!<
      double precision  function sGTOOvInt(a, b, R)
        implicit none
        double precision, intent(in) :: a,b,R
        double precision :: p, q

        p = a + b
        q = a * b / p
        sGTOOvInt = (4*q/p)**0.75d0 * exp(-q*R*R)

!c       Sanity check
!        if (sGTOOvInt .ge. 1.0d0 .or. sGTOOvInt .lt. 0.0d0) then
!           print *, "Error: Overlap integral exceeds bounds:
",sGTOOvInt
!           print *, a, b, R
!           stop
!        end if
      end function sGTOOvInt

!>
!! Computes derivative of Coulomb integral wrt R
!! \param a: Gaussian exponent of first atom in atomic units (inverse
squared Bohr)
!! \param b: Gaussian exponent of second atom in atomic units (inverse
squared Bohr)
!! \param R: internuclear distance in atomic units (bohr)
!> \return the derivative of the Coulomb potential energy integral
!<
      double precision  function sGTOCoulIntGrad(a, b, R)
```

```
        implicit none
        double precision , intent(in) :: a,b,R
        double precision, parameter :: pi =  3.141592653589793d0
        double precision, external :: sGTOCoulInt
        double precision :: p

        if (abs(R) .eq. 0) then
           print *, "FATAL ERROR: R = 0 in sGTOCoulIntGrad"
           stop
        end if

        p = sqrt(a * b / (a + b))
        sGTOCoulIntGrad = 2.0d0 * p / (R * sqrt(pi)) * exp(-(p*R)**2)
     &       - sGTOCoulInt(a,b,R) / R
      end function sGTOCoulIntGrad

!>
!! Computes gradient of overlap integral wrt R
!!
!! Computes the derivative of the overlap integral over two Gaussian-
type orbitals of s symmetry.
!! \param a: Gaussian exponent of first atom in atomic units (inverse
squared Bohr)
!! \param b: Gaussian exponent of second atom in atomic units (inverse
squared Bohr)
!! \param R: internuclear distance in atomic units (bohr)
!> \return the derivative of the sGTOOvInt integral
!<
      double precision function sGTOOvIntGrad(a,b,R)
        implicit none
        double precision, intent(in) :: a,b,R
        double precision, external :: sGTOOvInt

        sGTOOvIntGrad = -2 * (a*b)/(a+b)* R * sGTOOvInt(a,b,R)
      end function sGTOOvIntGrad
```

### A.9. io.f

```
!> Reads XYZ file
!!
!! loads an external file containing a XYZ geometry
!! \param fileName: name of the XYZ geometry file
!! \return A Molecule data structure
!<
      function loadXYZ(fileName)
        use Parameters
        implicit none
        character (len=*), intent(in) :: fileName
        type(Molecule) :: loadXYZ
        character (len=2) :: AtomSymbol
        integer :: j, l, stat
C       file handle
        integer :: fXYZ = 101
```

```
        logical :: isParameterized

        open(unit=fXYZ, status="old", action="read", iostat=stat,
     &       file=fileName)
        if (stat.ne.0) then
          print *,"Problem loading geometry file ", fileName
          stop
        end if
C       First line says how many atoms there are
        read (unit=fXYZ, fmt=*) loadXYZ%NumAtoms
C       Second line may contain a comment, skip it
        read (unit=fXYZ, fmt=*)

C       Allocate memory for atoms and coordinate data
        call NewAtoms(loadXYZ%Atoms, loadXYZ%NumAtoms)

        do j=1,loadXYZ%NumAtoms
          read (unit=fXYZ, fmt=*) AtomSymbol, loadXYZ%Atoms(j)%Position
C         Convert units from Angstroms to atomic units (Bohr)
          loadXYZ%Atoms(j)%Position = loadXYZ%Atoms(j)%Position
     &         * Angstrom
C         look up AtomSymbol to assign parameters
          isParameterized = .False.
          do l=1,numParameterizedAtoms
            if (AtomSymbol.eq.ParameterizedAtoms(l)%Symbol) then
              loadXYZ%Atoms(j)%Element = ParameterizedAtoms(l)
              isParameterized = .True.
            end if
          end do

C         assign basis set
          if (isParameterized) then
C           Assign a Gaussian basis
            call AssignsGTOBasis(loadXYZ%Atoms(j))
            call AssignFittedGTOBasis(loadXYZ%Atoms(j))
C           Replace with this line to assign STO
c            call AssignsSTOBasis(loadXYZ%Atoms(j))
          else
             print *, "Error: Unknown element type: ", AtomSymbol
             stop
          end if
        end do
c       By default, assign zero total charge
        loadXYZ%TotalCharge = 0.0d0
        close(fXYZ)
      end function loadXYZ

!>
!! Read in parameters from an external file
!<
      subroutine UpdateParameters(filename, Mol)
        use Parameters
        implicit none
        character (len=*), intent(in) :: fileName
        type(AtomData), dimension(:), allocatable :: ParameterSet
```

```
        integer :: i, j, N, stat

        type(Molecule), intent(inout), optional :: Mol
        logical :: isParameterized
        double precision, dimension(:), allocatable ::
     &     CustomGaussianExponent
        double precision, external:: sGTOFromHardness

C       file handle
        integer :: fPar = 1002

        open(unit=fPar, status="old", action="read", iostat=stat,
     &       file=fileName)
        if (stat.ne.0) then
          print *,"Problem loading parameter file ", fileName
          stop
        end if

C       First line says how many Parameters there are
        read (unit=fPar, fmt=*) N

c       Allocate
        if (allocated(ParameterSet)) deallocate(ParameterSet)
        allocate(ParameterSet(N))

        call NewVector(CustomGaussianExponent,N)

        do i=1,N
          read (unit=fPar, fmt=*) ParameterSet(i)%Symbol,
     &      ParameterSet(i)%Z, ParameterSet(i)%FormalCharge,
     &      ParameterSet(i)%Electronegativity,
ParameterSet(i)%Hardness,
     &      CustomGaussianExponent(i)
c     Assume units of electron volts are specified
          ParameterSet(i)%Electronegativity =
     ,         ParameterSet(i)%Electronegativity * eV
          ParameterSet(i)%Hardness = ParameterSet(i)%Hardness * eV
c     If exponent specified is zero, then calculate it automatically
from hardness relation
          if (abs(CustomGaussianExponent(i)) .lt. 1.0d-16) then
            CustomGaussianExponent(i) =
     &          sGTOFromHardness(ParameterSet(i)%Hardness)
            print *, "Automatically generated Gaussian exponent"
            print *, ParameterSet(i)%Symbol,
     &         ParameterSet(i)%Z, ParameterSet(i)%FormalCharge,
     &         CustomGaussianExponent(i)

          end if
        end do

        close(fPar)

c     If Molecule is specified, update its parameters
        if (present(Mol)) then
           do i=1,Mol%NumAtoms
```

```
C     look up AtomSymbol to assign parameters
              isParameterized = .False.
              do j=1,N
                 if (Mol%Atoms(i)%Element%Symbol
     &               .eq.ParameterSet(j)%Symbol) then
                    Mol%Atoms(i)%Element = ParameterSet(j)
                    Mol%Atoms(i)%Basis%zeta = CustomGaussianExponent(j)
                    isParameterized = .True.
                 end if
              end do

C     assign basis set
c              if (isParameterized) then
c                 call AssignsGTOBasis(Mol%Atoms(i))
c              else
c                 print *, "Warning, could not parameterize atom", i
c              end if
           end do
        end if
      end subroutine UpdateParameters

!>
!! dumps QTPIE calculation results
!! \param Mol: molecule data structure
!! \param fileName Name of the log file to write or append to
!<
      subroutine WriteLog(Mol, fileName)
        use Parameters
        character (len=*), intent(in) :: fileName
        type(Molecule), intent(in) :: Mol
C       file handle
        integer :: fXYZ = 101, stat

        open(unit=fXYZ, status="new", action="write", iostat=stat,
     &       file=fileName)
        if (stat.ne.0) then
c    gfortran's code is 17, pgf90's is 208, ifort's is 10
          if ((stat.eq.208) .or. (stat.eq.17) .or. (stat.eq.10)) then
C           File already exists
            open(unit=fXYZ, status="old", action="write", iostat=stat,
     &           position="append", file=fileName)
          else
            print *,"Problem opening log file"
            print *,"Status code = ", stat
            stop
          end if
        end if

C       write out charges
        write (unit=fXYZ, fmt=1) Mol%Energy, Mol%Atoms(:)%Charge
 1     format(99999f10.5)

        close(fXYZ)
      end subroutine writelog
```

```
!>
!! dumps molecular geometry from QTPIE in XYZ formal
!! \param Mol: molecule data structure
!! \param fileName: Name of geometry file to write or append to
!<
      subroutine WriteXYZ(Mol, fileName)
        use Parameters
        character (len=*), intent(in) :: fileName
        type(Molecule), intent(in) :: Mol
C       file handle
        integer :: fXYZ = 102, stat, j

        open(unit=fXYZ, status="new", action="write", iostat=stat,
     &       file=fileName)
        if (stat.ne.0) then
          if ((stat.eq.208) .or. (stat .eq. 17) .or. (stat.eq.10)) then
C           File already exists
            open(unit=fXYZ, status="old", action="write", iostat=stat,
     &           position="append", file=fileName)
          else
            print *,"Problem writing geometry file ",fileName
            print *,"Status code = ", stat
            stop
          end if
        end if
C       write file
        write (unit=fXYZ, fmt=*) Mol%NumAtoms
        write (unit=fXYZ, fmt=*) "Written by QTPIE : WriteXYZ()"
        do j=1,Mol%NumAtoms
          write (unit=fXYZ, fmt=2) Mol%Atoms(j)%Element%Symbol,
     &            Mol%Atoms(j)%Position/Angstrom
        end do
        close(fXYZ)

 2     format(a2,3f15.10)

       end subroutine WriteXYZ
```

## A.10. matrixutil.f

```
!>
!! Determines if matrix is diagonally dominant
!! \param A : A real (double precision) square matrix
!! \param N : dimension of matrix
!! \return .True. if matrix is diagonally dominant
!! \deprecated
!<
      logical function IsDiagonallyDominant(A, N)
        implicit none
        integer, intent(in) :: N
        double precision, dimension(N,N), intent(in) :: A

        integer :: i,j
```

```
        double precision :: maxElt, maxDiag, maxRowElt

        IsDiagonallyDominant = .true.
        maxDiag = 0.0d0
        maxElt = 0.0d0
        do i = 1,N
          maxRowElt = 0.0d0
          do j = 1,N
            if (i.ne.j) then
              if (abs(A(i,j)).gt.abs(maxRowElt)) then
                maxRowElt = A(i,j)
              else
                IsDiagonallyDominant = .false.
                goto 1
              end if
            end if
          end do

          print *, i, A(i,i), maxRowElt, A(i,i)/maxRowElt

          if (abs(maxDiag).lt.abs(A(i,i))) then
            maxDiag = A(i,i)
          end if

          if (abs(maxElt).lt.abs(maxRowElt)) then
            maxElt = maxRowElt
          end if
        end do
c        Final decision
        print *, maxDiag, maxElt
        if (abs(maxElt).gt.abs(maxDiag)) then
          IsDiagonallyDominant = .false.
        else
          IsDiagonallyDominant = .true.
        end if
 1    end function IsDiagonallyDominant
```

## A.11. onexyz.f

```
!>
!! Runs QTPIE for a single XYZ geometry
!<
      program onexyz
        use Parameters
        implicit none
        type(Molecule) :: Mol
        type(Molecule), external :: loadXYZ
        integer :: NumArgs
        intrinsic :: iargc
        character (len = 50) :: fileName, paramfilename
        external :: dipmom
        double precision, dimension(3) :: dm
        double precision, dimension(3,3) :: pol
        print *, "Single geometry mode"
```

```
          NumArgs = iargc()
          if (NumArgs.ge.1) then
             call getarg(1, fileName)
             print *, "Reading in file ", fileName
          else
             print *, "Reading default file name qtpie.xyz"
             fileName = "qtpie.xyz"
          end if

          Mol =  loadXYZ(fileName)

          if (NumArgs.ge.2) then
             call getarg(2, paramfileName)
             print *, "Reading in parameter file ", paramfileName
             call UpdateParameters(paramfileName, Mol)
          end if

          call DosGTOIntegrals(Mol)

          print *, "Using QTPIE"
          call QTPIE(Mol)
          print *, "QTPIE Energy is", Mol%Energy
!          print *, "Using QEq"
!          call QEq(Mol)
!          print *, "QEq Energy is", Mol%Energy

          call WriteLog(Mol, "qtpie.log")
          print *, "Calculated charges written to qtpie.log"

          call dipmom(Mol, dm)
          print *, "Dipole moment (Debyes)"
          print *, dm/Debye
          print *, "Norm = ",
     &     sqrt(dm(1)*dm(1)+dm(2)*dm(2)+dm(3)*dm(3))/Debye
          print *, "Dipole moment (atomic units)"
          print *, dm
          print *, "Polarizability (atomic units)"
          call polarizability(Mol, pol)
!          call polarizability_ff(Mol, pol)
          print *, pol(1,1:3)
          print *, pol(2,1:3)
          print *, pol(3,1:3)

          !call polarizability_ff(Mol, pol)
          !print *, pol(1,1:3)
          !print *, pol(2,1:3)
          !print *, pol(3,1:3)
          print *, Mol%Energy, dm(1), dm(2), dm(3),
     &    pol(1,1), pol(2,2), pol(3,3)

!        print *, "Numerical forces"
!        call DoGradientsByFiniteDifference(Mol)
!        print *, Mol%EGradient
!        print *, "Analytic forces"
!        call DoGradientsAnalytically(Mol)
```

```
!      print *, Mol%EGradient

      print *, "Analysis specific to this batch of data"
      print *, filename,
     & sqrt(dm(1)*dm(1)+dm(2)*dm(2)+dm(3)*dm(3))/Debye,
     & pol(2,2) / (Angstrom ** 3), pol(3,3) / (Angstrom**3)

!      print *, Mol%Energy, dm(1)/Debye, dm(2)/Debye,
!     & dm(3)/Debye
      end program onexyz
```

## A.12. parameters.f

```
!>
!! Stores parameters for our charge models
!<
      module Parameters
        use AtomicUnits
        use SparseMatrix
        implicit none
        save

        double precision, parameter :: pi =  3.141592653589793d0

!>
!! Parameters for a s-type Slater type orbital (STO) basis function
!!
!! \param n : principal quantum number
!! \param zeta : zeta exponent with dimensions of inverse length in
atomic units
!<
      type sSTO
        double precision, dimension(1:3) :: Position
        integer :: n
        double precision :: zeta
      end type

!> Parameters for a s-type Gaussian type orbital (GTO) basis function
!!
!! \param Position : an array of three double precisions describing
Cartesian coordinates
!! \param zeta: exponent with dimensions of inverse square length in
atomic units
!<
      type sGTO
        double precision, dimension(1:3) :: Position
        double precision :: zeta
      end type

!>
!! Atomic parameters
!! \param Symbol            : elemental symbol
!! \param Z                 : atomic number
!! \param FormalCharge      : formal charge, integers only
```

```
!! \param Electronegativity : Mulliken electronegativity in atomic
units
!! \param Hardness          : Parr-Pearson chemical hardness in atomic
units
!<
      type AtomData
        character (len = 2) :: Symbol
        integer             :: Z, FormalCharge
        double precision    :: Electronegativity, Hardness
      end type AtomData

!>
!! Describes an atom in a molecule
!!
!! \param  Element : type(AtomData) containing atomic parameters
!! \param Basis   : A basis function associated with the atom
!! \param Position: double precision(3) vector of Cartesian coordinates
describing spatial location
!! \param Charge  : double precision, result of charge model
calculation
!<
      type Atom
        type (AtomData) :: Element
        type (sSTO) :: Basis
        double precision, dimension(1:3) :: Position
        double precision :: Charge
      end type Atom

!>
!! Describes a molecular system
!!
!! \param Description: a text label of 132 characters
!! \param    NumAtoms: number of atoms (integer)
!! \param TotalCharge: total charge of system (double precision)
!! \param       Atoms: array of atoms
!! \param     Overlap: overlap matrix
!! \param      OvNorm: overlap norm vector (useful temporary variable)
!! \param     Coulomb: Coulomb matrix
!! \param      Energy: QTPIE contribution to the potential energy
!! \param   EGradient: Energy gradients
        type Molecule
          character (len=132) :: Description
          integer :: NumAtoms
          double precision :: TotalCharge
          double precision :: ChemicalPotential, SchurCoulomb
          Type(Atom), dimension(:), allocatable :: Atoms
! Use sparse datatype for Overlap!
          Type(CSRMatrix) :: Overlap
c          double precision, dimension(:,:), allocatable :: Overlap
          double precision, dimension(:), allocatable :: OvNorm
          double precision, dimension(:), allocatable :: Voltage
          double precision, dimension(:,:), allocatable :: Coulomb
          double precision :: Energy
          double precision, dimension(:,:), allocatable :: EGradient
        end type Molecule
```

```
C       Here are a bunch of predefined elements
C       As parameterized by Rappe and Goddard for QEq

        type(AtomData), parameter :: Hydrogen    =
     &       AtomData( "H",  1, 0, 4.528*eV, 13.890*eV)
        type(AtomData), parameter :: Lithium     =
     &       AtomData("Li",  3, 0, 3.006*eV,  4.772*eV)
        type(AtomData), parameter :: Carbon      =
     &       AtomData( "C",  6, 0, 5.343*eV, 10.126*eV)
        type(AtomData), parameter :: Nitrogen   =
     &       AtomData( "N",  7, 0, 7.139*eV, 12.844*eV)
        type(AtomData), parameter :: Oxygen     =
     &       AtomData( "O",  8, 0, 8.741*eV, 13.364*eV)
        type(AtomData), parameter :: Fluorine   =
     &       AtomData( "F",  9, 0,10.874*eV, 14.948*eV)
        type(AtomData), parameter :: Sodium     =
     &       AtomData("Na", 11, 0, 2.843*eV,  4.592*eV)
        type(AtomData), parameter :: Silicon    =
     &       AtomData("Si", 14, 0, 4.168*eV,  6.974*eV)
        type(AtomData), parameter :: Phosphorus =
     &       AtomData( "P", 15, 0, 5.463*eV,  8.000*eV)
        type(AtomData), parameter :: Sulphur    =
     &       AtomData( "S", 16, 0, 6.084*eV, 10.660*eV)
        type(AtomData), parameter :: Chlorine   =
     &       AtomData("Cl", 17, 0, 8.564*eV,  9.892*eV)
        type(AtomData), parameter :: Potassium  =
     &       AtomData( "K", 19, 0, 2.421*eV,  3.840*eV)
        type(AtomData), parameter :: Bromine    =
     &       AtomData("Br", 35, 0, 7.790*eV,  8.850*eV)
        type(AtomData), parameter :: Rubidium   =
     &       AtomData("Rb", 37, 0, 2.331*eV,  3.692*eV)
        type(AtomData), parameter :: Iodine =
     &       AtomData( "I", 53, 0, 6.822*eV,  7.524*eV)
        type(AtomData), parameter :: Cesium =
     &       AtomData("Cs", 55, 0, 2.183*eV,  3.422*eV)

        integer, parameter :: numParameterizedAtoms = 16 !< Number of
defined atomic parameters
        type(AtomData), parameter ::
     &   ParameterizedAtoms(numParameterizedAtoms) =
     &   (/Hydrogen, Lithium, Carbon, Nitrogen, Oxygen, Fluorine,
     &     Sodium, Silicon, Phosphorus, Sulphur, Chlorine,
     &     Potassium, Bromine, Rubidium, Iodine, Cesium /) !< Array of
defined atomic parameters

C       Parameters for cations. All experimental values!
!>
!!      Sodium cation
!<
        type(AtomData), parameter :: SodiumCation =
     &       AtomData("Na",11,+1,4562*kJ_mol, 5.13908*eV)
```

```
c       Data for newly parameterized Gaussian basis set
        double precision, parameter, dimension(numParameterizedAtoms)
::
     &       GaussianExponent =
     &   (/  0.534337523756312, 0.166838519142176, 0.206883838259186,
     &       0.221439796025873, 0.223967308625516, 0.231257590182828,
     &       0.095892938712585, 0.105219608142377, 0.108476721661715,
     &       0.115618357843499, 0.113714050615107, 0.060223294377778,
     &       0.070087547802259, 0.041999054745368, 0.068562697575073,
     &       0.030719481189777 /)

!>
!overlape!      Threshold for calculating overlap integrals
!<
        double precision, parameter :: OvIntThreshold = 1.0d-9

        double precision :: SmallestGaussianExponentInSystem = 1.0d40
!>
!!      Store pre-calculated thresholds for prescreening
!<
        double precision :: OvIntMaxR
!>
!!      Threshold for calculating Coulomb integrals
!<
        double precision, parameter :: CoulIntThreshold = 1.0d-9
!>
!!      Store pre-calculated thresholds for prescreening
!<
        double precision :: CoulIntMaxR
        contains

!>
!! A utility function for allocating dynamic memory for vectors and
matrices
!<
        subroutine NewVector(V, N)
        double precision, dimension(:), allocatable :: V
        integer, intent(in) :: N

        integer :: status

        if (allocated(V)) then
           deallocate(V, STAT=status)

           if (status .ne. 0) then
              print *,"FATAL ERROR: could not deallocate memory"
              stop
           end if
        end if

        allocate(V(N), STAT=status)

        if (status .ne. 0) then
           print *,"FATAL ERROR: could not allocate memory"
           stop
```

```
        end if

        end subroutine NewVector

!>
!! A utility function for allocating dynamic memory for matrices
!<
        subroutine NewMatrix(V, N, M)
        double precision, dimension(:,:), allocatable :: V
        integer, intent(in) :: N
        integer, intent(in), optional :: M

        integer :: status

        if (allocated(V)) then
           deallocate(V, STAT=status)

           if (status .ne. 0) then
              print *,"FATAL ERROR: could not deallocate memory"
              stop
           end if
        end if

        if (present(M)) then
           allocate(V(N, M), STAT=status)
        else
           allocate(V(N, N), STAT=status)
        end if

        if (status .ne. 0) then
           print *,"FATAL ERROR: could not allocate memory"
           stop
        end if

        end subroutine NewMatrix

!>
!! A utility function for allocating dynamic memory for a vector of
atoms
!<
        subroutine NewAtoms(Atoms, N)
        Type(Atom), dimension(:), allocatable :: Atoms
        integer, intent(in) :: N

        integer :: status

        if (allocated(Atoms)) then
           deallocate(Atoms, STAT=status)

           if (status .ne. 0) then
              print *,"FATAL ERROR: could not deallocate memory"
              stop
           end if
        end if
```

```
        allocate(Atoms(N), STAT=status)

        if (status .ne. 0) then
           print *,"FATAL ERROR: could not allocate memory"
           stop
        end if

        end subroutine NewAtoms

!>
!! Computes the expectation value of the radial distance over s-type
STOs
!! \param Basis: s-type STO basis function
!! \return the expectation value of the radial distance over s-type
STOs
!<
          double precision function ExpectR(Basis)
            implicit none
            type(sSTO), intent(in) :: Basis
            ExpectR = (Basis%n + 0.5) / Basis%zeta
          end function ExpectR

!>
!! Computes the principal quantum number of an atom given its atomic
number
!! \param theAtom Atom to determine principle quantum number for
!! \return the principal quantum number
!<
      integer function pqn(theAtom)
        implicit none
        integer :: j
        integer, parameter :: maxelectrons(7)=
     &    (/ 2, 10, 18, 36, 54, 86, 118 /) !< Lookup table for max
number of electrons for that quantum number
        type(AtomData), intent(in) :: theAtom
        pqn=1
C       work through each shell
        do j=1,7
          if (theAtom%Z.gt.maxelectrons(j)) then
             pqn=pqn+1
          end if
        end do
      end function pqn

      end module Parameters
```

### A.13. properties.f

```
!>
!! Computes the dipole moment
!! \param Mol The molecule
!! \param dm the dipole moment vector (size = 3)
!<
      subroutine dipmom(Mol, dm)
```

```
        use Parameters
        type(Molecule), intent(in) :: Mol
        double precision, dimension(3) :: dm
        double precision :: WeightedDistance
        integer :: i,j,k

        dm = 0.0d0

        do k=1,3
           do i=1,Mol%NumAtoms
              WeightedDistance = 0.0d0
              do j = Mol%Overlap%RowStart(i),
     &               Mol%Overlap%RowStart(i+1)-1
                 WeightedDistance = WeightedDistance +
     &                Mol%Overlap%Value(j) *
     &                ( Mol%Atoms(i)%Position(k)
     &                - Mol%Atoms(Mol%Overlap%ColIdx(j))%Position(k))
              end do
              WeightedDistance = WeightedDistance * Mol%OvNorm(i)
              dm(k) = dm(k) + WeightedDistance * Mol%Atoms(i)%Charge
           end do
        end do
      end subroutine dipmom

!>
!! Computes the dipole polarizability tensor
!! \param Mol The molecule
!! \param pol the dipole polarizability tensor (size = 3,3)
!! \todo Untested!
!<
      subroutine polarizability(Mol, pol)
        use Parameters
        type(Molecule), intent(in) :: Mol
        double precision, dimension(3,3), intent(out) :: pol
        integer, save :: N
        double precision, dimension(:,:), allocatable ::
     &      WeightedDistance, Temp
        double precision, dimension(:), allocatable :: Ones
        double precision :: TmpDist

c       Level 1 BLAS function for calculating scalar product of vectors
        double precision, external :: ddot

        integer :: i, j, mu, nu

        if (N .ne. Mol%NumAtoms) then
           N = Mol%NumAtoms
           call NewMatrix(WeightedDistance, N, 3)
           call NewMatrix(Temp, N, 3)
           call NewVector(Ones, N)
           Ones = 1.0d0
        end if

        pol = 0.0d0
```

```
        do nu=1,3
c          Calculate weighted distances
           do i=1,Mol%NumAtoms
              TmpDist = 0.0d0
              do j = Mol%Overlap%RowStart(i),
     &               Mol%Overlap%RowStart(i+1)-1
                 TmpDist = TmpDist + Mol%Overlap%Value(j) *
     &                ( Mol%Atoms(i)%Basis%Position(nu)
     &                - Mol%Atoms(Mol%Overlap%ColIdx(j))
     &                     %Basis%Position(nu))
              end do
              TmpDist = TmpDist * Mol%OvNorm(i)
              WeightedDistance(i,nu) = TmpDist
           end do
c       Solve Mol%Coulomb * Temp(nu) = WeightedDistance(nu)
c       for each spatial direction nu
           call solver(N, Mol%Coulomb, WeightedDistance(1:N, nu),
     &          Temp(1:N, nu))
        end do

c       Calculate elements of polarizability tensor
        do mu=1,3
           do nu=1,3
              pol(mu, nu) =
     &         ddot(N, WeightedDistance(1:N, mu), 1, Temp(1:N, nu), 1)
     &        -(ddot(N, Ones, 1, Temp(1:N, mu), 1)
     &         *ddot(N, Ones, 1, Temp(1:N, nu), 1))/Mol%SchurCoulomb
           end do
        end do

      end subroutine polarizability
!>
!! Computes the dipole polarizability tensor using the method of finite
fields
!! \param Mol The molecule
!! \param pol the dipole polarizability tensor (size = 3,3)
!! \todo Untested!
!<
      subroutine polarizability_ff(Mol, pol)
        use Parameters
        type(Molecule), intent(inout) :: Mol
        double precision, dimension(3,3), intent(out) :: pol
        double precision, dimension(-1:1,-1:1,-1:1) :: nrg
        integer :: i,j,k,n
        double precision, parameter :: FiniteFieldStrength = 1.0d-4
        integer, parameter :: x = 1, y = 2, z = 3
        nrg = 0.0d0
        do i = -1,1
           do j = -1,1
              do k = -1,1
                if (abs(i)+abs(j)+abs(k) .gt.2) exit
c               Perturb electronegativities
                do n = 1, Mol%NumAtoms
                     Mol%Atoms(n)%Element%Electronegativity
     &             = Mol%Atoms(n)%Element%Electronegativity
```

```
     &             - FiniteFieldStrength
     &             * ( Mol%Atoms(n)%Basis%Position(x) * i
     &               + Mol%Atoms(n)%Basis%Position(y) * j
     &               + Mol%Atoms(n)%Basis%Position(z) * k)
                end do
                call QTPIE(Mol)
!                call QEq(Mol)
                nrg(i,j,k) = Mol%Energy
                do n = 1, Mol%NumAtoms
                     Mol%Atoms(n)%Element%Electronegativity
     &             = Mol%Atoms(n)%Element%Electronegativity
     &             + FiniteFieldStrength
     &             * ( Mol%Atoms(n)%Basis%Position(x) * i
     &               + Mol%Atoms(n)%Basis%Position(y) * j
     &               + Mol%Atoms(n)%Basis%Position(z) * k)
                end do
              end do
           end do
        end do

      pol(x,x)=-(nrg(1,0,0)-2*nrg(0,0,0)+nrg(-1,0,0))
     &  *FiniteFieldStrength**(-2)
      pol(y,y)=-(nrg(0,1,0)-2*nrg(0,0,0)+nrg(0,-1,0))
     &  *FiniteFieldStrength**(-2)
      pol(z,z)=-(nrg(0,0,1)-2*nrg(0,0,0)+nrg(0,0,-1))
     &  *FiniteFieldStrength**(-2)

      pol(x,y)=-(nrg(1,1,0)-nrg(-1,1,0)-nrg(1,-1,0)+nrg(-1,-1,0))*0.25
     &  *FiniteFieldStrength**(-2)
      pol(x,z)=-(nrg(1,0,1)-nrg(-1,0,1)-nrg(1,0,-1)+nrg(-1,0,-1))*0.25
     &  *FiniteFieldStrength**(-2)
      pol(y,z)=-(nrg(0,1,1)-nrg(0,-1,1)-nrg(0,1,-1)+nrg(0,-1,-1))*0.25
     &  *FiniteFieldStrength**(-2)

      pol(y,x)=pol(x,y)
      pol(z,x)=pol(x,z)
      pol(z,y)=pol(y,z)

      end subroutine polarizability_ff
```

### A.14. qtpie.f

```
!>
!! Populates integral matrices in Mol data type
!!
!! Mol%Coulomb and Mol%Overlap are initialized
!! \param Mol : of the Molecule data type
!<
      subroutine DosGTOIntegrals(Mol)
        use Parameters
        implicit none
        double precision, external :: sGTOCoulInt, sGTOOvInt
        double precision, external :: Distance, InverseDistance
        logical, external :: isNear
```

```
        type(Molecule) :: Mol
        integer :: i1, i2, N, CSRIdx
        logical :: isFirstInRow
        save N

!       Temporary variables for caching
        double precision :: R !< Temporary distance
        double precision :: zeta1, zeta2 !< Scalar replacment variables
for exponents
        double precision :: Integral !< Temporary integrals
        double precision, dimension(3) :: Pos1, Pos2

        if (N .ne. Mol%NumAtoms) then
           N = Mol%NumAtoms
           call NewMatrix(Mol%Coulomb, N)
           call CSRNew(Mol%Overlap, N, N, N*N)
           call NewVector(Mol%OvNorm, N)
        end if

C       Calculate integral pre-screening thresholds
        do i1 = 1,N
          SmallestGaussianExponentInSystem = min(
     &       SmallestGaussianExponentInSystem,
     &       Mol%Atoms(i1)%Basis%zeta)
        end do

        OvIntMaxR = sqrt(
     M      log( (pi/(2*SmallestGaussianExponentInSystem)**3)
     E           / OvIntThreshold**2)
     M       /SmallestGaussianExponentInSystem)

!       An asymptotic expansion of erfc-1(x) gives this formula
        CoulIntMaxR = 2 * sqrt(-log(CoulIntThreshold)/
     &                    SmallestGaussianExponentInSystem)

C       Populate integral matrices
        CSRIdx = 0
        do i1 = 1, N
           Pos1 = Mol%Atoms(i1)%Basis%Position
           zeta1= Mol%Atoms(i1)%Basis%zeta
           isFirstInRow = .True.
           do i2 = 1, i1-1
c             Although appearing earlier in the code, this is the LOWER
c             triangle that is calculated LATER.
              Pos2 = Mol%Atoms(i2)%Basis%Position

              if (isNear(Pos1, Pos2, CoulIntMaxR)) then
                 zeta2= Mol%Atoms(i2)%Basis%zeta
                 R = Distance(Pos1, Pos2)
                 Integral = sGTOCoulInt(zeta1, zeta2, R)
              else
                 Integral = InverseDistance(Pos1, Pos2)
              end if

              Mol%Coulomb(i1, i2) = Integral
```

```
               Mol%Coulomb(i2, i1) = Integral

c              If Overlap integral is judged to be big enough, calculate
it
               if (isNear(Pos1, Pos2, OvIntMaxR)) then
                  zeta2= Mol%Atoms(i2)%Basis%zeta
                  R = Distance(Pos1, Pos2)

c                 The Overlap matrix is stored in CSR (compressed sparse
c                 row) format in lower triangular form. First increment
the
c                 CSR array index, then save the column index and the
data.

                  CSRIdx = CSRIdx + 1
                  Mol%Overlap%ColIdx(CSRIdx) = i2
                  Mol%Overlap%Value (CSRIdx) = sGTOOvInt(zeta1, zeta2,
R)

c                 If this is the first element in the matrix, also set
the
c                 row index value
                  if (isFirstInRow) then
                     Mol%Overlap%RowStart(i1) = CSRIdx
                     isFirstInRow = .False.
                  end if
               end if
            end do

C           For the diagonal elements, use hardness
            Mol%Coulomb(i1, i1) = Mol%Atoms(i1)%Element%Hardness
c           Diagonal element
            CSRIdx = CSRIdx + 1
            Mol%Overlap%ColIdx(CSRIdx) = i1
            Mol%Overlap%Value (CSRIdx) = ONE
            if (isFirstInRow) then
               Mol%Overlap%RowStart(i1) = CSRIdx
               isFirstInRow = .False.
            end if

c           For overlap matrix, the CSR format makes it easier to NOT
c           take advantage of symmetry
            do i2 = i1+1, N
               Pos2 = Mol%Atoms(i2)%Basis%Position
               if (isNear(Pos1, Pos2, OvIntMaxR)) then
                  zeta2= Mol%Atoms(i2)%Basis%zeta
                  R = Distance(Pos1, Pos2)

                  CSRIdx = CSRIdx + 1
                  Mol%Overlap%ColIdx(CSRIdx) = i2
                  Mol%Overlap%Value (CSRIdx) = sGTOOvInt(zeta1, zeta2,
R)
               end if
            end do
```

```
        end do

        Mol%Overlap%RowStart(N+1) = CSRIdx + 1
c       Calculate due normalization
        do i1 = 1, N
           Mol%OvNorm(i1) = 1.0d0/(SumRow(Mol%Overlap, i1))
        end do

      end subroutine DosGTOIntegrals

!>
!! Populates integral matrices in Mol data type
!!
!! Mol%Coulomb and Mol%Overlap are initialized
!! \param Mol : of the Molecule data type
!! \note This subroutine does NOT work since the Overlap matrix has
been
!! changed to a sparse format.
!<
      subroutine DosSTOIntegrals(Mol)
        use Parameters
        implicit none
        double precision, external :: sSTOCoulInt, sSTOOvInt, Distance
        type(Molecule) :: Mol
        integer :: i1, i2, N, stat

        double precision, dimension(:,:), allocatable :: RefOverlap

C       Check if memory for integral matrices have been allocated
        if (N .ne. Mol%NumAtoms) then
           N = Mol%NumAtoms
           call NewMatrix(Mol%Coulomb, N)
!           call NewMatrix(Mol%Overlap, N)
           call NewVector(Mol%OvNorm, N)
           call NewMatrix(RefOverlap, N)
        end if

C       Now compute Coulomb matrix
        do i1 = 1,Mol%NumAtoms
           do i2 = 1, i1-1
                 Mol%Coulomb(i1, i2) = sSTOCoulInt(
     &               Mol%Atoms(i1)%Basis%zeta,
Mol%Atoms(i2)%Basis%zeta,
     &               Mol%Atoms(i1)%Basis%n , Mol%Atoms(i2)%Basis%n,
     &               Distance(Mol%Atoms(i1)%Basis%Position,
     &                        Mol%Atoms(i2)%Basis%Position) )
c              print *, "Co", i1, i2, Mol%Coulomb(i1,i2)
C             Fill in the other triangle
              Mol%Coulomb(i2, i1) = Mol%Coulomb(i1,i2)
           end do
C          For the diagonal elements, use hardness
              Mol%Coulomb(i1, i1) = Mol%Atoms(i1)%Element%Hardness
        end do

C       Now compute Overlap and RefOverlap matrices
```

```
        do i1 = 1,Mol%NumAtoms
           do i2 = 1, i1-1
!              Mol%Overlap(i1, i2) = sSTOOvInt(
!     &            Mol%Atoms(i1)%Basis%zeta, Mol%Atoms(i2)%Basis%zeta,
!     &            Mol%Atoms(i1)%Basis%n , Mol%Atoms(i2)%Basis%n,
!     &            Distance(Mol%Atoms(i1)%Basis%Position,
!     &                     Mol%Atoms(i2)%Basis%Position) )

C             Calculate the same quantity but referenced to an
intrinsic
C             length scale
              RefOverlap(i1, i2) = sSTOOvInt(
     &            Mol%Atoms(i1)%Basis%zeta, Mol%Atoms(i2)%Basis%zeta,
     &            Mol%Atoms(i1)%Basis%n , Mol%Atoms(i2)%Basis%n,
     &            ExpectR(Mol%Atoms(i1)%Basis)
     &           +ExpectR(Mol%Atoms(i2)%Basis) )

C             Fill in the other triangle
c              print *, "Ov", i1, i2, Mol%Overlap(i1,i2)
!              Mol%Overlap(i2, i1) = Mol%Overlap(i1, i2)
              RefOverlap(i2, i1) = RefOverlap(i1, i2)
           end do
C          For the diagonal elements, the overlap is just the orbital
normalization
!           Mol%Overlap(i1, i1) = 1.0d0
           RefOverlap(i1, i1) = 1.0d0
        end do

C     Now compute normalization of Attenuation (overlap) matrix
      do i1 = 1,Mol%NumAtoms
        Mol%OvNorm(i1) = 0.0d0
        do i2 = 1,Mol%NumAtoms
          Mol%OvNorm(i1) = Mol%OvNorm(i1) + RefOverlap(i1, i2)
        end do
        Mol%OvNorm(i1) = Mol%OvNorm(i1) / Mol%NumAtoms
      end do

C     Deallocate temporary variables
      deallocate(RefOverlap, STAT=stat)
      end subroutine DosSTOIntegrals

!>
!! Populates atomic charges according to the QEq(-H) charge model
!! \param Mol : of the Molecule data type
!! Mol%Atoms(i)%Charge are computed
!> \note The model is described in the seminal paper below:
!!       "Charge equilibration for Molecular dynamics simulations"
!!       A. K. Rappe and W. A. Goddard, J. Phys. Chem., 1991, 95(8),
3358-3363
!!       doi:10.1021/j100161a070
!> \note This implementation does not do the additional procedure for H
atoms
!!       nor does it check for overly large charges that exceed the
principal
!!      quantum number of the given atom.
```

```
!<
      subroutine QEq(Mol)
        use Parameters
        implicit none
        type(Molecule) :: Mol
        integer :: i1, i2

        integer :: N !< size of problem

        external :: SolveConstrained

        N = Mol%NumAtoms

        call SolveConstrained(N, Mol%Coulomb,
     &       -Mol%Atoms(1:N)%Element%Electronegativity,
     &       Mol%Atoms(1:N)%Charge,
     & Mol%ChemicalPotential, Mol%SchurCoulomb)

*       Calculate energy
        Mol%Energy = 0.0d0
        do i1=1,N
           Mol%Energy = Mol%Energy + Mol%Atoms(i1)%Charge
     &          * Mol%Atoms(i1)%Element%Electronegativity
           do i2=1,N
c             Calculate the contribution to the electrostatic energy.
If
c             we are interfacing with TINKER, remember to turn off
c             corresponding calculation in TINKER to avoid double
c             counting
              Mol%Energy = Mol%Energy + 0.5d0 * Mol%Atoms(i1)%Charge
     &             * Mol%Atoms(i2)%Charge * Mol%Coulomb(i1, i2)
           end do
        end do
      end subroutine QEq

!>
!! Populates atomic charges according to the QTPIE charge model
!! \param Mol : of the Molecule data type
!! Mol%Atoms(i)%Charge are computed
!! \note The model is described in the paper below:
!!       J. Chen and T. J. Martinez, Chem. Phys. Lett., 438 (4-6),
2007, 315-320
!!       doi:10.1016/j.cplett.2007.02.065
!<
      subroutine QTPIE(Mol)
        use Parameters
        implicit none
        type(Molecule) :: Mol

        double precision :: ThisCharge !< Temporary atomic charge
variable
        double precision :: VoltageDifference
C       i1-i2 loop over atoms
        integer :: i1, i2
```

```
        integer :: N !< size of matrix problem
        save N

C       Wrapper for linear algebra solver
        external :: SolveConstrained

c       Check if memory needs to be allocated
        if (N .ne. Mol%NumAtoms) then
           N = Mol%NumAtoms
           call NewVector(Mol%Voltage, N)
        end if

C       Construct voltages

        do i1 = 1,N
           ThisCharge = ZERO

c          Calculate due normalization
           do i2= Mol%Overlap%RowStart(i1), Mol%Overlap%RowStart(i1+1)-
1
              VoltageDifference =
     &             ( Mol%Atoms(i1)%Element%Electronegativity
     &             - Mol%Atoms(Mol%Overlap%ColIdx(i2))
     &                  %Element%Electronegativity )
              if (VoltageDifference.ne.ZERO) then
                 ThisCharge = ThisCharge - VoltageDifference
     &                * Mol%Overlap%Value(i2)
              end if
           end do
           Mol%Voltage(i1) = ThisCharge * Mol%OvNorm(i1)
        end do

c       Print voltages
c        print *, "Voltages = "
c        do i1=1,N
c           print *, Mol%Voltage(i1)/eV
c        end do

        call SolveConstrained(N, Mol%Coulomb, Mol%Voltage,
     &       Mol%Atoms(1:N)%Charge,
     & Mol%ChemicalPotential, Mol%SchurCoulomb)

c       Calculate energy
c       This simplified formula is derived in the notes dated 2008-05-
04
       ThisCharge = 0.0d0
        do i1=1,N
           ThisCharge = ThisCharge
     &          + Mol%Atoms(i1)%Charge * Mol%Voltage(i1)
        end do
        Mol%Energy = -0.5d0 * ThisCharge

      end subroutine QTPIE

      subroutine SolveConstrained(N, A, b, x, mu, schurA)
```

```
        use Parameters
        implicit none
        integer, intent(in) :: N
        double precision, dimension(N, N), intent(in) :: A
        double precision, dimension(N), intent(in) :: b
        double precision, dimension(N), intent(inout) :: x
        double precision, intent(out) :: mu, schurA

        integer :: i
        double precision, dimension(:), allocatable :: Ones,
Constraints

        integer :: PrevSize
        save PrevSize, Ones, Constraints

        external solver

c       Check if memory needs to be allocated
        if (N .ne. PrevSize) then
           PrevSize = N

           call NewVector(Ones, N)
           Ones = 1.0d0

           call NewVector(Constraints, N)
           Constraints = 0.0d0
        end if

c       First solve the unconstrained problem
        call solver(N, A, b, x)

        mu = 0.0d0
        do i = 1,N
           mu = mu + x(i)
        end do

c       Now solve for contribution of constraints
        call solver(N, A, Ones, Constraints)

        schurA = 0.0d0
        do i = 1,N
           schurA = schurA + Constraints(i)
        end do


        mu = mu / schurA

c       Add in contribution of constraints
        do i = 1,N
           x(i) = x(i) - mu * Constraints(i)
        end do

      end subroutine SolveConstrained

!>
```

```
!! Computes energy gradients numerically
!!
!! Calculates energy gradients using the method of finite differences
!! using forward gradients
!! As you can imagine, this is pretty slow
!! You should not use this routine!
!! \param Mol : of the Molecule data type
!! Mol%EGradient is calculated
!<
      subroutine DoGradientsByFiniteDifference(Mol)
        use Parameters
        implicit none
        type(Molecule) :: Mol
        integer :: i1, i2, N
        double precision :: OriginalEnergy
        double precision, parameter :: Eps = 1.0d-4

C       Check if memory for gradient matrix has been allocated
        if (N .ne. Mol%NumAtoms) then
           N = Mol%NumAtoms
           call NewMatrix(Mol%EGradient, N, 3)
        end if

C       Save current energy
        OriginalEnergy = Mol%Energy
C       Calculate energy gradients
        do i1=1,N
          do i2=1,3
C           Perturb Geometry
            Mol%Atoms(i1)%Position(i2) =
     &      Mol%Atoms(i1)%Position(i2) + Eps
            Mol%Atoms(i1)%Basis%Position(i2) =
     &      Mol%Atoms(i1)%Basis%Position(i2) + Eps
C           Redo QTPIE
            call DosGTOIntegrals(Mol)
            call QTPIE(Mol)
C           Calculate gradient
            Mol%EGradient(i1, i2) =
     &           (Mol%Energy - OriginalEnergy)  / ( Eps)
C           Perturb Geometry
            Mol%Atoms(i1)%Position(i2) =
     &      Mol%Atoms(i1)%Position(i2) - Eps
            Mol%Atoms(i1)%Basis%Position(i2) =
     &      Mol%Atoms(i1)%Basis%Position(i2) - Eps
          end do
        end do
C       Redo integrals
        call DosGTOIntegrals(Mol)
      end subroutine DoGradientsByFiniteDifference

!>
!! Computes energy gradients analytically
!!
!! Calculates energy gradients using analytic derivatives
!! \param Mol : of the Molecule data type
```

```
!! Mol%EGradient is calculated
!<
      subroutine DoGradientsAnalytically(Mol)
        use Parameters
        implicit none
        type(Molecule) :: Mol
        double precision :: a,b,R, Force
        double precision, external :: Distance
        double precision, external :: sGTOOvIntGrad, sGTOCoulIntGrad
        integer :: i1, i2, i3, i4, N
        double precision, dimension(3) :: Pos1, Pos2

C       Check if memory for gradient matrix has been allocated
        if (N .ne. Mol%NumAtoms) then
           N = Mol%NumAtoms
           call NewMatrix(Mol%EGradient, N, 3)
        end if

c       Initialize gradients
        Mol%EGradient=0.0d0

        do i1=1,N
           a = Mol%Atoms(i1)%Basis%zeta
           Pos1 = Mol%Atoms(i1)%Basis%Position

c           Term1 = Mol%Atoms(i1)%Charge * Mol%OvNorm(i1)

           do i2 = Mol%Overlap%RowStart(i1),
     &             Mol%Overlap%RowStart(i1+1) - 1
              i3 = Mol%Overlap%ColIdx(i2)
              if (i1 .ne. i3) then
                 b = Mol%Atoms(i3)%Basis%zeta
                 Pos2 = Mol%Atoms(i3)%Basis%Position

                 R = Distance(Pos1, Pos2)
                 Force = Mol%Atoms(i3)%Charge * Mol%OvNorm(i3)
     &                *(Mol%Atoms(i3)%Element%Electronegativity
c                 Force = Term1
*(Mol%Atoms(i3)%Element%Electronegativity
     &                - Mol%Voltage(i3)
     &                - Mol%Atoms(i1)%Element%Electronegativity)
     &                * sGTOOvIntGrad(a,b,R)

                 Force = Force / R

c     Calculates projection onto direction vector
c     $Temp*\frac{\partial R_{i1,i2}}{\partial R_{k,i3}}
c     * (\delta_{i1,k} - \delta_{i2,k})$

                 do i4=1,3
                    Mol%EGradient(i1,i4)=Mol%EGradient(i1,i4) +
     &                   (Pos1(i4) - Pos2(i4))*Force
                 end do
              end if
           end do
```

```
c     Add in contribution of Coulomb term
           do i3=1,N
              if (i1 .ne. i3) then
                 b = Mol%Atoms(i3)%Basis%zeta
                 Pos2 = Mol%Atoms(i3)%Basis%Position

                 R = Distance(Pos1, Pos2)
                 Force = Force + Mol%Atoms(i1)%Charge
     &                * Mol%Atoms(i3)%Charge * sGTOCoulIntGrad(a,b,R)
                 Force = Force / R

c     Calculates projection onto direction vector
c     $Temp*\frac{\partial R_{i1,i2}}{\partial R_{k,i3}}
c     * (\delta_{i1,k} - \delta_{i2,k})$

                 do i4=1,3
                    Mol%EGradient(i1,i4)=Mol%EGradient(i1,i4) +
     &                   (Pos1(i4) - Pos2(i4))*Force
                 end do
              end if
           end do
        end do
!        do i1=1,N
c          Obtain basis set parameters for atom i1
!
c          Calculate contribution to gradient from voltage term
!           do i2=1,N
c          Diagonal part has no contribution to gradient
!           if (i1.ne.i2) then
c             Obtain basis set parameters for atom i2
!              b = Mol%Atoms(i2)%Basis%zeta
!
c             Calculate pairwise distance
!              R = Distance(Mol%Atoms(i1)%Basis%Position,
!     &                     Mol%Atoms(i2)%Basis%Position)
!
!              Force = 2 * Mol%Atoms(i1)%Charge
!     &          * ( Mol%Atoms(i1)%Element%Electronegativity
!     &            - Mol%Atoms(i2)%Element%Electronegativity )
!     &          * sGTOOvIntGrad(a,b,R) / Mol%OvNorm(i1)
!              Force = Force - Mol%Voltage(i1)/Mol%OvNorm(i1)
!     &          * sGTOOvIntGrad(a,b,R)
!              Force = Force - Mol%Atoms(i1)%Charge / Mol%NumAtoms
!     &          * ( Mol%Atoms(i1)%Element%Electronegativity
!     &            - Mol%Atoms(i2)%Element%Electronegativity )
!     &          * Mol%Overlap(i1,i2) / (Mol%OvNorm(i1) ** 2)
!     &          * sGTOOvIntGrad(a,b,R)
c     &          * (1/Mol%OvNorm(i1) - 1/Mol%OvNorm(i2))
!             Force = Force + Mol%Atoms(i1)%Charge *
Mol%Atoms(i2)%Charge
!     &                 * sGTOCoulIntGrad(a,b,R)
c             Calculates projection onto direction vector
c             $Temp*\frac{\partial R_{i1,i2}}{\partial R_{k,i3}}
c             * (\delta_{i1,k} - \delta_{i2,k})$
```

```
!               Force = Force / R
!               do i3=1,3
!                  Mol%EGradient(i1,i3)=Mol%EGradient(i1,i3) +
!     &                 (Mol%Atoms(i1)%Basis%Position(i3)-
!     &                  Mol%Atoms(i2)%Basis%Position(i3))*Force
!               end do
!            end if
!            end do
!         end do
      end subroutine DoGradientsAnalytically
```

## A.15. sparse.f

```
module SparseMatrix
      implicit none
!>
!! Data type for compressed sparse row matrix format
!! \note NO range bounds checking
!<
        type CSRMatrix
          integer :: RowDim, ColDim
          integer, dimension(:), allocatable :: RowStart
          integer, dimension(:), allocatable :: ColIdx
          double precision, dimension(:), allocatable :: Value
        end type CSRMatrix

      contains

!>
!!    Initializes a CSR matrix
!<
      subroutine CSRNew(A, RowDim, ColDim, MaxNumVals)
      implicit none
      integer, intent(in) :: RowDim, ColDim, MaxNumVals
      type(CSRMatrix), intent(inout) :: A

      integer :: stat

      call CSRDelete(A)

      A%RowDim = RowDim
      A%ColDim = ColDim

      allocate(A%RowStart(RowDim+1), A%ColIdx(MaxNumVals),
     &         A%Value(MaxNumVals), STAT=stat)

      if (stat .ne. 0) then
         print *, "CSRNew: Error allocating new sparse matrix"
         print *, "Error code =",stat
         stop
      end if

      A%RowStart = 0
```

```
      end subroutine CSRNew

!>
!!    Deletes a CSR matrix
!<
      subroutine CSRDelete(A)
      implicit none
      type(CSRMatrix), intent(inout) :: A
      integer :: stat

      if (allocated(A%Value)) then
         deallocate(A%RowStart, A%ColIdx, A%Value, STAT=stat)

         if (stat .ne. 0) then
            print *, "CSRDelete: Error deallocating sparse matrix"
            print *, "Error code =",stat
            stop
         end if
      end if

      end subroutine CSRDelete

!>
!!    Prints a CSR matrix
!<
      subroutine CSRPrint(A, rowidx)
      implicit none
      type(CSRMatrix), intent(inout) :: A
      integer, optional :: rowidx

      integer :: i, low, upp

      if (present(rowidx)) then
         low = rowidx
         upp = rowidx
      else
         low = 1
         upp = A%RowDim
      end if
      do i = low, upp
         print *, "Row", i, "runs from",A%RowStart(i),
     &        "to",A%RowStart(i+1)-1
         print *, " "
         print *, "Column indices"
         print *, "--------------"
         print *, A%ColIdx(A%RowStart(i):A%RowStart(i+1)-1)
         print *, "Matrix elements"
         print *, "---------------"
         print *, A%Value(A%RowStart(i):A%RowStart(i+1)-1)
      end do

      end subroutine CSRPrint
!>
!! Does sparse matrix-vector multiplies
!<
```

```
      subroutine MatrixVectorMultiply(N, A, x, y)
      implicit none
      integer, intent(in) :: N
      type(CSRMatrix), intent(in) :: A
      double precision, dimension(N), intent(in) :: x
      double precision, dimension(N), intent(out) :: y

      integer :: i, j
      double precision :: ThisElement

      do i = 1, N
         ThisElement = 0.0d0
         do j = A%RowStart(i), A%RowStart(i+1) - 1
            ThisElement = ThisElement + A%Value(j) * x(A%ColIdx(j))
         end do
         y(i) = ThisElement
      end do

      end subroutine MatrixVectorMultiply

!>
!! Sums over the entire row of a CSR.
!<
      double precision function SumRow(A, RowIdx)
      implicit none
      type(CSRMatrix), intent(in) :: A
      integer, intent(in) :: RowIdx !<Index to find sum of

      integer :: i

      SumRow = 0.0d0
      do i = A%RowStart(RowIdx), A%RowStart(RowIdx+1) - 1
         SumRow = SumRow + A%Value(i)
      end do

      end function SumRow

      end module SparseMatrix
```

### A.16. sto-int.f

```
!>
!! Calculates a Slater-type orbital exponent
!! based on the hardness parameters
!! \param Hardness: chemical hardness in atomic units
!! \param        n: principal quantum number
!! \note See research notes dated 2007-08-30
!<
      subroutine AssignsSTOBasis(theAtom)
        use Parameters
        implicit none
        type(Atom) :: theAtom
        double precision, external :: sSTOCoulInt
        integer :: n
```

```
C       Approximate the exact value of the constant of proportionality
C       by its value at a very small distance epsilon
C       since the exact R = 0 case has not be programmed
        double precision :: epsilon = 1.0d-8

C       Assign position
        theAtom%Basis%Position = theAtom%Position

C       Assign principal quantum number
        n = pqn(theAtom%Element)
        theAtom%Basis%n = n

C       Assign orbital exponent
        theAtom%Basis%zeta = (sSTOCoulInt(1.0d0, 1.0d0, n, n, epsilon)
     &      /theAtom%Element%Hardness)**(-1.0d0/(3.0d0 + 2.0d0*n))
      end subroutine AssignsSTOBasis

!>
!! Computes Rosen's Guillimer-Zener function A
!!
!! Computes Rosen's A integral, an auxiliary quantity needed to
!! compute integrals involving Slater-type orbitals of s symmetry.
!! \f[
!! A_n(\alpha) = \int_1^\infty x^n e^{-\alpha x}dx
!! = \frac{n! e^{-\alpha}}{\alpha^{n+1}}\sum_{\nu=0}^n
!! \frac{\alpha^\nu}{\nu!}
!! \f]
!! \param n - principal quantum number
!! \param alpha - Slater exponent
!! \return the value of Rosen's A integral
!! \note N. Rosen, Phys. Rev., 38 (1931), 255
!<
      double precision function RosenA(n,a)
        implicit none
        integer, intent(in) :: n
        double precision, intent(in) :: a
        double precision :: Term
        integer :: nu
        if (a.ne.0.0d0) then
          Term = 1.0d0
          RosenA = Term
          do nu = 1,n
            Term = a/nu*Term
            RosenA = RosenA + Term
          end do
          RosenA=RosenA/Term*exp(-a)/a
        else
          RosenA=0.0d0
        end if
      end function RosenA

!>
!! Computes Rosen's Guillimer-Zener function B
!!
```

```
!! Computes Rosen's B integral, an auxiliary quantity needed to
!! compute integrals involving Slater-type orbitals of s symmetry.
!! \f[
!! B_n(\alpha) = \int_{-1}^1 x^n e^{-\alpha x} dx
!!             = \frac{n!}{\alpha^{n+1}}
!!               \sum_{\nu=0}^n \frac{e^\alpha(-\alpha)^\nu
!!                 - e^{-\alpha} \alpha^\nu}{\nu!}
!! \f]
!! \param n - principal quantum number
!! \param alpha - Slater exponent
!! \return the value of Rosen's B integral
!! \note N. Rosen, Phys. Rev., 38 (1931), 255
!<
      double precision  function RosenB(n,alpha)
        implicit none
        integer, intent(in) :: n
        double precision , intent(in) :: alpha
        double precision :: TheSum, Term
        double precision :: PSinhRosenA, PCoshRosenA, PHyperRosenA
        integer :: nu
        logical :: IsPositive
        if (alpha.ne.0.0d0) then
          Term = 1.0d0
          TheSum = 1.0d0
          IsPositive = .True.
C         These two expressions are (up to constant factors) equivalent
C         to computing the hyperbolic sine and cosine of a respectively
C         The series consists of adding up these terms in an
C         alternating fashion
          PSinhRosenA =  exp(alpha) - exp(-alpha)
          PCoshRosenA = -exp(alpha) - exp(-alpha)
          TheSum=PSinhRosenA
          do nu = 1,n
            if (isPositive) then
              PHyperRosenA = PCoshRosenA
              isPositive = .False.
            else !term to add should be negative
              PHyperRosenA = PSinhRosenA
              isPositive = .True.
            end if
            Term=alpha/(1.0d0*nu)*Term
            TheSum=TheSum+Term*PHyperRosenA
          end do
          RosenB=TheSum/(alpha*Term)
        else
C         pathological case of a=0
          print *, "WARNING, a = 0 in RosenB"
          RosenB=(1.0d0-(-1.0d0)**n)/(n+1.0d0)
        end if
      end function RosenB

!>
!! Computes Rosen's D combinatorial factor
!!
!! Computes Rosen's D factor, an auxiliary quantity needed to
```

```
!! compute integrals involving Slater-type orbitals of s symmetry.
!! \f[
!! RosenD^{mn}_p = \sum_k (-1)^k \frac{m! n!}
!!                {(p-k)!(m-p+k)!(n-k)!k!}
!! \f]
!! \return the value of Rosen's D factor
!! \note N. Rosen, Phys. Rev., 38 (1931), 255
!<
      integer function RosenD(m,n,p)
        use Factorial
        implicit none
        integer, intent(in) :: m,n,p
        integer k
        RosenD = 0

        if (m+n+p.gt.maxFact) then
          print *, "Error, arguments exceed maximum factorial computed"
     &      , m+n+p,">",maxFact
          stop
        end if
        do k=max(p-m,0),min(n,p)
           if (mod(k,2).eq.0) then
           RosenD = RosenD + fact(m) / ( fact(p-k) *
     &       fact(m-p+k)) * fact(n) / (fact(n-k) * fact(k))
           else
           RosenD = RosenD - fact(m) / ( fact(p-k) *
     &       fact(m-p+k)) * fact(n) / (fact(n-k) * fact(k))
           end if
        end do
      end function RosenD

!>
!! Computes Coulomb integral analytically over s-type STOs
!!
!! Computes the two-center Coulomb integral over Slater-type
!! orbitals of s symmetry.
!! \param a: Slater zeta exponent of first atom in inverse Bohr (au)
!! \param b: Slater zeta exponent of second atom in inverse Bohr (au)
!! \param m: principal quantum number of first atom
!! \param n: principal quantum number of second atom
!! \param R: internuclear distance in atomic units (bohr)
!! \return value of the Coulomb potential energy integral
!! \note N. Rosen, Phys. Rev., 38 (1931), 255
!! \note In Rosen's paper, this integral is known as K2.
!<
      double precision  function sSTOCoulInt(a, b, m, n, R)
        use Factorial
        implicit none
        integer, intent(in) :: m,n
        double precision , intent(in) :: a,b,R
        double precision , external :: RosenA, RosenB
        integer, external :: RosenD
        integer :: nu, p
        double precision :: x, K2
        double precision :: Factor1, Factor2, Term, OneElectronTerm
```

```
        double precision :: eps, epsi

C       To speed up calculation, we terminate loop once contributions
C       to integral fall below the bound, epsilon
        double precision, parameter :: epsilon = 0.0d0

C       x is the argument of the auxiliary RosenA and RosenB functions
        x=2.0*a*R
C       First compute the two-electron component
        sSTOCoulInt = 0.0d0
        if (x.eq.0) then
C         Pathological case
          if ((a.eq.b).and.(m.eq.n)) then
            do nu = 0,2*n-1
              K2 = 0.0d0
              do p = 0, 2*n+m
                K2 = K2 + 1.0d0 / fact(p)
              end do
              sSTOCoulInt = sSTOCoulInt + fact(2*n+m)/fact(m)*K2
            end do
            sSTOCoulInt = 2*a/(n*fact(2*n))*sSTOCoulInt
          else
C          Not implemented
            print *, "ERROR, sSTOCoulInt cannot compute from arguments"
            print *, "a = ",a,"b = ",b,"m =",m,"n = ",n,"R = ",R
            stop
          end if
        else
          OneElectronTerm = 1.0d0/R + x**(2*m)/(fact(2*m)*R)*
     &             ((x-2*m)*RosenA(2*m-1,x)-exp(-x)) + sSTOCoulInt
          eps = epsilon / OneElectronTerm
          if (a.eq.b) then
C           Apply Rosen (48)
            Factor1 = -a*(a*R)**(2*m)/(n*fact(2*m))
            do nu=0,2*n-1
              Factor2 = (2.0d0*n-nu)/fact(nu)*(a*R)**nu
              epsi = eps / abs(Factor1 * Factor2)
              K2=0.0d0
              do p=0,m+(nu-1)/2
                Term = RosenD(2*m-1,nu,2*p)/(2.0d0*p+1.0d0)
     &              *RosenA(2*m+nu-1-2*p,x)
                K2=K2 + Term
                if ((Term.gt.0).and.(Term.lt.epsi)) then
                  goto 1
                end if
              end do
              sSTOCoulInt=sSTOCoulInt+K2*Factor2
            end do
 1          sSTOCoulInt=sSTOCoulInt*Factor1
          else
            Factor1 = -a*(a*R)**(2*m)/(2.0d0*n*fact(2*m))
            epsi = eps/abs(Factor1)
            if (b.eq.0.0d0) then
              print *, "WARNING: b = 0 in sSTOCoulInt"
            else
```

```
C           Apply Rosen (54)
            do nu=0,2*n-1
              K2=0
              do p=0,2*m+nu-1
                K2=K2+RosenD(2*m-1,nu,p)*RosenB(p,R*(a-b))
     &                  *RosenA(2*m+nu-1-p,R*(a+b))
              end do
              Term = K2*(2*n-nu)/fact(nu)*(b*R)**nu
              sSTOCoulInt=sSTOCoulInt+Term
              if (abs(Term) .lt. epsi) then
                goto 2
              end if
            end do
 2          sSTOCoulInt=sSTOCoulInt*Factor1
            end if
          end if
C         Now add the one-electron term from Rosen (47) = Rosen (53)
          sSTOCoulInt=sSTOCoulInt + OneElectronTerm
        end if
      end function sSTOCoulInt

!>
!! Computes overlap integral analytically over s-type STOs
!!
!!       Computes the overlap integral over two
!!      Slater-type orbitals of s symmetry.
!! \param a: Slater zeta exponent of first atom in inverse Bohr (au)
!! \param b: Slater zeta exponent of second atom in inverse Bohr (au)
!! \param m: principal quantum number of first atom
!! \param n: principal quantum number of second atom
!! \param R: internuclear distance in atomic units (bohr)
!! \return the value of the sSTOOvInt integral
!! \note N. Rosen, Phys. Rev., 38 (1931), 255
!! \note In the Rosen paper, this integral is known as I.
!<
      double precision  function sSTOOvInt(a,b,m,n,R)
        use Factorial
        implicit none
        integer, intent(in) :: m,n
        double precision , intent(in) :: a,b,R
        double precision , external :: RosenA, RosenB
        integer, external :: RosenD
        integer :: q

        double precision :: Factor, Term, eps

C       To speed up calculation, we terminate loop once contributions
C       to integral fall below the bound, epsilon
        double precision, parameter :: epsilon = 0.0d0

        sSTOOvInt=0.0d0

        if (a.eq.b) then
          Factor = (a*R)**(m+n+1)/sqrt(fact(2*m)*fact(2*n))
          eps = epsilon / abs(Factor)
```

```
          do q=0,(m+n)/2
            Term = RosenD(m,n,2*q)/(2.0d0*q+1.0d0)*RosenA(m+n-2*q,a*R)
            sSTOOvInt=sSTOOvInt+Term
            if (abs(Term).lt.eps) then
              exit
            end if
          end do
          sSTOOvInt=sSTOOvInt*Factor
        else
          Factor = 0.5d0*(a*R)**(m+0.5d0)*(b*R)**(n+0.5d0)
     &         /sqrt(fact(2*m)*fact(2*n))
          eps = epsilon / abs(Factor)
          do q=0,m+n
            Term = RosenD(m,n,q)*RosenB(q,R/2.0d0*(a-b))
     &            *RosenA(m+n-q,R/2.0d0*(a+b))
            sSTOOvInt=sSTOOvInt+Term
            if (abs(Term) .lt. eps) then
              exit
            end if
          end do
          sSTOOvInt=sSTOOvInt*Factor
        end if
      end function sSTOOvInt

!>
!! Computes kinetic energy integral analytically over s-type STOs
!!
!! Computes the overlap integral over two Slater-type orbitals of s
symmetry.
!! \param a: Slater zeta exponent of first atom in inverse Bohr (au)
!! \param b: Slater zeta exponent of second atom in inverse Bohr (au)
!! \param m: principal quantum number of first atom
!! \param n: principal quantum number of second atom
!! \param R: internuclear distance in atomic units (bohr)
!! \return the value of the kinetic energy integral
!! \note N. Rosen, Phys. Rev., 38 (1931), 255
!! \note untested
!<
      double precision  function KinInt(a,b,m,n,R)
        implicit none
        integer, intent(in) :: m,n
        double precision , intent(in) :: a,b,R
        double precision , external :: sSTOOvInt
        KinInt=-0.5*b*b*sSTOOvInt(a,b,m,n,R)
        if (n.gt.0) then
          KinInt=KinInt+b*b*(2*b/(2*b-1))**0.5*sSTOOvInt(a,b,m,n-1,R)
          if (n.gt.1) then
            KinInt=KinInt+(n*(n-1)/((n-0.5)*(n-1.5)))**0.5
     &                      *sSTOOvInt(a,b,m,n-2,R)
          end if
        end if
      end function

!>
```

```
!! Computes derivative of Coulomb integral with respect to the
interatomic distance
!!
!! Computes the two-center Coulomb integral over Slater-type orbitals
of s symmetry.
!! \param a: Slater zeta exponent of first atom in inverse Bohr (au)
!! \param b: Slater zeta exponent of second atom in inverse Bohr (au)
!! \param m: principal quantum number of first atom
!! \param n: principal quantum number of second atom
!! \param R: internuclear distance in atomic units (bohr)
!! \return the derivative of the Coulomb potential energy integral
!! \note Derived in QTPIE research notes, May 15 2007
!<
      double precision  function sSTOCoulIntGrad(a, b, m, n, R)
        use Factorial
        implicit none
        integer, intent(in) :: m,n
        double precision , intent(in) :: a,b,R
        double precision , external :: RosenA, RosenB, sSTOCoulInt
        integer, external :: RosenD
C       loop counters
        integer :: nu, p
C       temporary quantities
        double precision :: x, y, z, K2, TheSum
C       x is the argument of the auxiliary RosenA and RosenB functions
        x=2.0*a*R
C       First compute the two-electron component
        sSTOCoulIntGrad = 0.0d0
        if (x.eq.0) then
C         Pathological case
          print *, "WARNING: argument given to sSTOCoulIntGrad is 0"
          print *, "a = ", a, "R = ", R
        else
          if (a.eq.b) then
            TheSum = 0.0d0
            do nu=0,2*(n-1)
              K2 = 0.0d0
              do p=0,(m+nu)/2
                K2 = K2 + RosenD(2*m-1, nu+1, 2*p)/(2*p + 1.0d0)
     &                 * RosenA(2*m+nu-1-2*p, x)
              end do
              TheSum = TheSum + (2*n-nu-1)/fact(nu)*(a*R)**(nu) * K2
            end do
            sSTOCoulIntGrad = -a**(2*m+2)*R**(2*m)
     &                     /(n*fact(2*m))*TheSum
            TheSum = 0.0d0
            do nu=0,2*n-1
              K2 = 0.0d0
              do p=0,(m+nu-1)/2
                K2 = K2 + RosenD(2*m-1, nu, 2*p)/(2*p + 1.0d0)
     &                 * RosenA(2*m+nu-2*p, x)
              end do
              TheSum = TheSum + (2*n-nu)/fact(nu)*(a*R)**nu * K2
            end do
            sSTOCoulIntGrad = sSTOCoulIntGrad + 2*a**(2*m+2)*R**(2*m)
```

```
     &                                /(n*fact(2*m))*TheSum
          else
C           Slater exponents are different
C           First calculate some useful arguments
            y = R*(a+b)
            z = R*(a-b)
            TheSum = 0.0d0
            do nu=0,2*n-1
              K2 = 0.0d0
              do p=0,2*m+nu
                K2 = K2 + RosenD(2*m-1, nu+1, p)
     &                  * RosenB(p,z)*RosenA(2*m+nu-p, y)
              end do
              TheSum = TheSum + (2*n-nu-1)/fact(nu)*(b*R)**nu * K2
            end do
            sSTOCoulIntGrad = -b*a**(2*m+1)*R**(2*m)/
     &                       (2*n*fact(2*m))*TheSum
            TheSum = 0.0d0
            do nu=0,2*n
              K2 = 0.0d0
              do p=0,2*m-1+nu
                K2 = K2 + RosenD(2*m-1, nu, p)
     &             * ((a-b)*RosenB(p+1,z)*RosenA(2*m+nu-p-1, y)
     &               +(a+b)*RosenB(p  ,z)*RosenA(2*m+nu-p  , y))
              end do
              TheSum = TheSum + (2*n-nu)/fact(nu)*(b*R)**nu * K2
            end do
            sSTOCoulIntGrad = sSTOCoulIntGrad
     &           + a**(2*m+1)*R**(2*m)/(2*n*fact(2*m))*TheSum
          end if
C         Now add one-electron terms and common term
          sSTOCoulIntGrad = sSTOCoulIntGrad - (2.0d0*m+1.0d0)/R**2
     &                  + 2.0d0*m/R * sSTOCoulInt(a,b,m,n,R)
     &         +x**(2*m)/(fact(2*m)*R**2) * ((2.0d0*m+1.0d0)*exp(-x)
     &          +2.0d0*m*(1.0d0+2.0d0*m-x)*RosenA(2*m-1,x))
        end if
      end function sSTOCoulIntGrad

!> Computes gradient of overlap integral with respect to the
interatomic diatance
!!
!! Computes the derivative of the overlap integral over two Slater-type
orbitals of s symmetry.
!! \param a: Slater zeta exponent of first atom in inverse Bohr (au)
!! \param b: Slater zeta exponent of second atom in inverse Bohr (au)
!! \param m: principal quantum number of first atom
!! \param n: principal quantum number of second atom
!! \param R: internuclear distance in atomic units (bohr)
!! \return the derivative of the sSTOOvInt integral
!! \note Derived in QTPIE research notes, May 15 2007
!<
      double precision  function sSTOOvIntGrad(a,b,m,n,R)
        use Factorial
        implicit none
        integer, intent(in) :: m,n
```

```
        double precision, intent(in) :: a,b,R
        double precision, external :: RosenA, RosenB
        integer, external :: RosenD
        double precision, external :: sSTOOvInt
C       Useful temporary quantities
        double precision :: w, x, y, z, TheSum
C       Loop variable
        integer :: q

C       Calculate first term
        sSTOOvIntGrad=(m+n+1.0d0)/R * sSTOOvInt(a,b,m,n,R)
C       Calculate remaining terms; answers depend on exponents
        TheSum = 0.0d0
        x = a * R
        if (a.eq.b) then
          do q = 0,(m+n)/2
            TheSum = TheSum + RosenD(m,n,2*q) / (2*q + 1.0d0)
     &                * RosenA(m+n-2*q+1, x)
          end do
          sSTOOvIntGrad = sSTOOvIntGrad - a*x**(m+n+1)/
     &                    sqrt(fact(2*m)*fact(2*n))*TheSum
        else
C         Useful arguments
          w = b*R
          y = 0.5d0*R*(a+b)
          z = 0.5d0*R*(a-b)
          do q = 0,m+n
            TheSum = TheSum + RosenD(m,n,q) *
     &           ((a-b)*RosenB(q+1,z)*RosenA(m+n-q  ,y)
     &           +(a+b)*RosenB(q  ,z)*RosenA(m+n-q+1,y))
          end do
          sSTOOvIntGrad = sSTOOvIntGrad - 0.25d0*sqrt((x**(2*m+1)
     &      *w**(2*n+1))/(fact(2*m)*fact(2*n)))*TheSum

        end if
      end function sSTOOvIntGrad
```

## A.17. test.f

```
!>
!! A simple program to test the functions implemented with some test
values
!<
      program test
        use Parameters
        use factorial
        implicit none
        double precision, external :: RosenA, RosenB
        integer, external :: RosenD
        double precision, external :: sSTOCoulInt, sSTOOvInt
        double precision, external :: sSTOCoulIntGrad, sSTOOvIntGrad
        double precision, external :: sGTOCoulInt, sGTOOvInt
        double precision, external :: sGTOCoulIntGrad, sGTOOvIntGrad
        type(Molecule) :: Mol1, Mol2
```

```
      type(Molecule), external :: loadXYZ
      integer :: i
      double precision, parameter :: epsilon = 1.0d-6

      print *, "Testing mode"

      if (fact(6).eq.720) then
        print *, "Factorial correct"
      else
        print *, "FATAL ERROR: Factorials incorrectly computed"
        print *, "6! = ", fact(6), ", expected 720"
        print *, "There is an error in factorial.f"
        stop
      end if

      if (fact(10).eq.3628800) then
        print *, "Factorial correct"
      else
        print *, "FATAL ERROR: Factorials incorrectly computed"
        print *, "10! = ", fact(10), ", expected 3628800"
        print *, "There is an error in factorial.f"
        stop
      end if

      if (abs(RosenA(4,3.0d0)-8.05198d-2).lt.epsilon) then
        print *, "Rosen A integral correct"
      else
        print *, "FATAL ERROR: Rosen A integral incorrectly computed"
        print *, "RosenA(4,3.0) = ", RosenA(4,3.0d0),
     &         "expected 0.0805198"
        print *, "There is an error in sto-int.f"
        stop
      end if

      if (abs(RosenA(12,10.0d0)-3.79157d-5).lt.epsilon) then
        print *, "Rosen A integral correct"
      else
        print *, "FATAL ERROR: Rosen A integral incorrectly computed"
        print *, "RosenA(12,10.0) = ", RosenA(12,10.0d0),
     &           ", expected", 3.79157d-5
        print *, "There is an error in sto-int.f"
        stop
      end if

      if (abs(RosenB(4,3.0d0)-2.6471457).lt.epsilon) then
        print *, "Rosen B integral correct"
      else
        print *, "FATAL ERROR: Rosen B integral incorrectly computed"
        print *, "RosenB(4,3.0) = ", RosenB(4,3.0d0), ", expected",
     &      2.6471457
        print *, "There is an error in sto-int.f"
        stop
      end if

      if (abs(RosenB(8,3.0d0)-1.715602).lt.epsilon) then
```

```
        print *, "Rosen B integral correct"
      else
        print *, "FATAL ERROR: Rosen B integral incorrectly computed"
        print *, "RosenB(8,3.0) = ", RosenB(8,3.0d0), ", expected",
     &      3.75628-2.04068
        print *, "There is an error in sto-int.f"
        stop
      end if

      if (abs(RosenB(12,10.0d0)-9.759958896510301d2).lt.epsilon) then
        print *, "Rosen B integral correct"
      else
        print *, "FATAL ERROR: Rosen B integral incorrectly computed"
        print *, "RosenB(12,10.0) = ", RosenB(12,10.0d0),
     &         ", expected", 9.75996d2-3.79157d-5
        print *, "There is an error in sto-int.f"
        stop
      end if

      if (RosenD(1,2,3).eq.1) then
        print *, "Rosen D factor correct"
      else
        print *, "FATAL ERROR: Rosen D factor incorrectly computed"
        print *, "RosenD(1,2,3) = ", RosenD(1,2,3), ", expected 1"
        print *, "There is an error in sto-int.f"
        stop
      end if

      if (RosenD(5,4,3).eq.-4) then
        print *, "Rosen D factor correct"
      else
        print *, "FATAL ERROR: Rosen D factor incorrectly computed"
        print *, "RosenD(5,4,3) = ", RosenD(5,4,3), ", expected -4"
        print *, "There is an error in sto-int.f"
        stop
      end if

      if (RosenD(5,3,8).eq.-1) then
        print *, "Rosen D factor correct"
      else
        print *, "FATAL ERROR: Rosen D factor incorrectly computed"
        print *, "RosenD(5,3,8) = ", RosenD(5,3,8), ", expected -1"
        print *, "There is an error in sto-int.f"
        stop
      end if

      if (abs(sSTOCoulInt(1.0d0,2.0d0,3,4,5.0d0)-0.1903871)
     &       .lt.epsilon) then
        print *, "Coulomb integral correct"
      else
        print *, "FATAL ERROR: Coulomb integral incorrectly computed"
        print *, "Coulomb(1,2,3,4,5) = ",
     &         sSTOCoulInt(1.0d0,2.0d0,3,4,5.0d0),
     &      ", expected", 0.1903871
        print *, "There is an error in sto-int.f"
```

```
          stop
        end if

        if (abs(sSTOCoulInt(1.0d0,1.0d0,2,3,5.0d0)-0.1879457)
     &       .lt.epsilon) then
          print *, "Coulomb integral correct"
        else
          print *, "FATAL ERROR: Coulomb integral incorrectly computed"
          print *, "Coulomb(1,1,2,3,5) = ",
     &         sSTOCoulInt(1.0d0,1.0d0,2,3,5.0d0),
     &         ", expected", 0.1879457
          print *, "There is an error in sto-int.f"
          stop
        end if

        if (abs(sSTOCoulInt(5.0d0,4.0d0,3,2,1.0d0)-0.9135013)
     &       .lt.epsilon) then
          print *, "Coulomb integral correct"
        else
          print *, "FATAL ERROR: Coulomb integral incorrectly computed"
          print *, "Coulomb(5,4,3,2,1) = ",
     &         sSTOCoulInt(5.0d0,4.0d0,3,2,1.0d0),
     &         ", expected", 0.9135013
          print *, "There is an error in sto-int.f"
          stop
        end if

        if (abs(sSTOOvInt(1.0d0,2.0d0,3,4,5.0d0)-0.3145446)
     &       .lt.epsilon) then
          print *, "Overlap integral correct"
        else
          print *, "FATAL ERROR: Overlap integral incorrectly computed"
          print *, "Overlap(1,2,3,4,5) = ",
     &         sSTOOvInt(1.0d0,2.0d0,3,4,5.0d0),
     &      ", expected", 0.3145446
          print *, "There is an error in sto-int.f"
          stop
        end if

        if (abs(sSTOOvInt(1.0d0,1.0d0,2,3,5.0d0)-0.3991235)
     &       .lt.epsilon) then
          print *, "Overlap integral correct"
        else
          print *, "FATAL ERROR: Overlap integral incorrectly computed"
          print *, "Overlap(1,1,2,3,5) = ",
     &         sSTOOvInt(1.0d0,1.0d0,2,3,5.0d0),
     &      ", expected", 0.3991235
          print *, "There is an error in sto-int.f"
          stop
        end if

        Mol1 =  loadXYZ("../test/nacl.xyz")
        print *, "Load XYZ successful"

c        call DosSTOIntegrals(Mol1)
```

```
        call DosGTOIntegrals(Mol1)
        call QEq(Mol1)

        if (abs(Mol1%Atoms(1)%Charge - 1.3895802392931).lt.epsilon)
then
          print *, "QEq Charge calculation for sodium chloride correct"
        else
          print *, "FATAL ERROR: QEq charges incorrectly computed"
          print *, "Charge(1) = ", Mol1%Atoms(1)%Charge,
     &         " expected", 1.3895802392931755
          print *, "There is an error in qtpie.f"
c          stop
        end if

        call QTPIE(Mol1)
        if (abs(Mol1%Atoms(1)%Charge - 0.7252290067905).lt.epsilon)
then
          print *, "QTPIE Charge calculation for ",
     &          "sodium chloride correct"
        else
          print *, "FATAL ERROR: QTPIE charges incorrectly computed"
          print *, "Charge(1) = ", Mol1%Atoms(1)%Charge,
     &         " expected", 0.72522900679059155
          print *, "There is an error in qtpie.f"
c          stop
        end if

        Mol2 =  loadXYZ("../test/h2o.xyz")
        print *, "Load XYZ successful"

c        call DosSTOIntegrals(Mol2)
        call DosGTOIntegrals(Mol2)
        call QEq(Mol2)

        if ((abs(Mol2%Atoms(1)%Charge + 0.98965172663).lt.epsilon)
.and.
     &      (abs(Mol2%Atoms(2)%Charge - 0.4943811799).lt. epsilon))
then
          print *, "QEq Charge calculation for water correct"
        else
          print *, "FATAL ERROR: QEq charges incorrectly computed"
          print *, "Charge(1) = ", Mol2%Atoms(1)%Charge,
     &         " expected", -0.98965172663781498
          print *, "Charge(2) = ", Mol2%Atoms(2)%Charge,
     &         " expected",  0.49438117990925540
          print *, "There is an error in qtpie.f"
c          stop
        end if

        call QTPIE(Mol2)

        if ((abs(Mol2%Atoms(1)%Charge + 0.81213640965).lt.epsilon)
.and.
     &      (abs(Mol2%Atoms(2)%Charge - 0.4055667959).lt. epsilon))
then
```

```
            print *, "QTPIE Charge calculation for water correct"
          else
            print *, "FATAL ERROR: QTPIE charges incorrectly computed"
            print *, "Charge(1) = ", Mol2%Atoms(1)%Charge,
     &          " expected", -0.81213640965
            print *, "Charge(2) = ", Mol2%Atoms(2)%Charge,
     &          " expected",  0.40556679590604666
            print *, "There is an error in qtpie.f"
c            stop
          end if

*      Now compare numerical and analytic gradients

       print *, "Testing gradients for Slater orbitals"

       if (abs(sSTOOvIntGrad(1.0d0,1.0d0,3,4,5.0d0) -
     &     (sSTOOvInt(1.0d0,1.0d0,3,4,5.0d0+epsilon)
     &     -sSTOOvInt(1.0d0,1.0d0,3,4,5.0d0-epsilon))
     &     /(2*epsilon)).lt.epsilon) then
          print *, "Overlap gradient correct"
       else
          print *, "FATAL ERROR: Overlap gradients don't match"
          print *, "sSTOOvIntGrad(1.0, 1.0, 3, 4, 5.0)"
          print *, "Analytic = ", sSTOOvIntGrad(1.0d0,1.0d0,3,4,5.0d0)
          print *, "Numerical= ",
     &         (sSTOOvInt(1.0d0,1.0d0,3,4,5.0d0+epsilon)
     &         -sSTOOvInt(1.0d0,1.0d0,3,4,5.0d0-epsilon))/(2*epsilon)
           print *, "There is an error in sto-int.f"
       end if

       if (abs(sSTOOvIntGrad(1.0d0,2.0d0,3,4,5.0d0) -
     &     (sSTOOvInt(1.0d0,2.0d0,3,4,5.0d0+epsilon)
     &     -sSTOOvInt(1.0d0,2.0d0,3,4,5.0d0-epsilon))
     &     /(2*epsilon)).lt.epsilon) then
          print *, "Overlap gradient correct"
       else
          print *, "FATAL ERROR: Overlap gradients don't match"
          print *, "sSTOOvIntGrad(1.0, 2.0, 3, 4, 5.0)"
          print *, "Analytic = ", sSTOOvIntGrad(1.0d0,2.0d0,3,4,5.0d0)
          print *, "Numerical= ",
     &         (sSTOOvInt(1.0d0,2.0d0,3,4,5.0d0+epsilon)
     &         -sSTOOvInt(1.0d0,2.0d0,3,4,5.0d0-epsilon))/(2*epsilon)
           print *, "There is an error in sto-int.f"
       end if

       if (abs(sSTOCoulIntGrad(4.0d0,4.0d0,2,2,5.0d0) -
     &     (sSTOCoulInt(4.0d0,4.0d0,2,2,5.0d0+epsilon)
     &     -sSTOCoulInt(4.0d0,4.0d0,2,2,5.0d0-epsilon))
     &     /(2*epsilon)).lt.epsilon) then
          print *, "Coulomb gradient correct"
       else
          print *, "FATAL ERROR: Coulomb gradients don't match"
          print *, "sSTOCoulIntGrad(4.0, 4.0, 2, 2, 5.0)"
          print *, "Analytic = ", sSTOCoulIntGrad(4.0d0,4.0d0,2,2,5.0d0)
          print *, "Numerical= ",
```

```
&         (sSTOCoulInt(4.0d0,4.0d0,2,2,5.0d0+epsilon)
&         -sSTOCoulInt(4.0d0,4.0d0,2,2,5.0d0-epsilon))/(2*epsilon)
     print *, "There is an error in sto-int.f"
 end if

 if (abs(sSTOCoulIntGrad(4.0d0,4.0d0,3,4,5.0d0) -
&     (sSTOCoulInt(4.0d0,4.0d0,3,4,5.0d0+epsilon)
&     -sSTOCoulInt(4.0d0,4.0d0,3,4,5.0d0-epsilon))
&     /(2*epsilon)).lt.epsilon) then
    print *, "Coulomb gradient correct"
 else
    print *, "FATAL ERROR: Coulomb gradients don't match"
    print *, "sSTOCoulIntGrad(4.0, 4.0, 3, 4, 5.0)"
    print *, "Analytic = ", sSTOCoulIntGrad(4.0d0,4.0d0,3,4,5.0d0)
    print *, "Numerical= ",
&         (sSTOCoulInt(4.0d0,4.0d0,3,4,5.0d0+epsilon)
&         -sSTOCoulInt(4.0d0,4.0d0,3,4,5.0d0-epsilon))/(2*epsilon)
     print *, "There is an error in sto-int.f"
 end if

 if (abs(sSTOCoulIntGrad(1.0d0,2.0d0,3,4,5.0d0) -
&     (sSTOCoulInt(1.0d0,2.0d0,3,4,5.0d0+epsilon)
&     -sSTOCoulInt(1.0d0,2.0d0,3,4,5.0d0-epsilon))
&     /(2*epsilon)).lt.epsilon) then
    print *, "Coulomb gradient correct"
 else
    print *, "FATAL ERROR: Coulomb gradients don't match"
    print *, "sSTOCoulIntGrad(1.0, 2.0, 3, 4, 5.0)"
    print *, "Analytic = ", sSTOCoulIntGrad(1.0d0,2.0d0,3,4,5.0d0)
    print *, "Numerical= ",
&         (sSTOCoulInt(1.0d0,2.0d0,3,4,5.0d0+epsilon)
&         -sSTOCoulInt(1.0d0,2.0d0,3,4,5.0d0-epsilon))/(2*epsilon)
     print *, "There is an error in sto-int.f"
 end if

 print *, "Testing gradients for Gaussian orbitals"

 if (abs(sGTOOvIntGrad(1.0d0,1.0d0,2.0d0) -
&     (sGTOOvInt(1.0d0,1.0d0,2.0d0+epsilon)
&     -sGTOOvInt(1.0d0,1.0d0,2.0d0-epsilon))
&     /(2*epsilon)).lt.epsilon) then
    print *, "Overlap gradient correct"
 else
    print *, "FATAL ERROR: Overlap gradients don't match"
    stop
 end if

 if (abs(sGTOCoulIntGrad(1.0d0,1.0d0,2.0d0) -
&     (sGTOCoulInt(1.0d0,1.0d0,2.0d0+epsilon)
&     -sGTOCoulInt(1.0d0,1.0d0,2.0d0-epsilon))
&     /(2*epsilon)).lt.epsilon) then
    print *, "Coulomb gradient correct"
 else
    print *, "FATAL ERROR: Coulomb gradients don't match"
    print *, sGTOCoulIntGrad(1.0d0, 1.0d0, 2.0d0)
```

```
    print *, (sGTOCoulInt(1.0d0,1.0d0,2.0d0+epsilon)
&     -sGTOCoulInt(1.0d0,1.0d0,2.0d0-epsilon))
&     /(2*epsilon)
    stop
 end if

 print *, "forces for NaCl"
 print *, "Numerical forces"
 call DoGradientsByFiniteDifference(Mol1)
 do i = 1,2
   print *, Mol1%EGradient(i,:)
 end do
 print *, "Analytic forces"
 call DoGradientsAnalytically(Mol1)
 do i = 1,2
   print *, Mol1%EGradient(i,:)
 end do
 print *, "forces for water"
 print *, "Numerical forces"
 call DoGradientsByFiniteDifference(Mol2)
 do i = 1,3
   print *, Mol2%EGradient(i,:)
 end do
 print *, "Analytic forces"
 call DoGradientsAnalytically(Mol2)
 do i = 1,3
   print *, Mol2%EGradient(i,:)
 end do
 end program test
```

## Author's biography

Jiahao Chen was born September 1, 1981 in Singapore. In April 2000, he received a Singapore Government Scholarship from the Public Service Commission of Singapore to pursue an undergraduate degree in chemistry at the University of Illinois at Urbana-Champaign. Jiahao graduated *magna cum laude* in May 2002, receiving a BS in Chemistry with Highest Departmental Distinction.

After receiving his bachelor's degree, Jiahao returned to Singapore where as part of his military service, he spent the next two years at the Applied Materials Laboratory at DSO National Laboratories where he synthesized and characterized organic and inorganic materials for nonlinear optical testing purposes. He was also instrumental in developing an automated multispectral materials characterization system using tunable optical parametric oscillator lasers. Jiahao also contributed to projects in the Applied Physics Laboratory at DSO with numerical simulations of quantum computation devices, as well as remote sensing using infrared lasers.

Having completed his military service, Jiahao returned to the University of Illinois at Urbana-Champaign in August 2004 to begin his PhD in chemical physics with computational science and engineering option. In May 2008, Jiahao received a MS in Applied Mathematics with Applications to the Sciences. Jiahao is grateful to have received assistantships from the Department of Chemistry for funding his graduate studies, as well as the 2004 Roger Adams Predoctoral Fellow, 2006 Robert F. Carr Predoctoral Fellow, and 2007 Lester E. & Kathleen A. Coleman Predoctoral Fellow.